# The 80-year development of Vietnam mathematical research: Preliminary insights from the SciMath database on mathematicians, their works and their networks

Ngo Bao Chau

Vuong Quan Hoang

La Viet Phuong

Le Tuan Hoa

Le Minh Ha

Trinh Thi Thuy Giang

Pham Hung Hiep

Nguyen Thanh Thanh Huyen

Nguyen Thanh Dung

Nguyen Thi Linh

Tran Trung

Nguyen Minh Hoang

Ho Manh Toan

# The 80-year development of Vietnam mathematical research: Preliminary insights from the SciMath database on mathematicians, their works and their networks

Ngo Bao Chau, Vuong Quan Hoang, La Viet Phuong, Le Tuan Hoa, Le Minh Ha, Trinh Thi Thuy Giang, Pham Hung Hiep, Nguyen Thanh Thanh Huyen, Nguyen Thanh Dung, Nguyen Thi Linh, Tran Trung, Nguyen Minh Hoang, Ho Manh Toan

## Abstract:

Starting with the first international publication of Le Van Thiem (Lê Văn Thiêm) in 1947, modern mathematics in Vietnam is a longstanding research field. However, what is known about its development usually comes from discrete essays such as anecdotes or interviews of renowned mathematicians. We introduce SciMath—a database on publications of Vietnamese mathematicians. To ensure this database covers as many publications as possible, data entries are manually collected from scientists' publication records, journals' websites, universities, and research institutions. Collected data went through various verification steps to ensure data quality and minimize errors. At the time of this report, the database covered 8372 publications, profiles of 1566 Vietnamese, and 1492 foreign authors since 1947. We found a growing capability in mathematics research in Vietnam in various aspects: scientific output, publications on influential journals, or collaboration. The database and preliminary results were presented to the Scientific Council of Vietnam Institute for Advanced Study in Mathematics (VIASM) on November 13th, 2020.

## The 80-year development of Vietnam mathematical research: Preliminary insights from the SciMath database on mathematicians, their works and their networks

### 1. An overview of the SciMath database

The project ***A Database of Vietnam Mathematics*** was initiated in 2019 by the Vietnam Institute for Advanced Study in Mathematics (VIASM), under the leadership of Prof. Ngo Bao Chau (Ngô Bảo Châu) and VIASM Executive Director Le Minh Ha, following a long-term discussion between Prof. Ngo Bao Chau and Dr. Vuong Quan Hoang (Phenikaa University, and AI for Social Data Lab – AISDL). In August 2019, the SciMath database was constructed. The website for data input and storage (accessed here: http://SciMath.aisdl.com) was later launched and tested on 16th December 2019. The first data lines were officially inputted into the database on 23rd December 2019.

At the beginning of 2020, the project officially started. Although the project underwent a hard time due to the COVID-19, the team continued working during the social distancing period to ensure the project stay on track. The data team and representatives from VIASM had two meetings in September and October 2020 to discuss the project's progress and amendments to improve the database system for future use. Pictures of the meetings are in Appendix 1.

For being inputted into the SciMath database, a publication must meet the following two criteria:

- At least one author of the publication is Vietnamese.
- The publication must be indexed by at least one of these databases: MathSciNet, zbMATH, Scopus, or Web of Science.

#### *1.1. The structure of the SciMath database*

There are three major components in the SciMath system (data curation, data analysis, and data presentation), which are constructed and set up for different administrative roles:

- The data curation system includes tools that automatically collect data of publications from scientific literature crawlers, tools for data inputting, editing, filtering, and verification.
- The data analysis system includes tools for searching, extracting, and producing results for statistical reports.
- The data presentation system consists of a website for public use and the Application Programming Interface (API) that provides statistical information to authorized agencies or organizations.

Currently, the curation system is being operated to ensure the data collection process progresses according to the projected timeline. The other two systems are in the construction and testing phase.

This is the master database that stores all information related to authors, affiliations, publications, sources, etc., which meet the criteria for being included in the SciMath database.

The data are gathered from various sources, such as scientific databases (Google Scholar, zbMATH, or MathSciNet), records contributed by specialists, and records automatically extracted from online databases, then edited, cleaned, and stored. The database structure is illustrated in Figure 1.

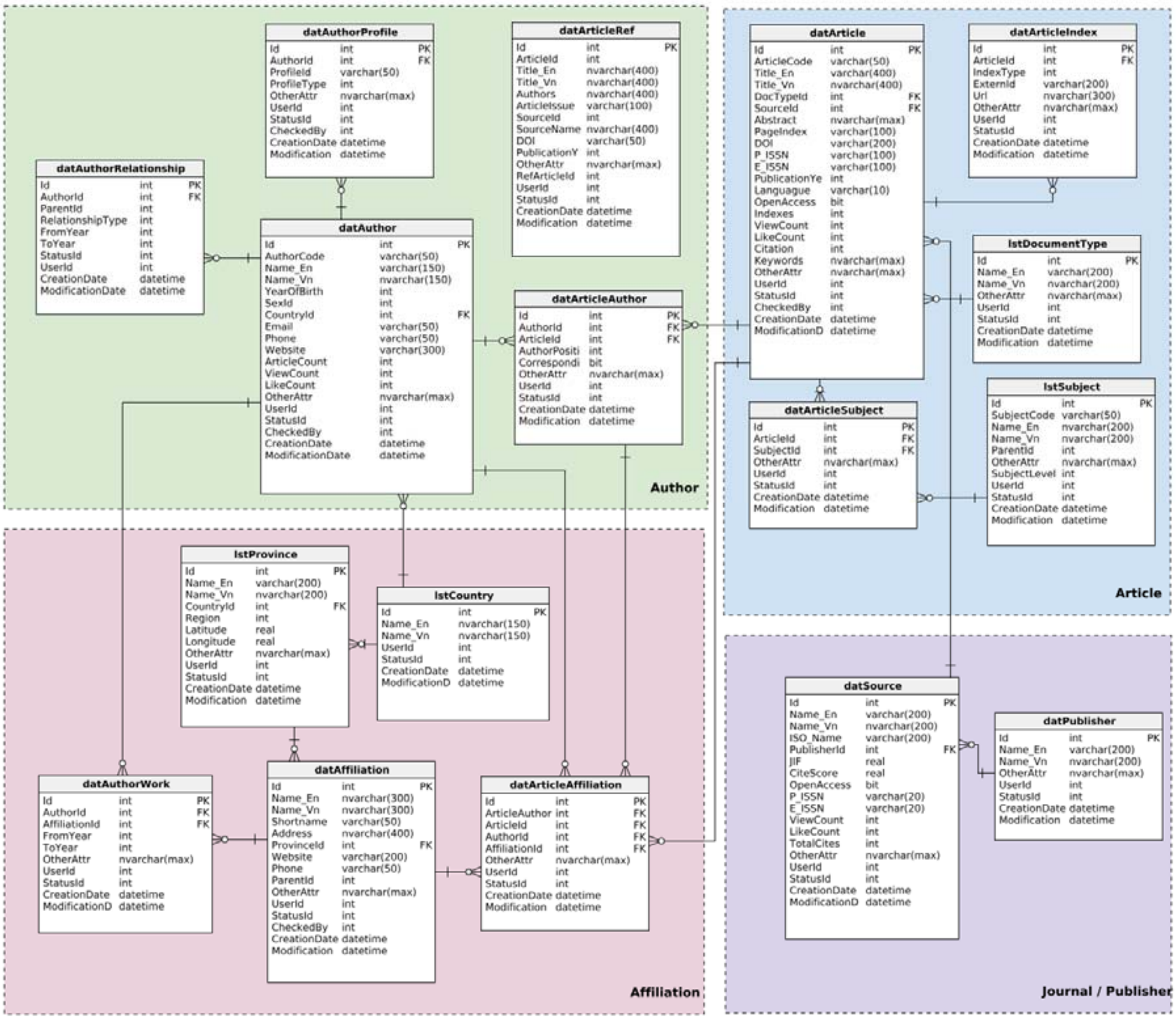

**Figure 1. The structure of the SciMath database**

The SciMath database is constructed with four primary groups of data:

- **Article**. This group includes the following fields: title, types of publications (journal article, book, proceedings), year of publication, source, DOI/link, abstract, the keyword(s), subject(s), author(s), and their affiliation(s). Among those fields, title, year of publication, and source must be filled during the data inputting phase.
- **Author**. This group includes the following fields: author's name, gender, nationality, affiliation, and other information such as website, phone number, and email. Each author is assigned to an ID based on their nationality, gender, and order of entry. For example, an author who is a Vietnamese woman and is the 999$^{th}$ author recorded into the database is given ID as vf.999 (v: Vietnamese; f: female; 999: the 999$^{th}$ author in the database). Similarly, the 800$^{th}$ author in the database, a foreign male, is given the ID as fm.800 (f: foreign; m: male; 800: the 800$^{th}$ author in the database). Data collectors use information, such as website, email, and phone number, to distinguish authors who have similar names or whose names are written in abbreviation.
- **Affiliation**. This group includes information about the organization where an author works
- **Source/ Publisher**. This group includes information about the journal/ book/ proceeding and publisher of the article.

For an article to be officially recorded into the database, its information corresponding to all four groups must be filled. Among the four groups, the **Article** group is the central one around which the information in other groups is collected, constructed, and connected.

### *1.2. Procedure*

<u>a) Data collection phases</u>

The data collection process consists of 2 phases.

*Phase 1: Input data from Reference Data storage*

All data recorded at this phase are based on 4500 data lines, which were automatically collected and stored in Reference Data storage. Every publication's information is verified using materials from online sources (e.g., publisher's website, Google Scholar, MathSciNet, zbMATH, etc.) before entering the database.

During this phase, publications recorded are only those published before 2009, thanks to the great manual data effort by Prof. Le Tuan Hoa (Hanoi-based Institute of Mathematics, part of Vietnam Academy of Science and Technology).

*Phase 2: Manual search*

Data collectors manually search for author profiles and their publications by accessing universities and mathematics research institutions' websites. For ensuring the database can cover as many publications of Vietnamese mathematicians as possible, online databases (Google Scholar, zbMATH, MathSciNet, Scopus, etc.), specialists' references, author blogs/websites, publications' reference lists are utilized. All sites are stored in a computer worksheet (Excel) for the next verification step.

Moreover, data collectors also use information gathered from authors whose profiles have already been recorded in the database since the first phase to update their publication records.

Data collectors prioritize information that is visible and convenient to be updated at the current stage, so the database has not yet covered all eligible publications.

b) Procedure

The data collection procedure is illustrated in the following figure (see Figure 2):

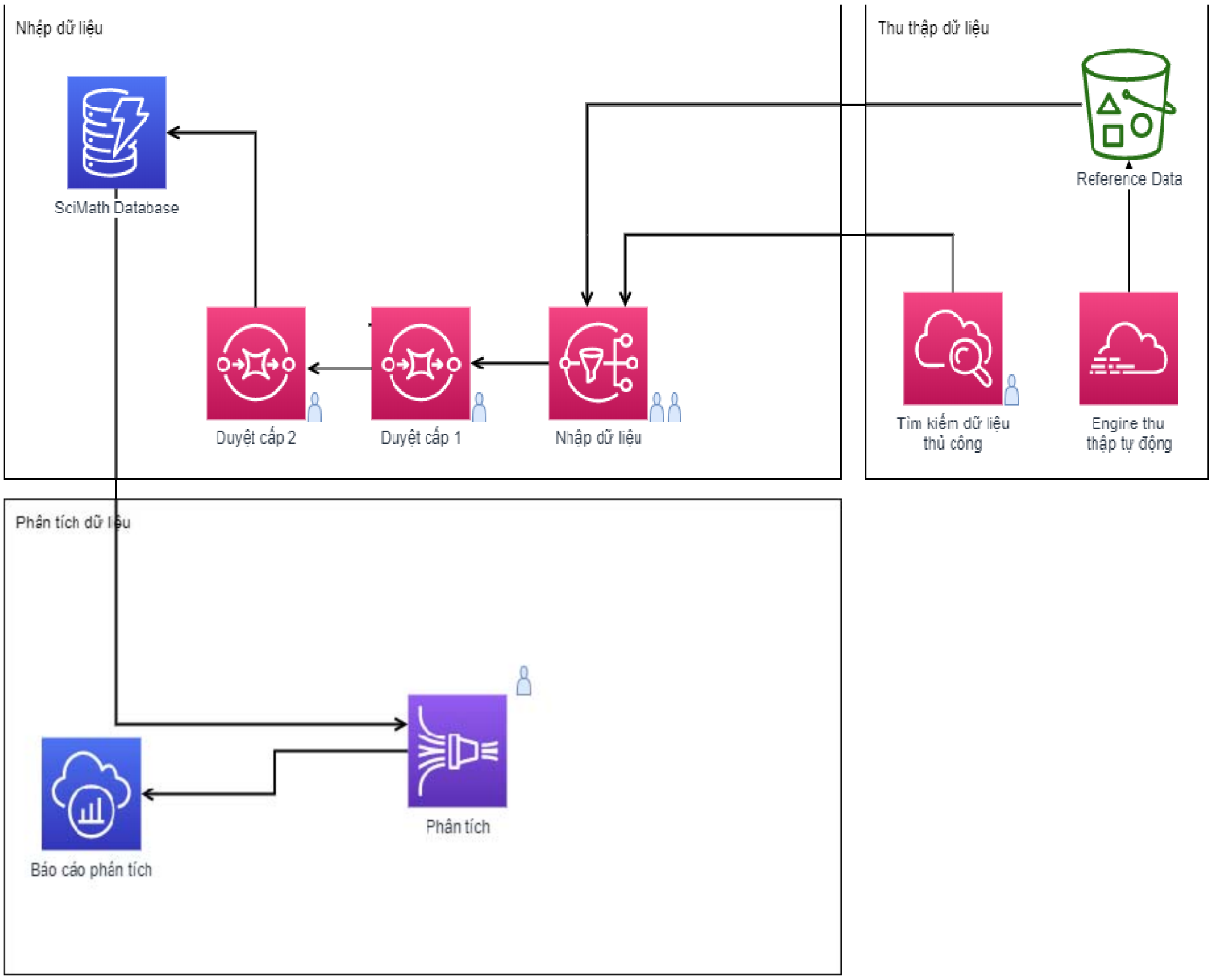

**Figure 2. Data collection procedure**

*Step 1: Data collection*

- Data about publications of Vietnamese mathematicians are automatically collected from zbMATH, MathSciNet, and Google Scholar and stored in Reference Data storage.
- Data are manually searched from the publications' reference lists, websites of universities and research institutions, and online databases (MathSciNet, Google Scholar, zbMATH, Scopus, etc.)

*Step 2: Data input*

- Data collectors check and input data into the database.
- Data collectors create and update the author's profile (publication, affiliation, and other information).

*Step 3: Data verification:*

- The database employs a two-level data verification system (level 1: data collectors check before inputting; level 2: verifying after the data are inputted).
- Data collectors check the eligibility of data.
- Data collectors check whether there are duplicates among publications and authors.
- Verified data will be recorded in the database.

*Step 4: Data analysis*

- Tables and charts can be generated using the recorded data and extracted for scientific purposes.

Although the procedure looks complicated, it serves two primary purposes: (i) to collect as many authors' profiles as possible; (ii) to ensure data accuracy. The two-level data verification process serves the second purpose. Before inputting data, data collectors must check whether they have already existed in the database. At level 2, the inputted data go through an automatic and a manual check. For example, a new title is always checked by the SciMath semi-automatic system when being inputted. This system will give the data collector a warning message if the title they have just inputted resembles at least 90% of an existing one. The second step is manually conducted. Data collectors will cross-check information collected by each other and delete/merge/correct them when duplicates or false/ineligible information is found. For example, because duplicates might be inputted due to mathematical symbols in the article's title, manual check allows the data team to search for articles using non-mathematical words. Hence, the data team can correct them quickly if any mistake is found. The same procedure is implemented in other data groups.

## 2. Preliminary results

Data used in this report are collected from 23rd December 2019 to 5th November 2020. The database recorded 8,372 publications, 3,058 authors (including 1,566 Vietnamese and 1,492 foreign authors), and 1,267 journals/books/proceedings. Notably, the results were extracted from the database on 5th November 2020, when the database was still in progress, so there might be a certain number of publications not recorded.

### *2.1. Scientific publications*

In general, the number of scientific publications before the 1960s was relatively low. From 1947 to 1960, there were only nine articles written by three authors. Among these articles, five articles were written by Le Van Thiem before and during 1950. Le Van Thiem also published the first article in 1947 (Le, 1947). This article was written in German and published in *Commentarii Mathematici Helvetici* of the Swiss Mathematical Society (see Figure 4). Besides English, mathematics articles before the 2000s were published in many different languages, such as German, French, and Russian.

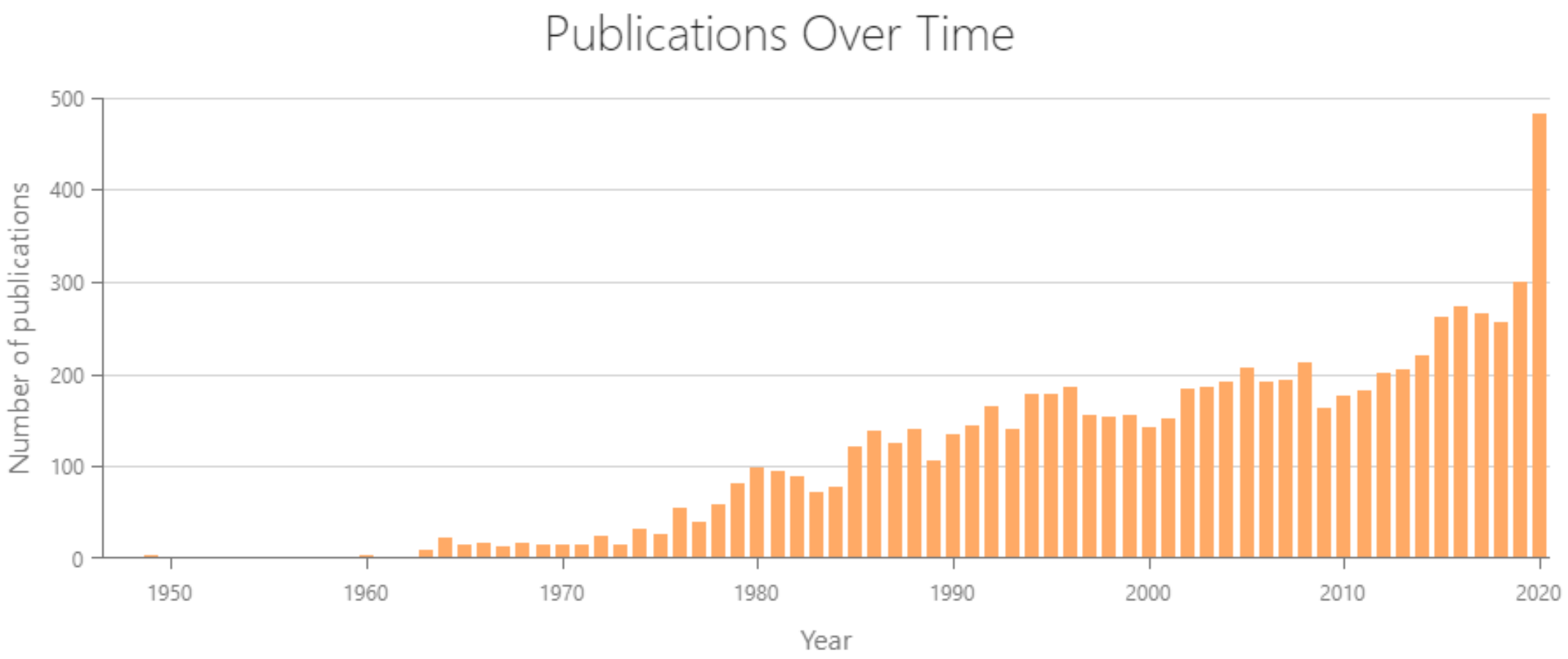

**Figure 3. The annual number of publications**

Other four articles were published in the late 1950s and early 60s by Dang Dinh Ang and Nguyen Dinh Ngoc. Since 1970, the Mathematics had witnessed rapid and steady growth in the number of scientific publications. The output during the 2008-2010 period was lower than what we expected. The low scientific production during this period might be due to the database's incompleteness. Or it might be a noteworthy observation, which will require further effort to understand fully.

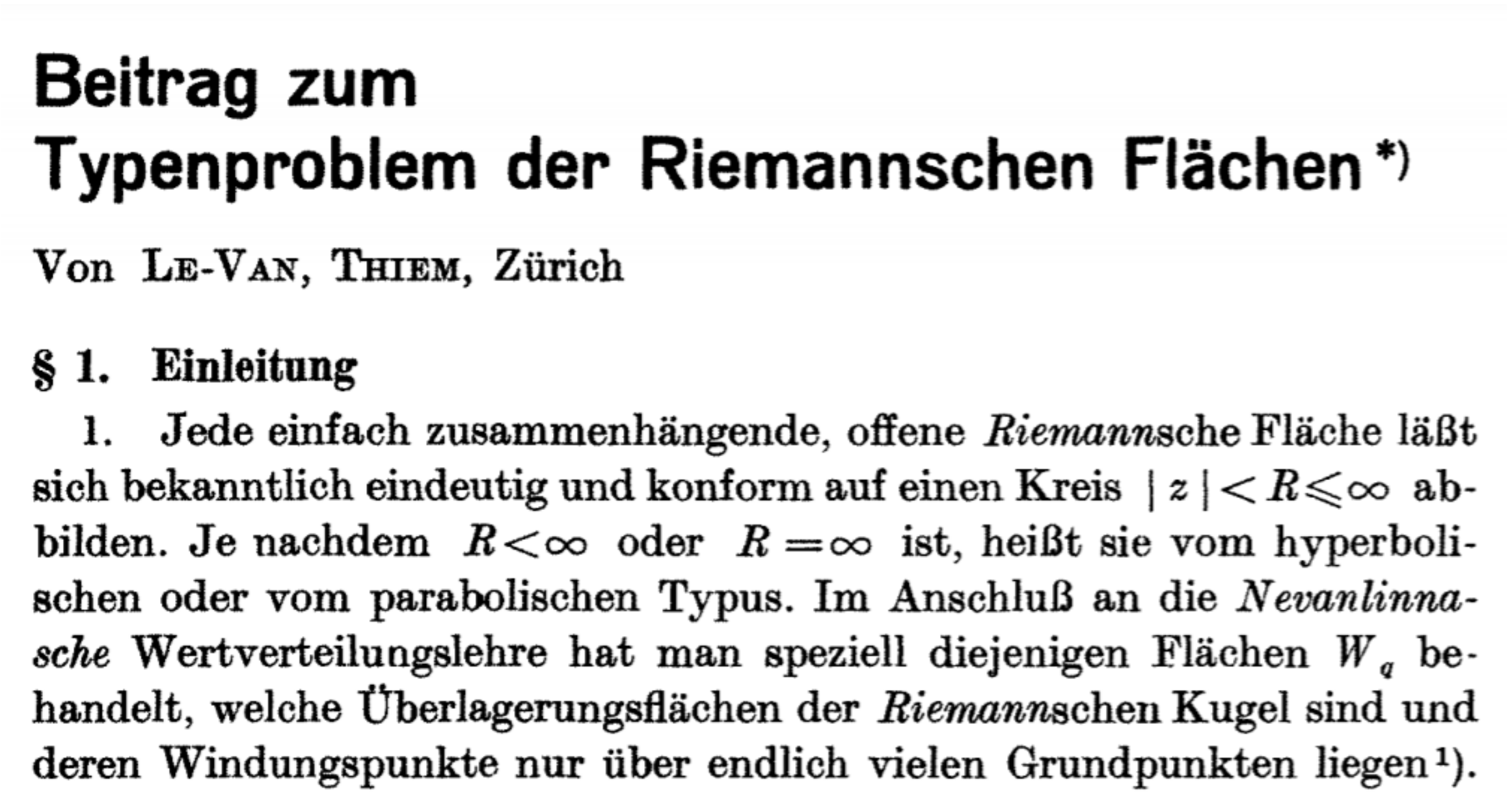
**Beitrag zum Typenproblem der Riemannschen Flächen *)**

Von LE-VAN, THIEM, Zürich

**§ 1. Einleitung**

1. Jede einfach zusammenhängende, offene *Riemann*sche Fläche läßt sich bekanntlich eindeutig und konform auf einen Kreis $|z| < R \leqslant \infty$ abbilden. Je nachdem $R < \infty$ oder $R = \infty$ ist, heißt sie vom hyperbolischen oder vom parabolischen Typus. Im Anschluß an die *Nevanlinna*sche Wertverteilungslehre hat man speziell diejenigen Flächen $W_q$ behandelt, welche Überlagerungsflächen der *Riemann*schen Kugel sind und deren Windungspunkte nur über endlich vielen Grundpunkten liegen[1]).

**Figure 4. The first international journal article by a Vietnamese author, Le Van Thiem (Le, 1947)**

Regarding types of publications, a majority of documents were published in the form of a journal article. Journal article was also the type of publication that grew steadily until 2019. In 2020, there were 483 publications recorded, whereas less than 300 articles per year were recorded in the previous years. One possibility to explain this sudden increase is that a significant number of past articles have not been recorded in previous decades. Nonetheless, VIASM experts have discussed and concluded that this is hardly the case. Therefore, the spurt is more likely attributed to a surge in the number of new authors desperately seeking to publish in “just good enough journals.”

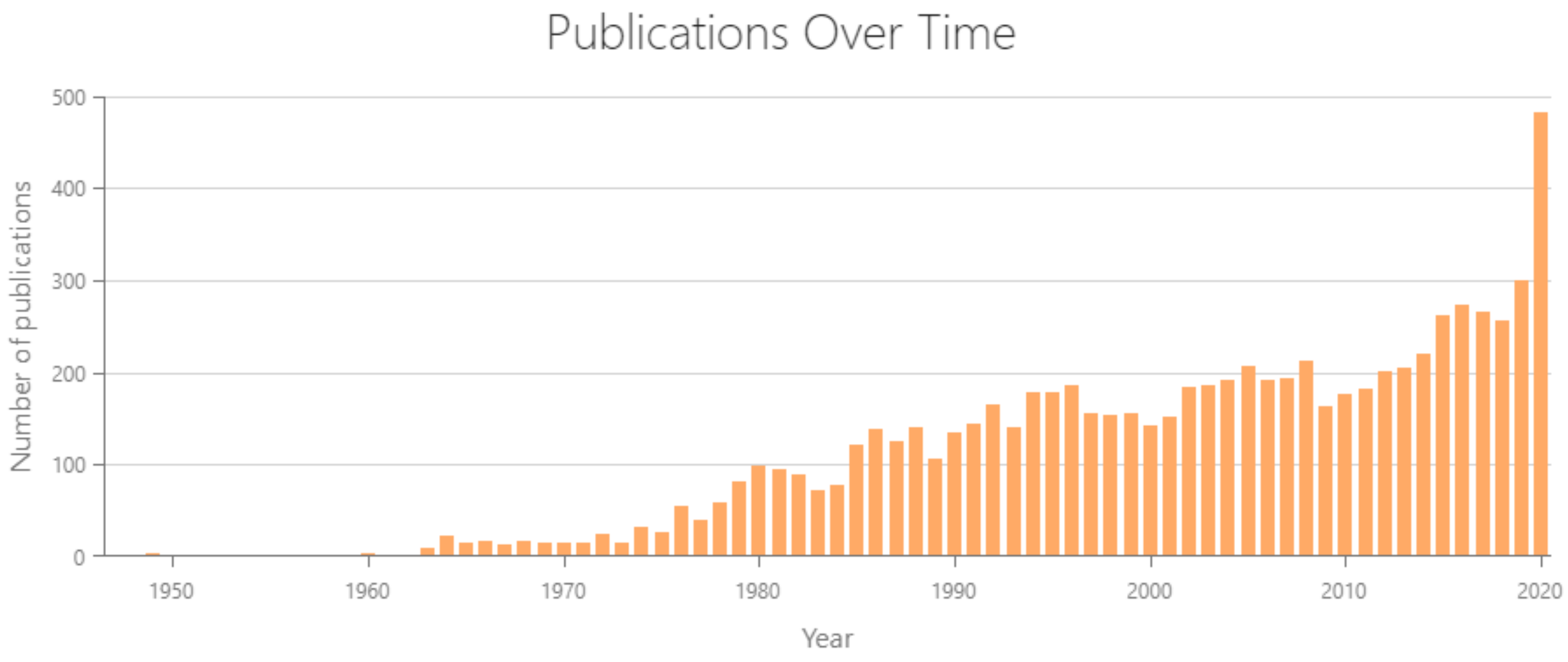

**Figure 5. The annual number of journal articles**

Regarding books, Vietnamese authors contributed both books and book chapters. By the time of extracting data, there were 201 books/book chapters recorded. The number of books/book chapters was not stable throughout the years (see Figure 6). The 2007-2013 period was the most productive period, with at least six books/book chapters published annually. The first book was "*Lectures on the theory of symmetry of elementary particles. Part I. Theory of groups. Compact symmetry groups,*" written by Nguyen Van Hieu (Nguyễn Văn Hiệu) in 1966, while working at Dubna Nuclear Research Institute, the former Soviet Union. Besides, there were two books published in Vietnamese, namely *"Derivative Equations"* and *"Numerical Analysis."* Both were edited by Nguyen Minh Chuong (Institute of Mathematics – VAST) and recorded on zbMATH. Because the total number of books/book chapters was low (the annual book/book chapter in several years was only one), we anticipated that the actual number would likely be significantly higher than what was reported.

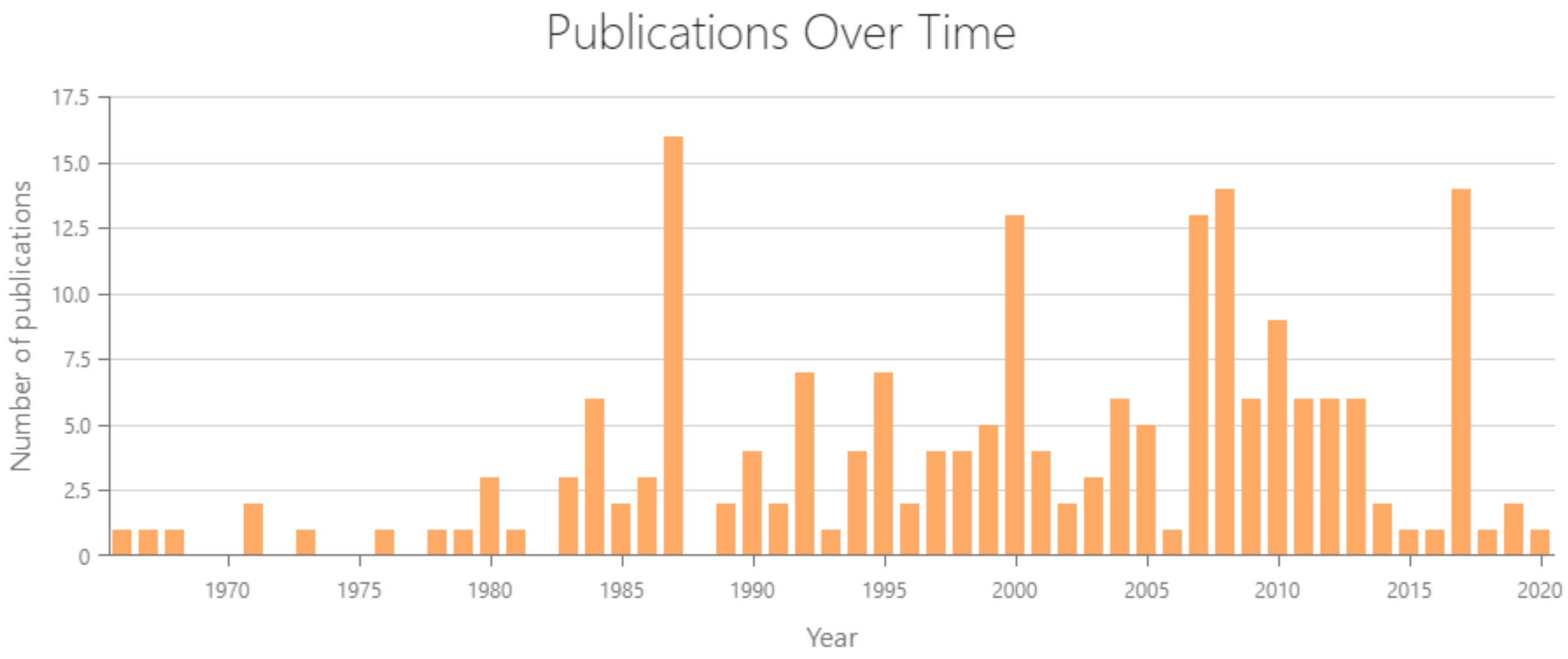

**Figure 6. The annual number of books/book chapters**

Until 5th November 2020, the database had recorded 368 proceeding papers. The number of proceeding papers had not changed much over the years, with an average of ten proceeding papers published annually. The number of proceeding papers also surged in several years (30 proceedings and more in 2005 and 1995, respectively). Overall, the number of proceeding paper was relatively stable.

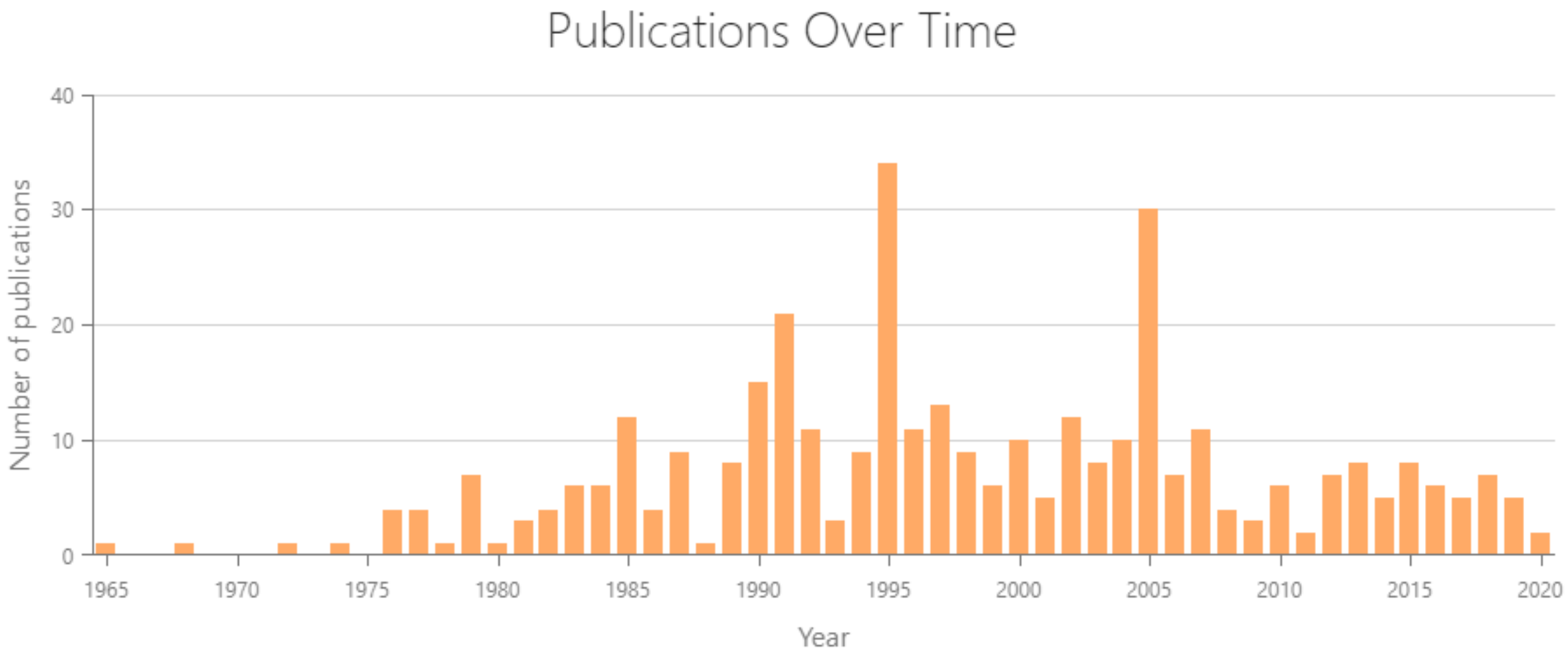

**Figure 7. The annual number of proceeding paper**

### *2.2. Sources (journal, book, and proceedings)*

The database had recorded 1,267 sources, among which there were 738 journals, 148 books, and 281 conference proceedings. The top three journals with the highest numbers of publications by Vietnamese authors are *Acta Mathematica Vietnamica*, *Vietnam Journal of Mathematics*, and *Journal of Optimization Theory and Applications* (see Table 1).

From the data given in Table 1, *Acta Mathematica Vietnamica* and *Vietnam Journal of Mathematics* published more articles than the remaining journals in the top 10. The top two journals' dominance might be attributed to the fact that they were operated by Vietnamese institutions, which made them widely known by Vietnamese authors due to peers or mentors' introduction or efficiently disseminating calls for papers among the national mathematics researchers.

**Table 1. Top 10 most relevant journals**

| No | Journal | Articles | Scimago Quartile (2019) |
|---|---|---|---|
| 1 | *Acta Mathematica Vietnamica* | 838 | Q3 |
| 2 | *Vietnam Journal of Mathematics* | 695 | Q3 |
| 3 | *Journal of Optimization Theory and Applications* | 158 | Q1/Q2 |
| 4 | *Journal of Mathematical Analysis and Applications* | 157 | Q1/Q2 |
| 5 | *Journal of Algebra* | 151 | Q1 |
| 6 | *Optimization* | 135 | Q1/Q2 |
| 7 | *Proceedings of the American Mathematical Society* | 113 | Q1 |
| 8 | *Nonlinear Analysis* | 108 | Q1 |
| 9 | *Ukrainian Mathematical Journal* | 91 | Q2 |
| 10 | *Journal of Global Optimization* | 90 | Q1/Q2 |

Vietnamese authors had also appeared in many prestigious Mathematics journals. Table 2 shows several journals with the highest Mathematical Citation Quotient (MCQ) in 2019 and their number of articles by Vietnamese authors.

**Table 2. Number of articles on some prestigious journals**

| No | Journal | Articles | MCQ 2019 |
|---|---|---|---|
| 1 | *Annals of Mathematics* | 13 | 5.24 (97% cited) |
| 2 | *Acta Mathematica* | 6 | 3.97 (98% cited) |
| 3 | *Annals of PDE* | 4 | 2.98 (100% cited) |
| 4 | *Journal of the American Mathematical Society* | 2 | 5.14 (99% cited) |
| 5 | *Forum of Mathematics, Pi* | 2 | 4.48 (100% cited) |

For a better understanding of Vietnamese authors' research capability, we display the annual number of publications by Vietnamese authors in the ten most influential journals according to the University of Copenhagen's ranking (see Figure 8). Among 8,372 recorded publications, only 97 publications (approximately 1% of the total records) were published in those ten journals. The annual number of publications had increased dramatically in many journals since 2010, which shows Vietnamese authors' tireless efforts to reach the top. Moreover, the substantial rise might be the result of Vietnam's two important mathematics events, both taking place in 2010.

The first event is when Prof. Ngo Bao Chau (Ngô Bảo Châu) became the first Vietnamese to be awarded the world's most prestigious prize in mathematics, Fields Medal, in 2010. The success story of Ngo Bao Chau was a tremendous inspiration for Vietnamese mathematicians to try harder to reach the top. The second event was the launch of the National Program for the Development of Mathematics, which created an environment supporting Vietnam's mathematical development on a national scale. (Details of the published works and authors are presented in Tables A, B, and C in Appendix 2.)

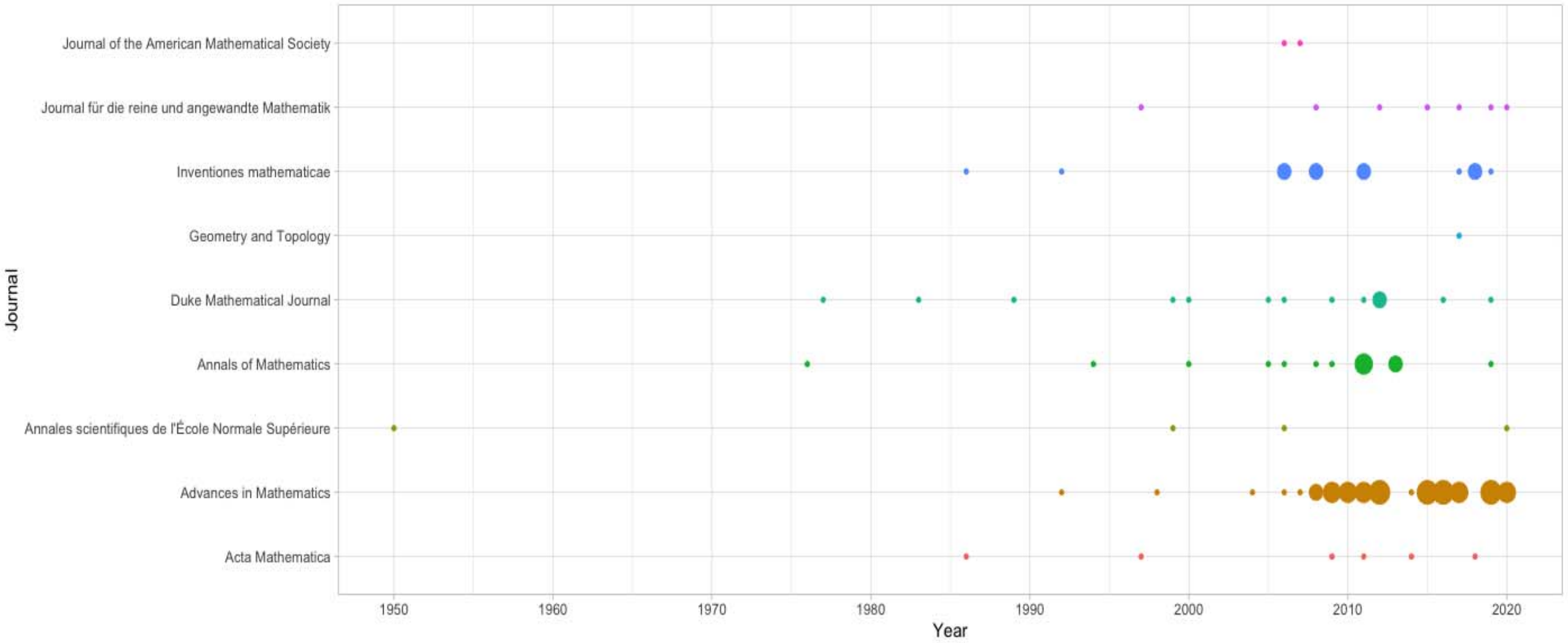

**Figure 8. Number of articles on top journals over time**

In addition, there is also a change of origin and language of journals over time. This is largely due to the change in countries where authors earn a degree, collaborate as well as a new global trend. Early generation mathematicians were usually trained and worked in the former Soviet Union; they could speak Russian fluently thus published many articles in Russian journals such as *Differentsial'nye Uravneniya* (42 articles), *Vestnik Moskovskogo Universiteta. Seriya 1. Matematika. Mekhanika* (30 articles). Publications are also available in French, German and English. Meanwhile, it has been rare to see articles in languages other than English in recent years, and there is an absence of Russian journals.

### *2.3. Authors*

There are 3058 authors, of which 1566 are Vietnamese authors and 1492 foreign authors in the SciMath database. The authors analysis showed two remarkable results. First, there is an increasing trend in the number of post authors and new authors over the years. Secondly, the data show an improvement in Vietnamese author research ability through leading and independently conducting research.

Figure 9 shows the number of Vietnamese authors having published research by year. According to this, we can see a growing trend of mathematical research in Vietnam.

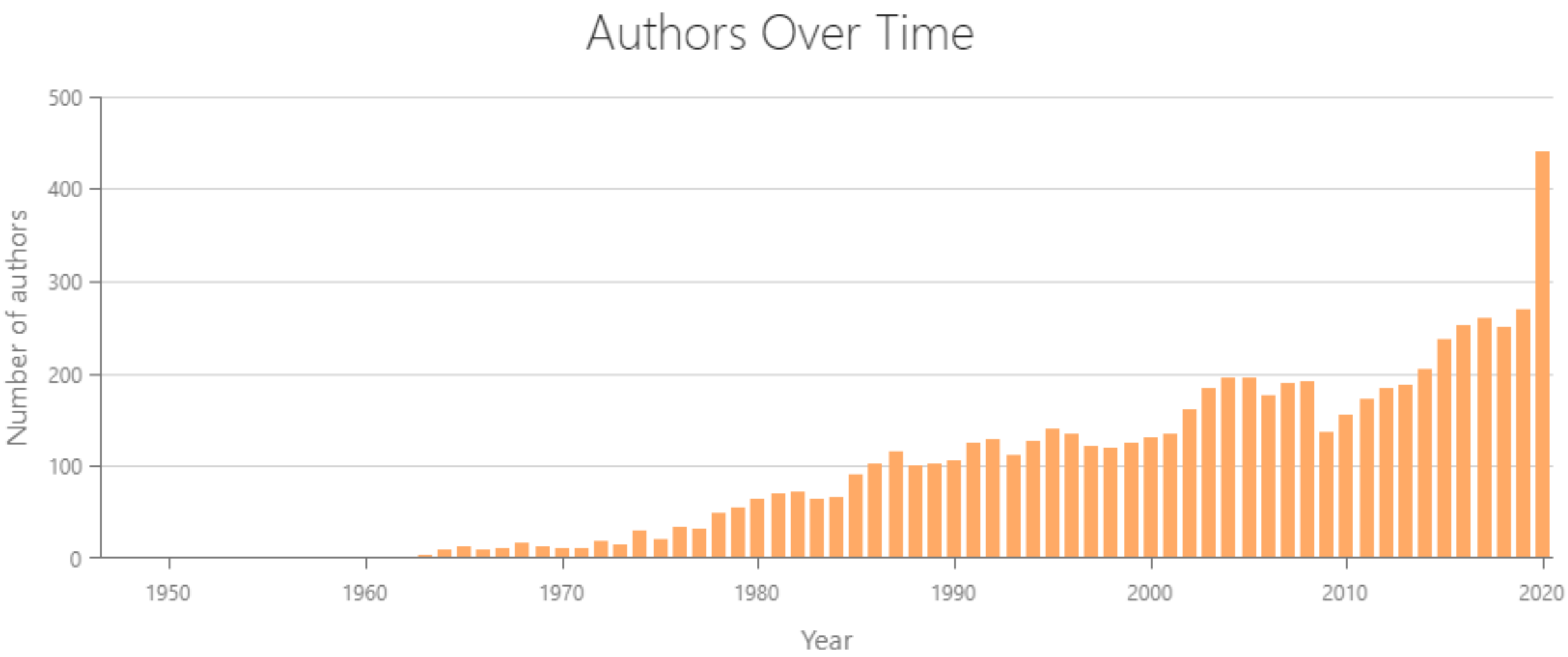

**Figure 9. Number of Vietnamese authors by year**

About publishing productivity by gender, it is found that men are the major contributor to mathematical research since they contribute to the majority of the annual publications. However, the remarkable increase in the number of articles by female authors is also a sign that their research capability is improving over the years.

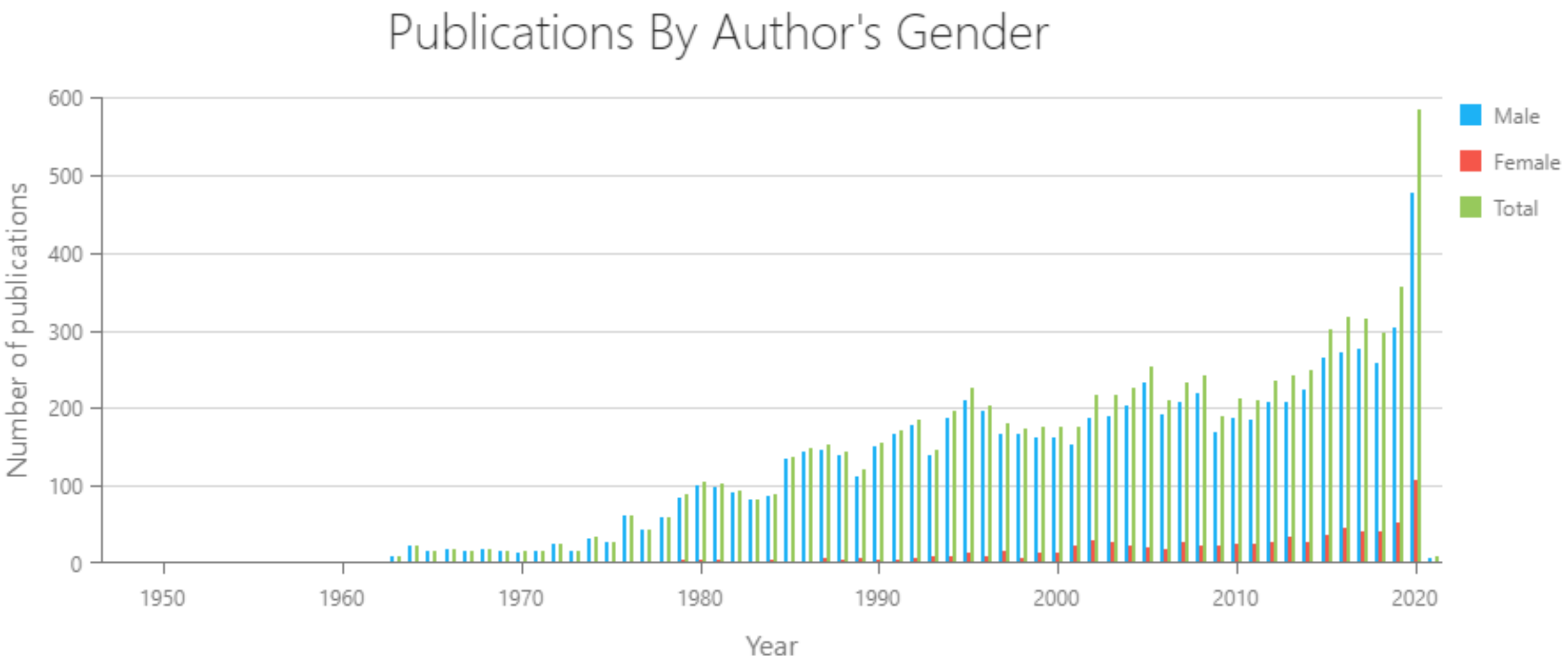

**Figure 10. Number of articles by gender**

Figure 11 shows the number of new authors[1]. Although the number of new male authors' surge largely drives the increasing number of new authors, the representation of female mathematicians is improving. While the annual number of new female authors is few before 2000, the number of new female authors increased steadily in the 2000s and sharply increased after 2010 (see Figure 11). It is noteworthy that this trend only appeared in the last 20 years. However, the history of female mathematicians in Vietnam began in 1968 with the article *"Behavior of the diffraction peak for particles with arbitrary spins"* by Nguyen Thi Hong in *Nuclear Physics B* (T. H. Nguyen, 1968). This article was written while the author was working at Dubna Nuclear Research Institute, the former Soviet Union.

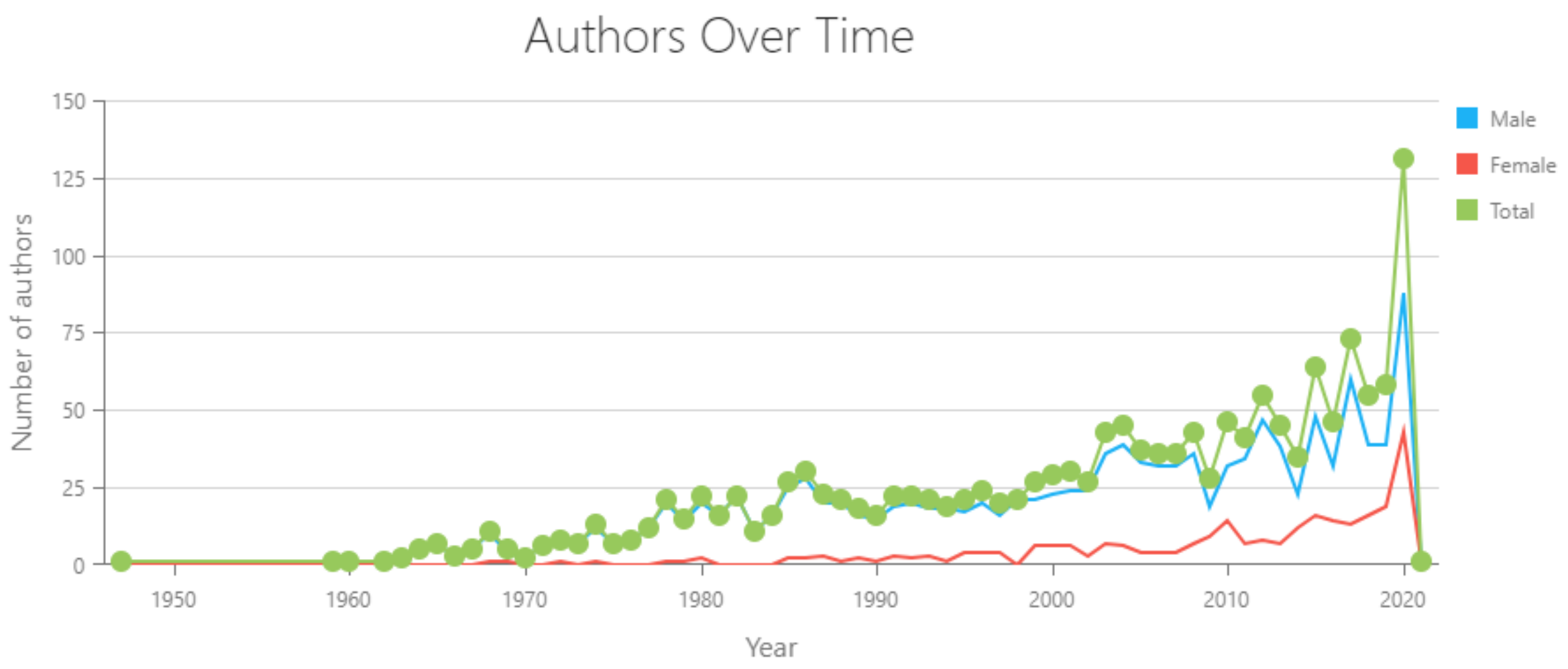

**Figure 11. Number of new authors by gender over year**

The ability to lead a research group or write solo work is another concern. Figure 12 shows that Vietnamese mathematicians tend to work solo or in small groups. We look at the total of 8372 published articles to know that a major portion of them are single-authored papers, more specifically 3400 articles. And, pairs of mathematicians wrote another 3000 articles. These numbers tell us that most works were performed by just an individual or two mathematicians in mathematics research.

[1] An author is new in the year when he/she had the first publication

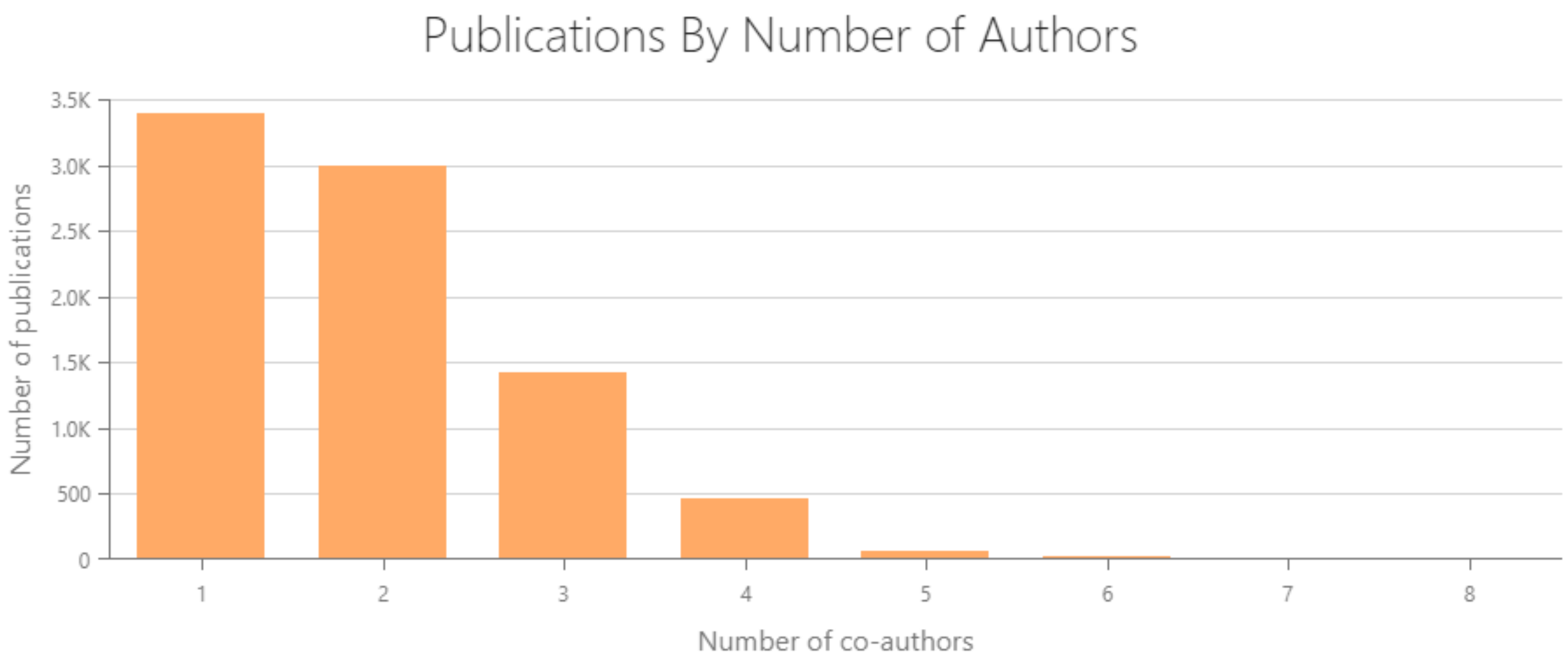

**Figure 12. Number of articles by the size of the author group**

The annual number of Vietnamese authors serving as senior authors in official publications is also an indication of an improvement in research capability. And this trend is given in Figure 13. Comparing with Figure 7 (about the annual number of Vietnamese authors), we see that the annual number of Vietnamese lead authors is about half of Vietnamese authors' total annual number.

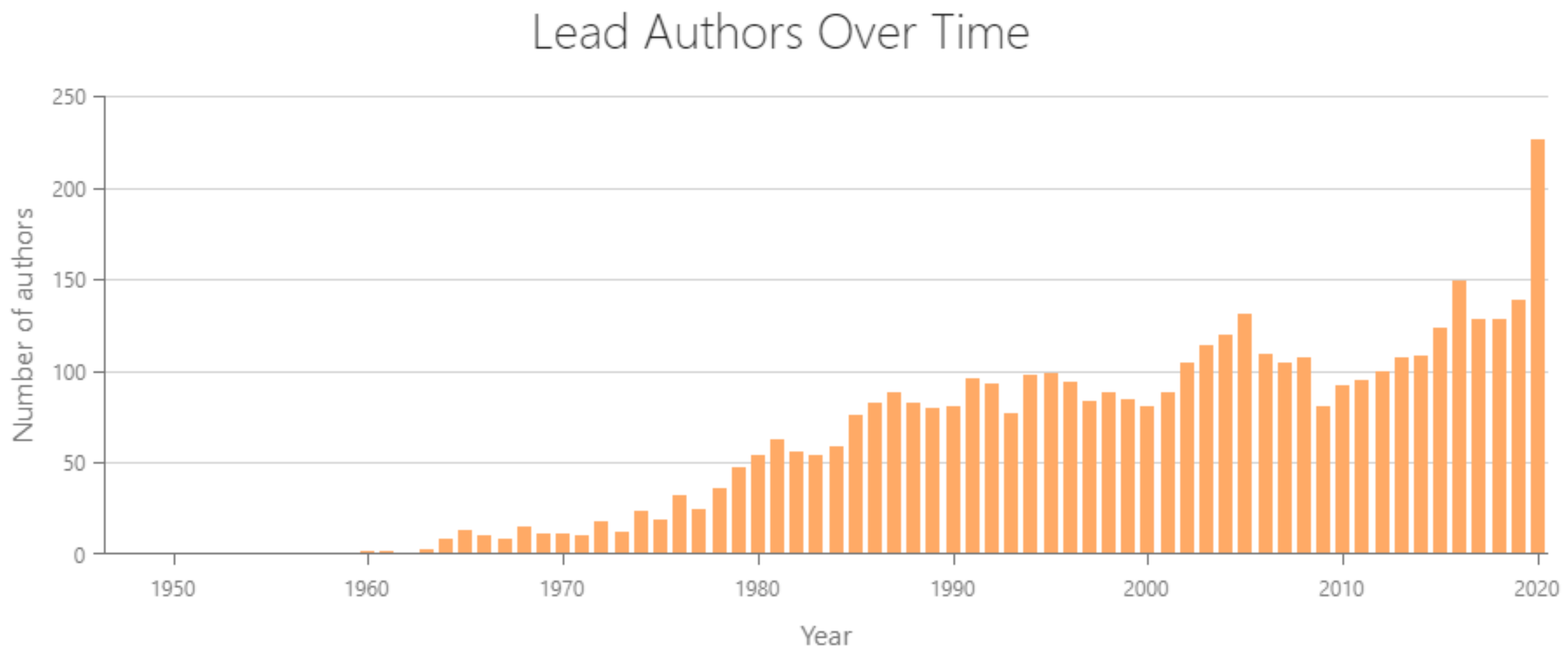

**Figure 13. Number of lead authors by year**

Mathematics is a field with a skewed distribution of research productivity. Of the 1566 Vietnamese authors, the ten most productive authors have collectively published 1232 articles, representing 14.3% of the total number of publications. The 100 most productive authors published 5261 articles (62.4%). Also, out of the 100 authors above, only four are female, with 140 articles. Table 3 shows the five most productive authors (this is a temporary list according to available data).

**Table 3. Five most productive Vietnamese**

| No | Author | Vietnamese name | Number of articles | Publication time |
|---|---|---|---|---|
| 1 | Pham Huu Tiep | Phạm Hữu Tiệp | 160 | 1986-2020 |
| 2 | Vu Ha Van | Vũ Hà Văn | 144 | 1996-2020 |
| 3 | Vu Ngoc Phat | Vũ Ngọc Phát | 144 | 1980-2020 |
| 4 | Hoang Tuy | Hoàng Tụy | 140 | 1964-2012 |
| 5 | Phan Quoc Khanh | Phan Quốc Khánh | 127 | 1979-2020 |

*2.4. Classification*

SciMath database uses the Mathematics Subject Classification System — MSC, currently used by reputable databases in the industry such as MathSciNet or zbMATH. Currently, the SciMath database records that the works of all Vietnamese mathematicians have covered 63 MSC subjects. Table 4 presents 63 subjects by the number of articles and authors.

**Table 4. Number of articles and authors by subject**

| Subjects | Number of articles | Number of authors |
|---|---|---|
| General mathematics | 3 | 7 |
| History and biography | 2 | 1 |
| Mathematical logic and foundations | 13 | 22 |
| Combinatorics | 147 | 123 |
| Order, lattices, ordered algebraic structures | 26 | 27 |
| General algebraic systems | 4 | 9 |
| Number theory | 134 | 85 |
| Field theory and polynomials | 22 | 18 |
| Commutative algebra | 309 | 183 |
| Algebraic geometry | 222 | 171 |
| Linear and multilinear algebra; matrix theory | 56 | 60 |
| Associative rings and algebras | 165 | 113 |
| Nonassociative rings and algebras | 23 | 25 |
| Category theory, homological algebra | 55 | 48 |
| $K$-theory | 12 | 8 |
| Group theory and generalizations | 300 | 168 |
| Topological groups, Lie groups | 36 | 36 |
| Real functions | 162 | 152 |
| Measure and integration | 27 | 32 |
| Functions of a complex variable | 163 | 123 |
| Potential theory | 37 | 48 |
| Several complex variables and analytic spaces | 354 | 190 |
| Special functions | 23 | 35 |
| Ordinary differential equations | 323 | 271 |
| Partial differential equations | 741 | 519 |
| Dynamical systems and ergodic theory | 116 | 107 |
| Difference and functional equations | 53 | 61 |
| Sequences, series, summability | 4 | 9 |
| Approximation and expansions | 63 | 66 |
| Harmonic analysis on Euclidean spaces | 121 | 99 |
| Abstract harmonic analysis | 11 | 18 |
| Integral transforms, operational calculus | 42 | 44 |
| Integral equations | 91 | 100 |
| Functional analysis | 246 | 200 |
| Operator theory | 472 | 380 |
| Calculus of variations and optimal control; optimization | 489 | 331 |
| Geometry | 11 | 22 |
| Convex and discrete geometry | 72 | 70 |
| Differential geometry | 76 | 71 |
| General topology | 77 | 79 |
| Algebraic topology | 67 | 57 |
| Manifolds and cell complexes | 35 | 30 |
| Global analysis, analysis on manifolds | 90 | 107 |
| Probability theory and stochastic processes | 241 | 193 |
| Statistics | 50 | 61 |
| Numerical analysis | 348 | 385 |
| Computer science | 73 | 95 |
| Mechanics of particles and systems | 8 | 10 |
| Mechanics of deformable solids | 32 | 41 |
| Fluid mechanics | 69 | 83 |
| Optics, electromagnetic theory | 43 | 40 |
| Classical thermodynamics, heat transfer | 16 | 31 |
| Quantum Theory | 44 | 26 |
| Statistical mechanics, structure of matter | 19 | 24 |
| Relativity and gravitational theory | 2 | 2 |
| Geophysics | 3 | 6 |
| Operations research, mathematical programming | 560 | 336 |
| Game theory, economics, social and behavioral sciences | 92 | 121 |
| Biology and other natural sciences | 44 | 67 |
| Systems theory; control | 191 | 144 |
| Information and communication, circuits | 18 | 24 |
| Mathematics education | 3 | 4 |

Currently, the number of articles and authors by subjects should be used only for reference. Each article's subjects are specified by each article, not determined by the author's specialization on the scientific profile. Therefore, an author can contribute to more than one subject because an article has an unlimited number of subjects. Besides, the initial classification encountered many difficulties due to the inability to access large databases such as MathSciNet or zbMATH. We can only input information specified in the PDF or be accessed for free on zbMATH. Only recently do we have access to MathSciNet to improve our ability to update this data. We will continue to add and edit the subject data to be more accurate in the coming time.

### *2.5. Collaboration*

As mentioned, the database has a total of 3058 authors, of which 1566 are Vietnamese authors and 1492 foreign authors. The small gap between the number of domestic authors and foreign authors shows that foreign authors have played a large role in Vietnamese mathematics development.

This preliminary report uses the collaborative network in the 2000-2008 period (Figure 14). Each author is a point, and a line connecting two points represents the collaboration relationship. The color of the points indicates the gender and nationality: blue for the male Vietnamese author, green for the female Vietnamese author, and yellow color for the foreign author. The size of the points depends on the number of collaborated articles.

Preliminary analysis shows that the relationship networks are widespread among the authors. This is shown in the fact that small groups are linked. Increasing collaboration strengths among small-sized groups in mathematics research may indicate many highly capable research groups in familiar subjects exist in Vietnam mathematics. In fact, many mathematicians can work independently. Typical examples can be the collaboration between the groups led by well-known mathematicians such as Hoang Xuan Phu, Nguyen Dong Yen, Nguyen Dinh, Dinh The Luc, and Phan Quoc Khanh, or among the groups of Ngo Viet Trung, Ha Huy Tai, and Le Tuan Hoa.

Some networks are independent of other networks, for example, Pham Xuan Huyen's network or Dinh Van Huynh's. It is possible that two authors worked abroad in the period 2000-2008, so their network includes only or mostly foreign authors and is separate from other groups. In addition, there are some solo authors like Le Anh Vinh.

Even in big research networks, the publication is usually done by small groups, the most common being two or three people.

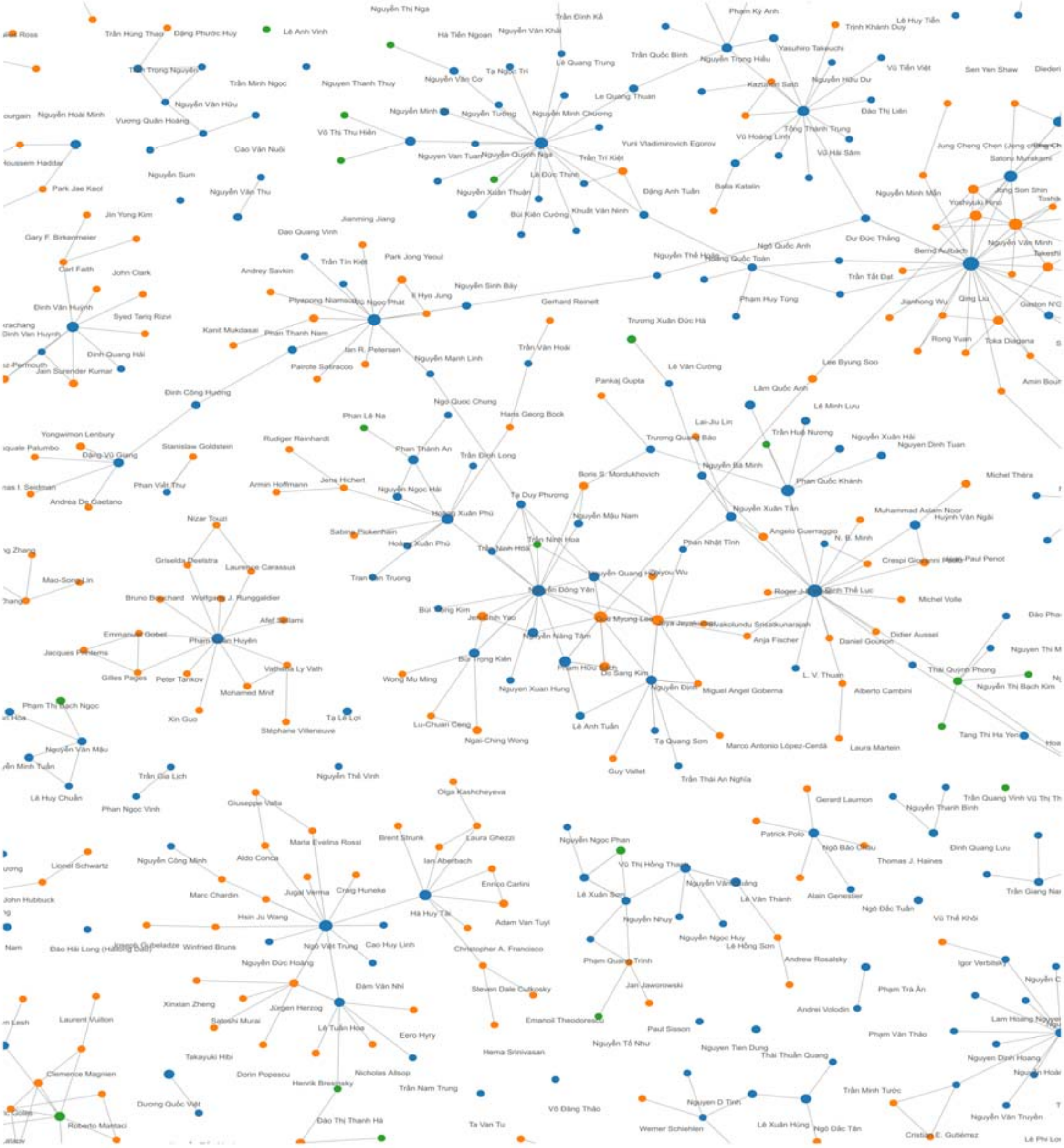

**Figure 14. A part of the collaboration network in the 2000-2008 period**

Also, we compare two networks in 1947-1960 and 1960-1970 (see Figures 13 and 14) to see the collaboration trend. Compared to the cooperation network in 2000-2008, the two periods 1947-1960 and 1960-1970 had a much smaller and simpler network. The rapid change between the period 1947-1960 and 1960-1970 also reflected the collaboration trend over time.

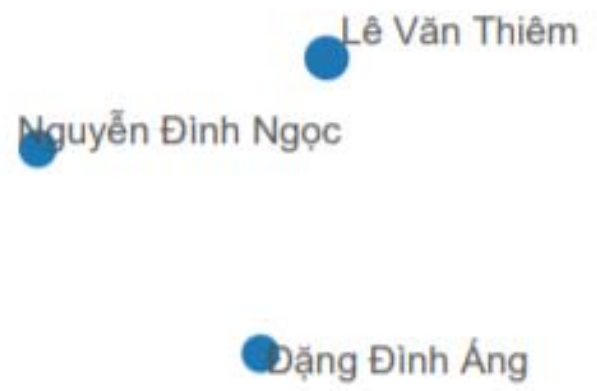

**Figure 15. The collaboration network in the 1947-1960 period**

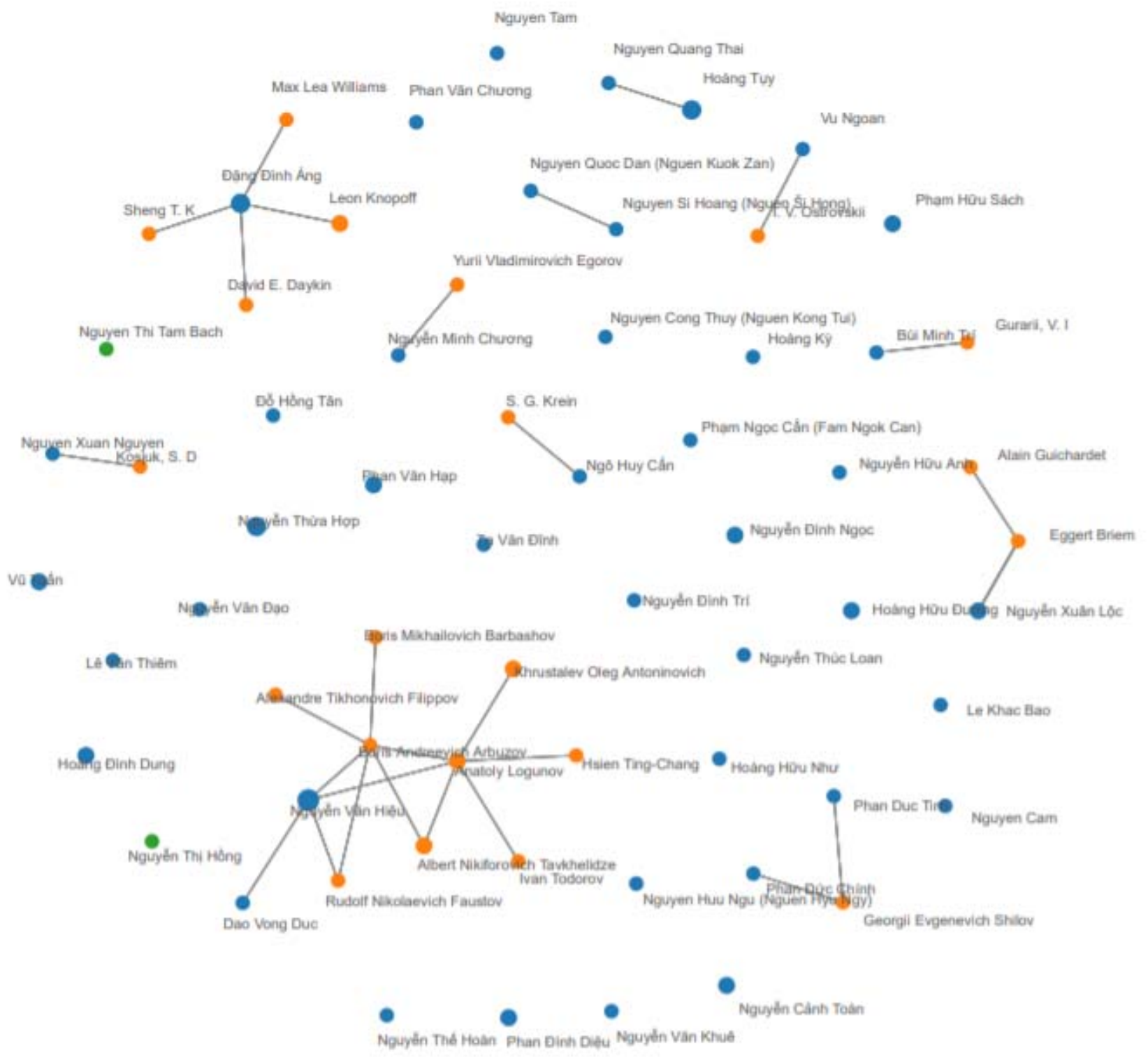

**Figure 16. The collaboration network in the 1960-1970 period**

### *2.6. Affiliation*

From 1947, the affiliations with an impressive number of publications are research institutions with a long history in Vietnam. The most productive is the Institute of Mathematics - VAST with 2696 publications. This position can be difficult to change in a short time because the number of papers of the Institute of Mathematics is four times as many as that of the second-ranked institution, Hanoi University of Science - Vietnam National University Hanoi. Table 5 lists the top 5 affiliations with the highest number of publications (based on available data).

**Table 5. Five most productive affiliations**

| No | Institution | Number of publications |
|---|---|---|
| 1 | Institute of Mathematics - VAST | 2696 |
| 2 | Hanoi University of Science - Vietnam National University Hanoi | 596 |
| 3 | Department of Mathematics and Informatics - Hanoi National University of Education | 522 |
| 4 | University of Science - Vietnam National University Ho Chi Minh City | 494 |
| 5 | Institute of Information Technology - VAST | 239 |

## 3. Important events

Mathematics is the first scientific discipline in Vietnam to publish internationally. The first generation of mathematicians had successfully passed on the tradition of excellence to successive ones. It also successfully spread this trend to other fields such as physics and computer science (T. T. H. Nguyen et al., 2020). Mathematics is one of the few disciplines with an international reputation, built upon great contributions by renowned mathematicians such as Hoang Tuy, Ngo Bao Chau, or Vu Ha Van, to name a few. About its long tradition and world-class achievements, the data team would like to note several important milestones and interesting events we learned during the data collection process:

In 1947, the first Vietnamese international journal article, *"Beitrag zum Typenproblem der Riemannschen Flächen"* by Le Van Thiem, was published on *Commentarii Mathematici Helvetici.* Along with Hoàng Tụy, the late professor Le Van Thiem has long been regarded as the founding father of Vietnam modern mathematics. In total, he wrote 13 publications in 7 journals in the 1947-1981 period. These publications were written in 4 languages (English, French, Russian, and German), and French was used in 8 articles.

Professor Hoàng Tụy introduced his famous "Tuy's cut" in 1964, which established a new mathematics subject, "global optimization." According to the current dataset, his writing career lasted for 48 years, from 1964 to 2012. During which, he published 140 documents and became one of the most productive mathematicians in Vietnam. His works are diverse in types of publications and languages (English, French, and Russian).

On February 5, 1969, the Institute of Mathematics (now part of VAST) was established. During its 51 years of history, the institute has become the leading institution in international publications in Vietnam. The Institute of Mathematics's 276 authors have published 2696 works on 585 different sources, diverse in types of publications and languages (Figure 17). Based on the subjects of publications, the database shows that strong research fields of the institute could be Operations research, mathematical programming (MSC 90); Calculus of variations and optimal control; optimization (MCS 49); Commutative algebra (MSC 13); and Algebraic geometry (MSC 14). Researchers of the Institute contributed in terms of scientific output and have published in prestigious journals. For example, two researchers of the institute, Nguyen Dang Hop and Ngo Viet Trung, recently had their new article *"Depth functions of symbolic powers of homogeneous ideals"* published on *Inventiones mathematicae* (MCQ = 3.56, rank 7$^{th}$ on MathSciNet in 2019) (H. D. Nguyen & Trung, 2019).

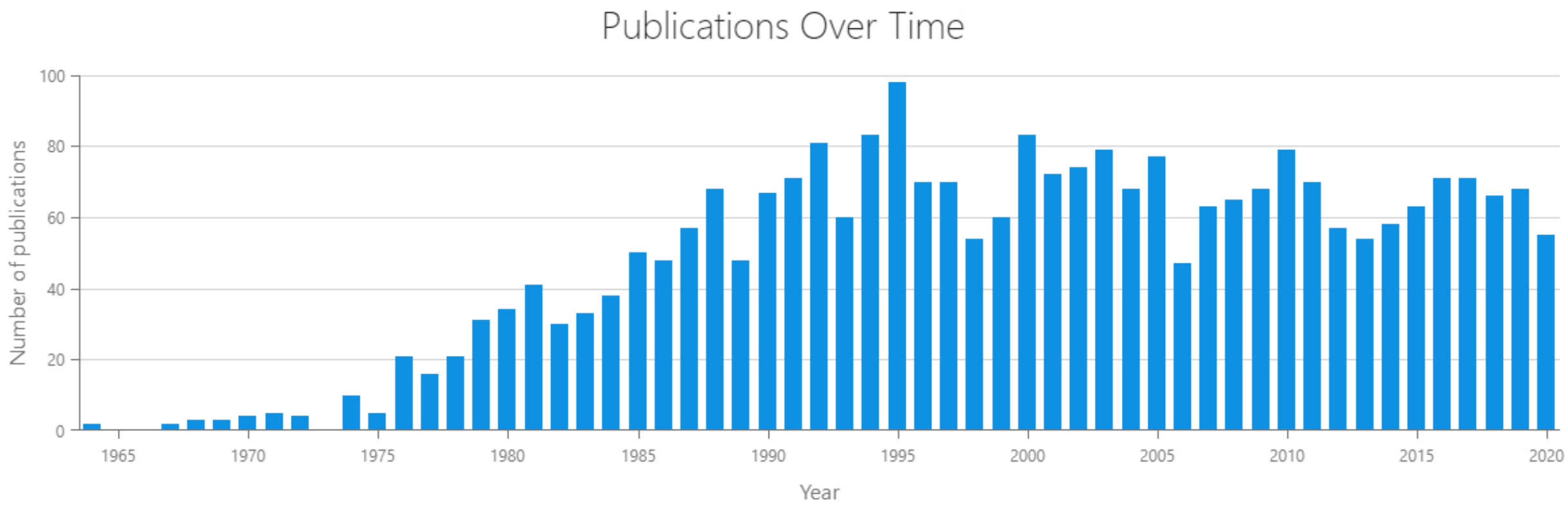

**Figure 17. The scientific output of Institute of Mathematics-VAST over years**

In 1976, Nguyen Huu Anh published *"Lie groups with square integrable representations"* in *Annals of mathematics,* one of the most prestigious journals in mathematics (Anh, 1976). Interestingly, he has thus far been the only Vietnamese author to publish in this journal while working in a university in Vietnam (Hanoi University of Science and Technology).

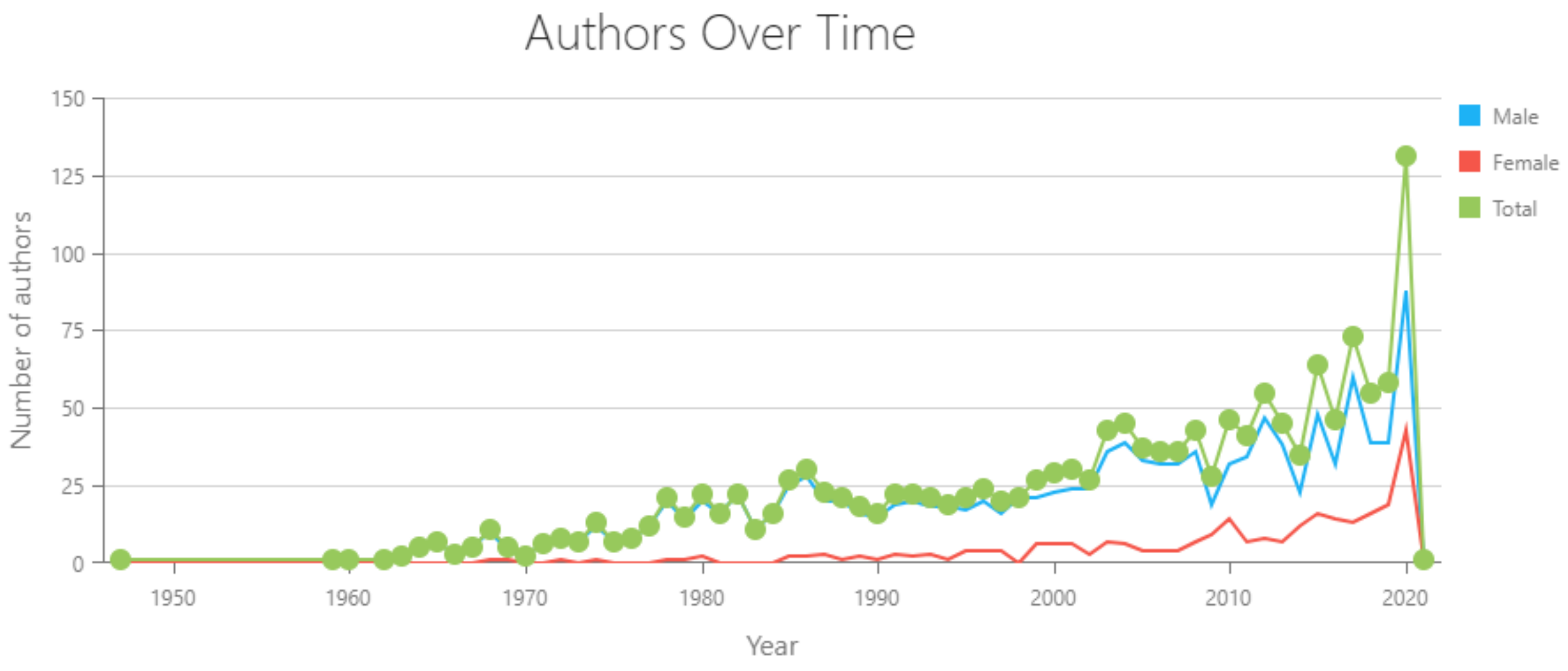

**Figure 18. The annual number of new authors.**

In 2004, Ngo Bao Chau received the Clay Research Awards (Clay Mathematics Institute) with Gérard Laumon. In 2010, he received the Fields Medal, which is usually referred to as the Nobel prize in Mathematics. Prof. Ngo Bao Chau has, up until now, been the only Vietnamese ever received these awards. His notable works include *"Fibration de Hitchin et endoscopie," "Le lemme fondamental pour les algèbres de Lie," và "Le lemme fondamental pour les groupes unitaires"* (Laumon et al., 2008; Ngô, 2006, 2010).

Perhaps, his achievement has inspired younger generations of Vietnamese mathematicians. The number of new authors increased fast after 2010 (Figure 18). In addition, the average age and maximum age of new authors went down slowly over time (Figure 19).

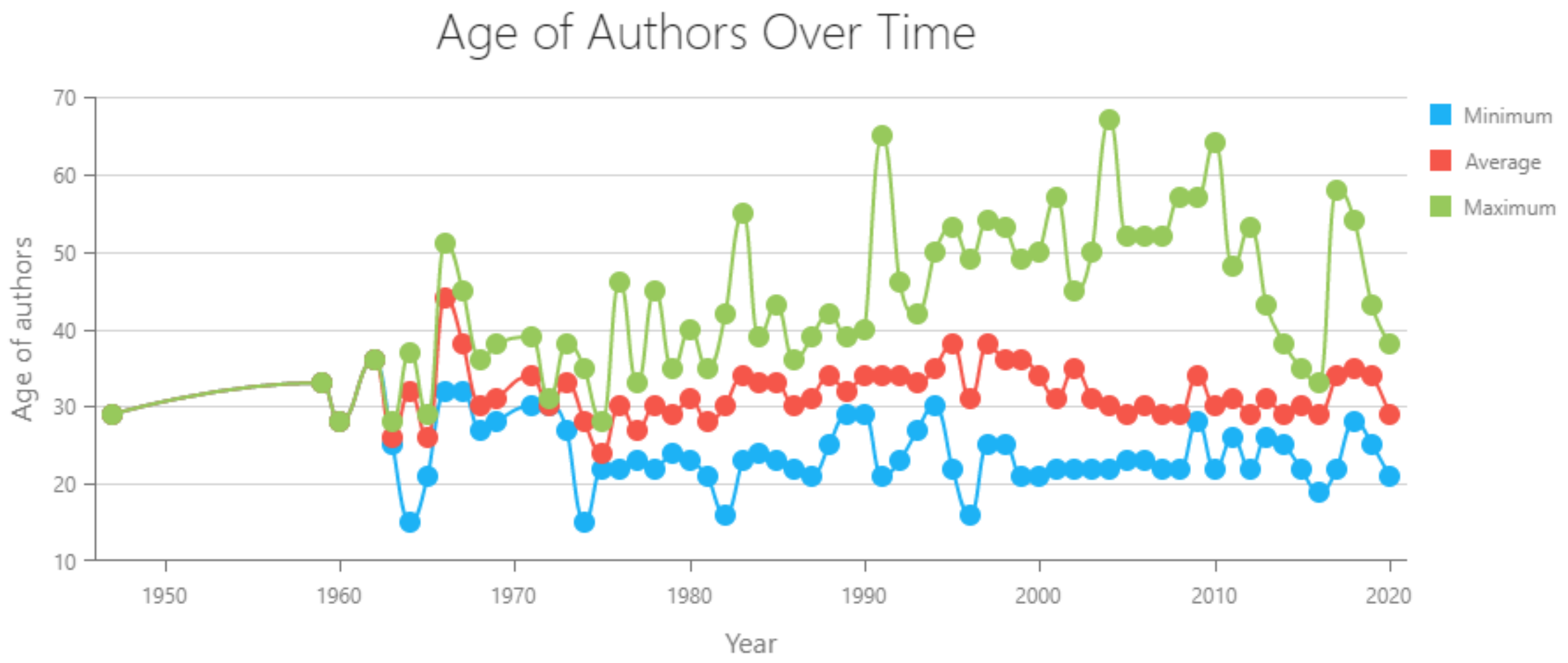

**Figure 19. Age of new authors over the years**

The development of Vietnam mathematics is attributable to the relentless endeavor and dedication of previous generations. The founding fathers of Vietnam modern mathematics - Le Van Thiem and Hoang Tuy - built a firm scientific foundation by not just integrating Vietnam mathematics with the world through international publication. They became the models to younger generations for their spirit to study and teach mathematics, which could not even be beaten by fire and guns during the France and American Wars in Vietnam. Professor Hoang Tuy wrote the first geometry textbook for high school students when he was in the war zone in 1949 (Koblitz & Tuy, 1990).

Vietnamese mathematicians' passion for studying, exploring, and talking about mathematics never dimmed even by the long and brutal attack of foreign armies (Koblitz, 2013). Thanks to these brave mathematicians' dedication, Vietnam – a poor and underdeveloped country – could compete against a science and technology giant like the United States in the battle of communications and tactical and strategic secrets (Hiệu & Koblitz, 2017).

Thanks to Neal Koblitz's (University of Washington) documents about the early days of Vietnam mathematics and his interviews (Koblitz, 1979), we could have insightful perspectives about the development of Vietnam mathematics and Vietnam's modern history. His recent interview with Ngo Bao Chau inspires younger generations of Vietnamese scientists. Passing on the passion for learning and exploring, has been the will of many Vietnamese scientists and educators, as well as of professor Ngo Bao Chau, who is in charge of this database project.

## 4. Conclusion and future directions

The preliminary report of the SciMath database shows positive results. The statistical analysis of this database has given insightful perspectives about the development of mathematics in Vietnam. In the future, this database will contribute to more bibliometric research and will be open to the community.

However, there are certain shortcomings that this database needs to solve. The data team expects many publications that have not been recorded into the SciMath database regarding data collection. Based on the number of mathematics publications' current growth, we estimate the number of mathematics works published during 1947-2020 to exceed the 10,000-paper milestone. The data team would commit its full capacity to record this amount of data until the end of 2020.

The data collection process is still in its second stage. Data collectors look for information from convenient sources such as Google Scholar pages, websites of universities, and research institutions, and verify them with a reliable database such as MathSciNet, zbMATH, Web of Science, and Scopus. Since many information sources and authors do not regularly update their publications, this creates difficulties when data collectors look for information. Besides, even though data are checked with reputable databases such as MathSciNet, zbMATH, Web of Science, and Scopus, there is a rising concern over low-quality research, which is still indexed in common databases. The genealogy tree also needs to be updated. This information needs consultation with specialists to assure its accuracy and integrity so that the knowledge passed on to the next generations and the public is both useful and reliable.

During the process of updating and making amendments to the database, the preliminary results show the potential of the SciMath database to become a useful and informational source for academia and the public. For those interested in mathematics, this database provides information, which is collected systematically and reliably and can be used to study scientific productivity and collaboration network in mathematics research. This is the first step for studying the impact of Vietnam mathematics research through citations or journal's impact, which brings in more insights about the history of Vietnam mathematics.

The plan of this project has four goals: (1) continue searching and updating data; (2) integrate tools for analysis such as genealogy, research network; (3) use this database to serve the scientific community and the public with publications, books, and website for the public (Vuong, 2018); (4) prepare communication channels for scientists to share their opinions/ comments or information related to data. We expect to publish a monograph on 80 years of Vietnam's mathematics history with the third goal.

To complete these goals needs a sufficient and high-quality database, which requires the data team's contribution. Hence, we are grateful for any opinion and information shared by specialists, scientists, and the mathematics community about publications, authors, genealogy, etc. This will be a valuable source for building an updated and reliable database for the mathematics community and the public.

**Acknowledgment:** This paper evolves from Technical Report No. VIASM-AISDL-20.01 "***Dự án Phát triển bộ cơ sở dữ liệu khoa học ngành toán của Việt Nam: Một số kết quả sơ bộ ban đầu từ CSDL SciMath***" used in the meeting of VIASM's Scientific Council on November 13, 2020. We would like to thank for the comments and suggestions of Ho Tu Bao (VIASM, Japan Advanced Institute of Science and Technology, John von Neumann Institute - Vietnam National University); Dinh Tien Cuong (National University of Singapore); Nguyen Huu Du (Hanoi University of Science - Hanoi Vietnam National University); Phung Ho Hai (Institute of Mathematics - VAST); Nguyen Xuan Hung (HUTECH University of Technology); Vu Hoang Linh (Hanoi University of Science - Hanoi Vietnam National University); Pham Tien Son (Da Lat University); Tran Van Tan (Hanoi University of National Education); Pham Huu Tiep (Rutgers University); Dang Duc Trong (University of Science - Vietnam National University in Ho Chi Minh City); Vu Ha Van (Yale University). The project is supported by the Phenikaa Innovation Foundation and the National Program for the Development of Mathematics 2010-2020.

## *Reference*

Anh, N. H. (1976). Lie Groups With Square Integrable Representations. *Annals of Mathematics, 104*(3), 431-458. doi:10.2307/1970965

Hiệu, P. D., & Koblitz, N. (2017). Cryptography during the French and American Wars in Vietnam. *Cryptologia, 41*(6), 491-511. doi:10.1080/01611194.2017.1292825

Koblitz, N. (1979). A Mathematical Visit to Hanoi. *The Mathematical Intelligencer, 2*(1), 38-42. doi:10.1007/BF03024385

Koblitz, N. (2013). Grothendieck's 1967 Lectures in the Forest in Vietnam. *The Mathematical Intelligencer, 35*(2), 32-34. doi:10.1007/s00283-013-9368-6

Koblitz, N., & Tuy, H. (1990). Recollections of mathematics in a country under siege. *The Mathematical Intelligencer, 12*(3), 16-34. doi:10.1007/BF03024014

Laumon, G., xe, rard, Ng, xf, Ch, B., & xe. (2008). Le Lemme Fondamental Pour Les Groupes Unitaires. *Annals of Mathematics, 168*(2), 477-573.

Le, V. T. (1947). Beitrag zum Typenproblem der Riemannschen Flächen. *Commentarii Mathematici Helvetici, 20*(1), 270-287. doi:10.1007/BF02568134

Ngô, B. C. (2006). Fibration de Hitchin et endoscopie. *Inventiones mathematicae, 164*(2), 399-453. doi:10.1007/s00222-005-0483-7

Ngô, B. C. (2010). Le lemme fondamental pour les algèbres de Lie. *Publications mathématiques de l'IHÉS, 111*(1), 1-169. doi:10.1007/s10240-010-0026-7

Ngo, B., Vuong, Q., La, V., Le, T., Le, M., Giang, T. T. T., … Ho, M. (2020). Dự án Phát triển bộ cơ sở dữ liệu khoa học ngành toán của Việt Nam: Một số kết quả sơ bộ ban đầu từ CSDL SciMath. *OSF Preprint*; doi:10.31219/osf.io/9d4bv

Nguyen, H. D., & Trung, N. V. (2019). Depth functions of symbolic powers of homogeneous ideals. *Inventiones mathematicae*, *218*(3), 779-827.

Nguyen, T. H. (1968). Behaviour of the diffraction peak for particles with arbitrary spins. *Nuclear Physics B, 8*(2), 303-310. doi:https://doi.org/10.1016/0550-3213(68)90243-5

Nguyen, T. T. H., Pham, H.-H., Vuong, Q.-H., Cao, Q.-T., Dinh, V.-H., & Nguyen, D. D. (2020). The adoption of international publishing within Vietnamese academia from 1986 to 2020: A review. *Learned Publishing*. doi:https://doi.org/10.1002/leap.1340

Vuong, Q. H. (2018). The (ir)rational consideration of the cost of science in transition economies. *Nature, 2*(1), 5.

# APPENDIXES

## Appendix 1

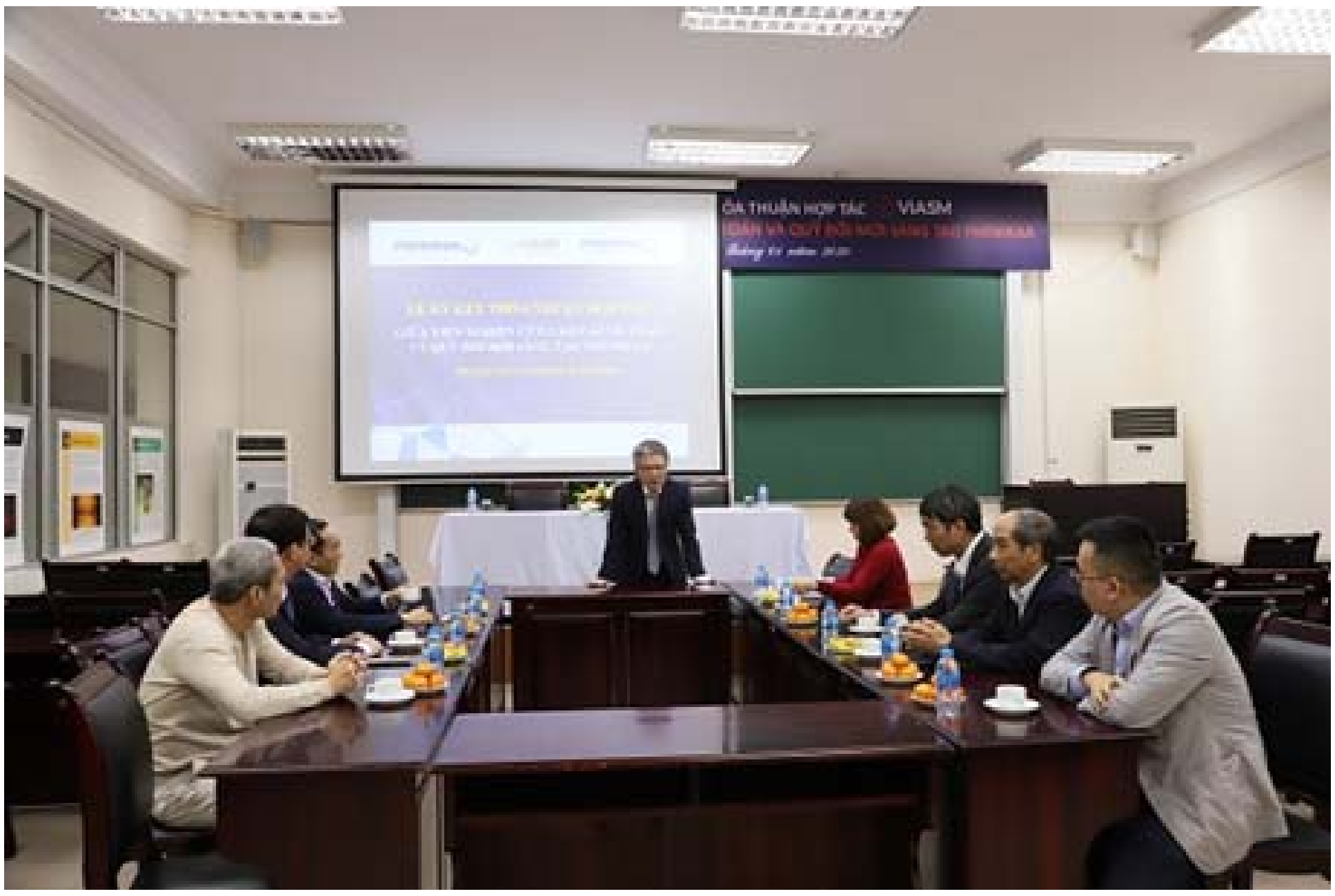

**Picture A.** The initiation of the SciMath project

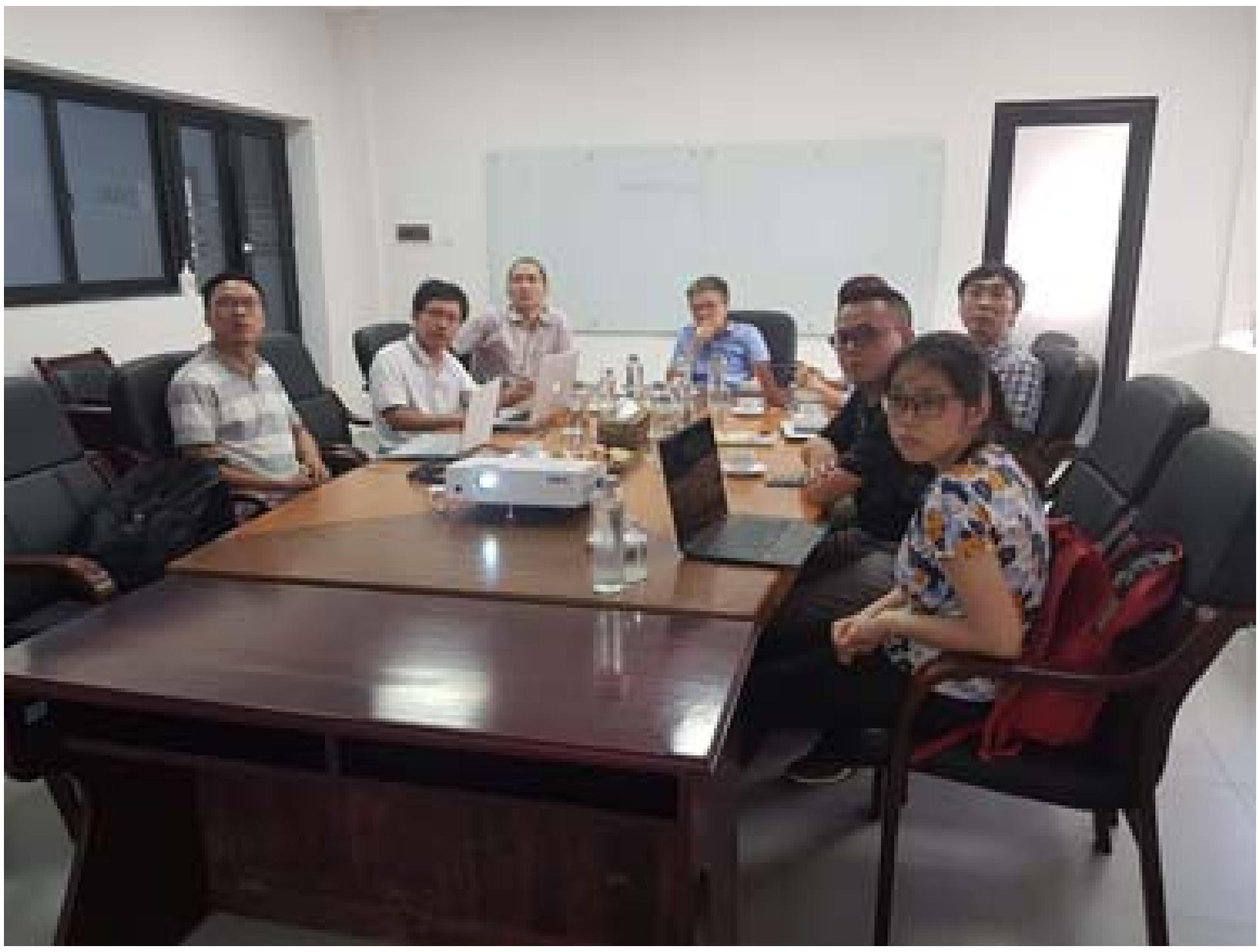

**Picture B.** The first meeting to report the progress of the SciMath project at VIASM Headquarters (September 16, 2020)

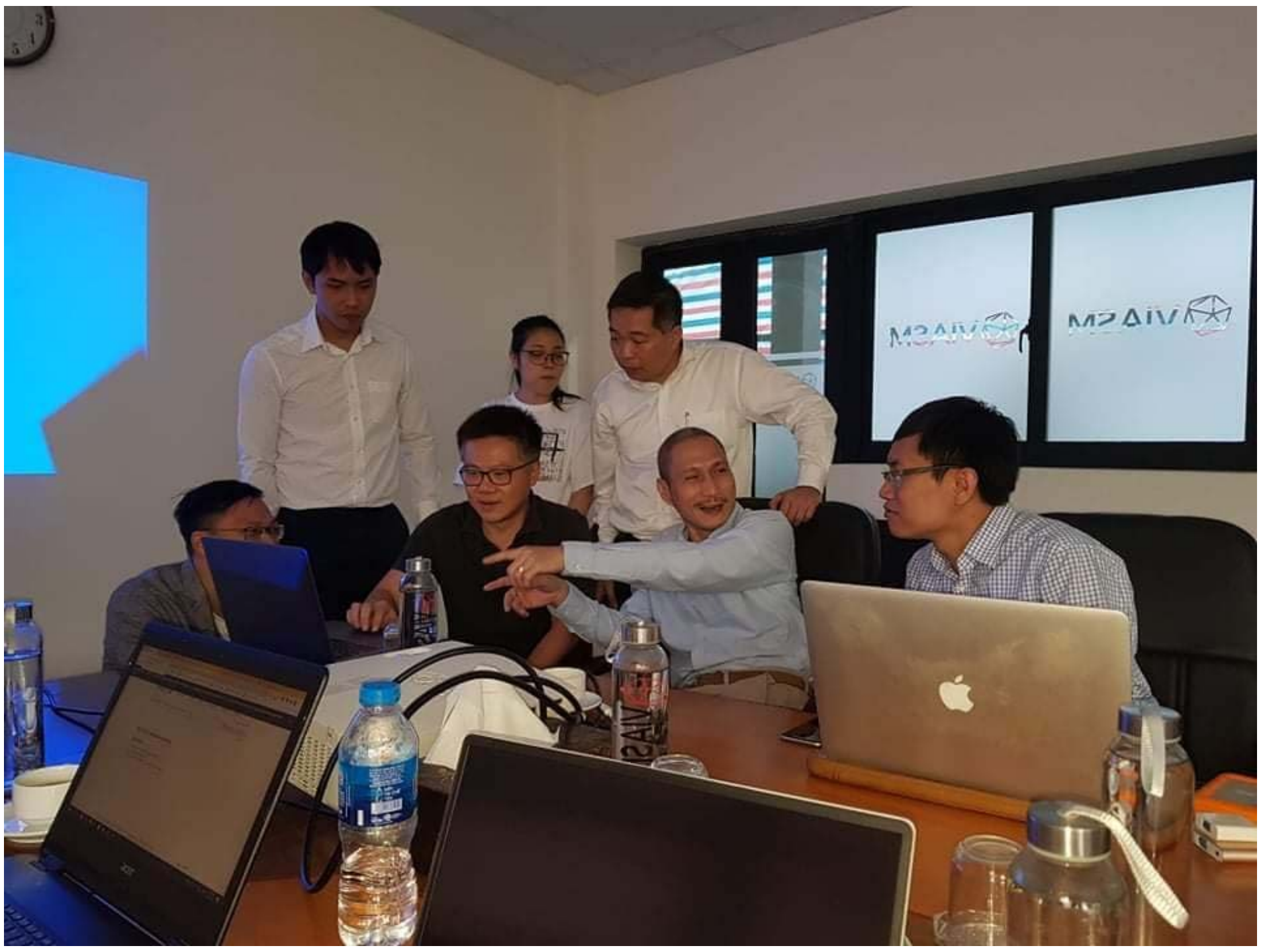

**Picture C.** Second meeting to report the progress of SciMath project at VIASM Headquarters (October 26, 2020)

**Appendix 2**

**Table A.** List of Mathematics Authors

| Name | Year of birth | Subjects | Number of publications | Year of first publication |
|---|---|---|---|---|
| Alain Pham Ngoc Dinh | | 35; 65; 42 | 45 | 1986 |
| Bach Hung Khang | 1942 | 68; 62 | 8 | 1974 |
| Ban D. V | | | 1 | 1998 |
| Banh Duc Dung | | 16 | 1 | 2008 |
| Bui Cong Cuong | 1939 | 47; 91 | 8 | 1976 |
| Bui Dac Tac | 1950 | 32; 46 | 9 | 1990 |
| Bui Doan Khanh | 1942 | | 3 | 1995 |
| Bui Duc Tien | 1963 | | 2 | 1992 |
| Bui Huy Bach | | 35; 76; 93 | 3 | 2018 |
| Bui Khoi Dam | 1951 | 60 | 13 | 1981 |
| Bui Kien Cuong | 1970 | 65; 35; 41 | 3 | 2001 |
| Bui Kim My | | 35 | 5 | 2016 |
| Bui Le Trong Thanh | | | 1 | 2020 |
| Bui Minh Phong | | | 1 | 1999 |
| Bui Minh Tri | 1939 | | 9 | 1969 |
| Bui Ngoc Muoi | | 90; 49 | 1 | 2017 |
| Bui Nguyen Thao Nguyen | 1989 | 32; 14 | 2 | 2014 |
| Bui Quang Nam | 1972 | 60 | 1 | 2017 |
| Bui Quoc Hoan | | 32; 46 | 5 | 2003 |
| Bui Quoc Trung | 1985 | | 2 | 2017 |
| Bui Tang Bao Ngoc | 1983 | | 1 | 2008 |
| Bui Thanh Duy | | 35; 26; 65 | 4 | 2015 |
| Bui The Anh | 1980 | 93; 35; 42 | 22 | 2006 |
| Bui The Hung | 1983 | 90; 49; 91 | 3 | 2011 |
| Bui The Quan | 1978 | 93; 34; 35 | 4 | 2011 |
| Bui The Tam | 1953 | 90; 49 | 10 | 1977 |
| Bui Thi Hoa | | 90; 49 | 2 | 2018 |
| Bui Thi Kieu Oanh | | 32 | 1 | 2015 |
| Bui Thu Lam | | | 3 | 2012 |
| Bui Tien Dung | | 46; 34 | 3 | 2000 |
| Bui Trinh Khanh | | 30 | 1 | 2008 |
| Bui Trong Kien | 1971 | 49; 47; 90 | 36 | 2001 |
| Bui Trong Kim | 1974 | 49; 47 | 1 | 2001 |
| Bui Tuong Tri | | | 2 | 1980 |
| Bui Van Chien | | | 4 | 2013 |
| Bui Van Dinh | | 90; 65; 49 | 7 | 2011 |
| Bui Van Thanh | | | 3 | 1990 |
| Bui Viet Huong | 1984 | 65; 46; 35 | 4 | 2010 |
| Bui Vu Quang | 1980 | 60; 62 | 1 | 2008 |
| Bui Xuan Dieu | | 47; 34 | 2 | 2015 |
| Bui Xuan Hai | | 16; 20 | 16 | 1989 |
| Bui Xuan Quang | | 35; 34; 37 | 3 | 2016 |
| Buu Huu Thai | | 65; 90; 68 | 2 | 2020 |
| Can Van Hao | 1989 | 82; 05; 60 | 10 | 2015 |
| Can Van Tuat | | | 4 | 1984 |
| Can Xuan Hien | | | 1 | 1973 |
| Cao Hoang Tru | | | 2 | 2010 |
| Cao Huu Hoa | | 35 | 2 | 2011 |
| Cao Huy Linh | 1965 | 13; 14 | 4 | 2005 |
| Cao Thanh Tinh | | 34 | 1 | 2016 |
| Cao Van Nuoi | 1970 | 60; 47; 43 | 2 | 2002 |
| Cao Xuan Phuong | | 62 | 4 | 2015 |
| Chau Thi Ha | | 76; 80 | 1 | 1989 |
| Chu Binh Minh | | 93 | 1 | 2017 |
| Chu Duc Khanh | | 74; 35 | 3 | 1995 |
| Chu Trong Thanh | | | 1 | 1997 |
| Chu Van Dong | | 06 | 3 | 1985 |
| Cung The Anh | 1977 | 35; 76; 37 | 98 | 2004 |
| D. N. Quynh | | 49; 90 | 1 | 2001 |
| D.V. Thanh (Do Van Thanh) | | 16 | 1 | 1991 |
| Dam Van Nhi | | 13; 14 | 4 | 1999 |
| Dang Anh Tuan | 1981 | 26; 35; 30 | 9 | 2003 |
| Dang Dinh Ang | 1926 | 35; 45; 74 | 90 | 1959 |
| Dang Dinh Chau | 1953 | | 2 | 2005 |
| Dang Dinh Hai | | 45; 47; 34 | 14 | 1986 |
| Dang Duc Trong | 1964 | 35; 65; 47 | 67 | 1994 |
| Dang Hai Long | 1988 | 49; 90; 65 | 7 | 2018 |
| Dang Hoa | | 90 | 2 | 2002 |
| Dang Hung Thang | 1955 | 60; 47; 37 | 36 | 1979 |
| Dang Huu Dao | 1943 | | 9 | 1978 |
| Dang Huy Cuong (Cuong Dang) | 1979 | | 1 | 2018 |
| Dang Huy Ruan | 1939 | 68 | 5 | 1973 |
| Dang Khai | | | 1 | 1985 |
| Dang Khanh Hoi | 1948 | 58; 35 | 6 | 1988 |
| Dang Ngoc Hoang Thanh | | | 2 | 2020 |
| Dang Phuoc Huy | | 60; 92 | 3 | 2001 |
| Dang Quang A | 1950 | 65; 35; 30 | 16 | 1988 |
| Dang Thanh Ha | 1968 | 68 | 1 | 2004 |
| Dang Thanh Hai | | | 1 | 2005 |
| Dang Thanh Son | | 35; 37; 76 | 4 | 2013 |
| Dang Thi My Van | | 90; 49; 65 | 4 | 2009 |
| Dang Thi Oanh | 1969 | 65 | 2 | 2017 |
| Dang Thi Phuong Thanh | | 35; 45; 76 | 7 | 2014 |
| Dang Van Cuong | | 90; 49; 53 | 6 | 2013 |
| Dang Van Hieu | 1983 | 65; 47; 91 | 19 | 2014 |
| Dang Van Hung | 1950 | | 13 | 1979 |
| Dang Vo Phuc | | 55 | 4 | 2015 |
| Dang Vu Giang | 1965 | 42; 46; 26 | 37 | 1991 |
| Dang Vu Huyen | | | 1 | 1978 |
| Dang Vu Phuong Ha | | 16 | 1 | 2010 |
| Dang Xuan Cuong | | 54 | 1 | 2008 |
| Dang Xuan Son | 1981 | 47; 65; 90 | 2 | 2017 |
| Danh Hua Quoc Nam | | 35; 47 | 4 | 2019 |
| Dao Bao Dung | | 35; 47 | 1 | 2003 |
| Dao Hai Long (Hailong Dao) | 1975 | 13; 14; 05 | 43 | 2007 |
| Dao Huu Ho | 1944 | 60 | 2 | 1994 |
| Dao Huy Cuong | | 65; 76; 35 | 2 | 2018 |
| Dao Manh Khang | | 65 | 1 | 2020 |
| Dao Ngoc Minh | | 90; 47; 49 | 16 | 2015 |
| Dao Nguyen Anh | | 65; 42; 43 | 6 | 2017 |
| Dao Nguyen Van Anh | | 20; 11 | 2 | 2020 |
| Dao Phan Vu | | 90 | 1 | 2006 |
| Dao Phuong Bac | 1980 | 20; 14; 13 | 7 | 2005 |
| Dao Quang Khai | 1985 | 76; 35 | 8 | 2013 |
| Dao Quang Tuyen | | 60; 62 | 7 | 1992 |
| Dao Quang Vinh | | | 1 | 2007 |
| Dao Thanh Tinh | 1962 | 35; 37 | 3 | 1995 |
| Dao Thi Lien | | | 1 | 2003 |
| Dao Thi Thanh Ha | | 13; 14; 32 | 3 | 2008 |
| Dao Thi Thu Ha | | | 1 | 2014 |
| Dao Trong Quyet | | 35; 37 | 6 | 2012 |
| Dao Trong Thi | 1951 | 49; 53; 58 | 21 | 1978 |
| Dao Tuan Anh | | 35 | 1 | 2020 |
| Dao Van Duong | | 42; 47 | 7 | 2013 |
| Dao Van Tra | 1953 | 32; 22; 57 | 5 | 1975 |
| Dao Vong Duc | | | 9 | 1967 |
| Dau Hoang Hung | 1979 | 32; 31; 41 | 4 | 2005 |
| Dau Hong Quan | | 30; 32 | 1 | 2018 |
| Dau Son Hoang | | | 1 | 2006 |
| Dau The Cap | 1952 | | 5 | 1986 |
| Dau The Phiet | | 32 | 1 | 2020 |
| Dau Xuan Luong | | 47; 49; 65 | 2 | 2010 |
| Dinh Cao Duy Thien Vu | | 35 | 1 | 2011 |
| Dinh Cong Huong | | 39; 92; 93 | 14 | 2005 |
| Dinh Dieu Hang | 1984 | 90; 49; 45 | 5 | 2014 |
| Dinh Duc Tai | | 16 | 1 | 2010 |
| Dinh Dung | 1951 | 41; 42; 46 | 72 | 1979 |
| Dinh Hoang Anh | 1964 | | 1 | 1986 |
| Dinh Huy Hoang | 1956 | 32; 46 | 4 | 1996 |
| Dinh Ngoc Quy | | 90; 49; 58 | 5 | 2010 |
| Dinh Ngoc Thanh | 1955 | 35; 45; 44 | 20 | 1981 |
| Dinh Nguyen Duy Hai | | 35; 65; 47 | 10 | 2017 |
| Dinh Nho Hao | 1960 | 65; 35; 49 | 88 | 1987 |
| Dinh Phu Bong | | | 5 | 1986 |
| Dinh Quang Hai | 1979 | 16 | 4 | 2003 |
| Dinh Quang Luu | 1947 | 60; 41; 91 | 48 | 1980 |
| Dinh Si Tiep | 1980 | 14; 32; 58 | 10 | 2009 |
| Dinh Sy Dai | 1948 | | 3 | 1987 |
| Dinh Thanh Duc | 1960 | 44; 42; 46 | 7 | 1993 |
| Dinh Thanh Giang | 1988 | 52; 65; 68 | 3 | 2010 |
| Dinh Thanh Trung | | 14; 13 | 1 | 2018 |
| Dinh The Luc | 1952 | 90; 49; 52 | 97 | 1978 |
| Dinh Thi Ngoc Thanh | | | 2 | 1986 |
| Dinh Tien Cuong | | 32; 37; 14 | 32 | 2003 |
| Dinh Trung Hoa | 1970 | 46; 15; 47 | 10 | 2013 |
| Dinh Van Huynh | 1950 | 16 | 89 | 1974 |
| Dinh Van Ruy | | 65; 45 | 5 | 1995 |
| Dinh Xuan Khanh | | 35; 37; 34 | 1 | 2016 |
| Do Anh Tuan | | 11 | 1 | 2017 |
| Do Ba Khang | 1958 | | 3 | 1982 |
| Do Cong Khanh | 1949 | 34; 93; 47 | 19 | 1982 |
| Do Duc Dong | 1981 | | 1 | 2014 |
| Do Duc Hung | | | 3 | 1987 |
| Do Duc Tan | | 42; 35; 47 | 3 | 2020 |
| Do Duc Thai | 1961 | 32; 30; 46 | 47 | 1991 |
| Do Duc Thuan | | 34; 65; 47 | 13 | 2008 |
| Do Duy Chinh | | | 2 | 1982 |
| Do Duy Hieu | | | 3 | 2013 |
| Do Duy Thanh | | 65; 90 | 1 | 2015 |
| Do Hoai Vu | | 45; 47; 34 | 1 | 2012 |
| Do Hoang Giang | | 13; 14 | 4 | 2010 |
| Do Hoang Son | 1988 | 32; 35; 31 | 8 | 2016 |
| Do Hoang Viet | | 20; 05 | 1 | 2020 |
| Do Hong Nhat | | | 1 | 1988 |
| Do Hong Tan | 1937 | 47; 54; 55 | 29 | 1968 |
| Do Huy Hoang | | 34 | 3 | 2016 |
| Do Lan | | 35; 47; 34 | 6 | 2010 |
| Do Long Van | 1947 | 68; 94; 20 | 39 | 1974 |
| Do Ngoc Diep | 1950 | 22; 19; 46 | 50 | 1974 |
| Do Phi Nga | | 14 | 1 | 2001 |
| Do Phuong An | | 32; 14; 30 | 4 | 2013 |
| Do Quang Yen | | 42; 37; 30 | 15 | 2011 |
| Do Tan Si | 1942 | | 1 | 1978 |
| Do Thai Duong | 1994 | 31; 32 | 3 | 2019 |
| Do Thi Phuong Quynh | | 22; 11 | 3 | 2015 |
| Do Trong Hoang | 1985 | 13; 05 | 7 | 2013 |
| Do Van Kien | | 13 | 1 | 2018 |
| Do Van Loi | | 47; 35; 34 | 2 | 2016 |
| Do Van Luu | 1944 | 90; 49; 26 | 51 | 1982 |
| Do Van Thuan | | 16 | 1 | 2020 |
| Do Viet Bach | | | 1 | 2017 |
| Do Viet Hung | 1981 | 57 | 1 | 2017 |
| Do Xuan Duong | | 90 | 1 | 2003 |
| Do Xuan Tho | | | 4 | 1985 |
| Do Xuan Tung | 1983 | | 2 | 2017 |
| Doan Cong Dinh | | 35; 30; 15 | 3 | 2020 |
| Doan Hai An | | | 1 | 1998 |
| Doan Minh Luan | | 30; 47; 68 | 6 | 2015 |
| Doan Quang Manh | | 30; 11; 14 | 4 | 2002 |
| Doan Thai Son | 1984 | 37; 34; 26 | 43 | 2007 |
| Doan The Hieu | 1960 | 53; 49; 58 | 9 | 1996 |
| Doan Thi Thu Ha | | | 1 | 2015 |
| Doan Trung Cuong | 1981 | 13; 14; 11 | 13 | 2003 |
| Doan Van Ngoc | | 35 | 3 | 1984 |
| Du Duc Thang | 1975 | 47; 26 | 2 | 2005 |
| Du Thi Hoa Binh | 1987 | 47; 52; 15 | 3 | 2014 |
| Duc Ho | | 05; 13 | 1 | 2015 |
| Duong Anh Tuan | 1984 | 35; 34; 47 | 7 | 2008 |
| Duong Dang Xuan Thanh | 1982 | 34; 93; 47 | 11 | 2006 |
| Duong Hoang Dung | 1985 | 16 | 1 | 2014 |
| Duong Hong Phong | 1953 | 81; 53; 32 | 93 | 1977 |
| Duong Luong Son | | | 1 | 1998 |
| Duong Manh Hong | 1983 | 30; 35; 26 | 2 | 2019 |
| Duong Minh Duc | 1951 | 35; 47; 58 | 35 | 1981 |
| Duong Minh Thanh | | 17 | 1 | 2013 |
| Duong Quang Hoa | | 46; 20; 22 | 1 | 2010 |
| Duong Quoc Huy | | 42; 47; 43 | 4 | 2016 |
| Duong Quoc Viet | 1954 | 13; 14 | 15 | 1992 |
| Duong Thi Anh Tuyet | | | 1 | 2015 |
| Duong Thi Hong | | | 1 | 2019 |
| Duong Thi Huong | 1983 | 13 | 2 | 2019 |
| Duong Thi Kim Huyen | 1990 | 90; 49 | 3 | 2016 |
| Duong Thi Thanh Binh | | 35; 45; 65 | 6 | 1997 |
| Duong Thi Viet An | 1989 | 49; 90; 93 | 6 | 2015 |
| Duong Trong Luyen | 1984 | 35 | 8 | 2012 |
| Duong Trong Nhan | | | 5 | 1980 |
| Duong Viet Thong | | 47; 65; 68 | 19 | 2012 |
| Duong Xuan Thinh | | 35; 42 | 3 | 2016 |
| Duong Xuan Vinh | | 35; 76; 34 | 2 | 2020 |
| Giang Hoang (Hoang Ton Nu Huong Giang) | | | 1 | 2015 |
| Ha Binh Minh | | 93 | 1 | 2017 |
| Ha Binh Minh | | 34 | 1 | 2004 |
| Ha Dai Ton | | 14; 22 | 1 | 2009 |
| Ha Dang Cao Tung | | | 1 | 1997 |
| Ha Duc Vuong | | 46; 54 | 3 | 2001 |
| Ha Duy Hung | | 42; 47; 46 | 12 | 2000 |
| Ha Duyen Trung | | | 1 | 2012 |
| Ha Hoang Hop | | | 2 | 1988 |
| Ha Huong Giang | | 30; 32 | 6 | 2012 |
| Ha Huy Bang | 1960 | 46; 26; 42 | 77 | 1982 |
| Ha Huy Khoai | 1946 | 30; 11; 32 | 38 | 1973 |
| Ha Huy Tai | 1973 | 13; 05; 14 | 51 | 1998 |
| Ha Huy Vui | 1950 | 14; 32; 11 | 46 | 1977 |
| Ha Le Anh | | 51 | 2 | 1981 |
| Ha Manh Linh | | 90; 49; 54 | 1 | 2020 |
| Ha Minh Hoang | | | 13 | 2015 |
| Ha Minh Lam | 1979 | 13; 05; 14 | 6 | 2007 |
| Ha Minh Long | | | 1 | 2013 |
| Ha Quang Minh | 1992 | | 1 | 2020 |
| Ha Thi Ngoc Yen | 1980 | 39; 34; 35 | 3 | 2004 |
| Ha Thi Thanh Tam | 1985 | 47; 46; 03 | 1 | 2017 |
| Ha Thi Thu Hien | | 13 | 2 | 2015 |
| Ha Tien Ngoan | 1951 | 35; 53; 47 | 19 | 1977 |
| Ha Tran Phuong | 1971 | 32; 30 | 7 | 2007 |
| Ha Vinh Tan | | | 3 | 1989 |
| Ho Dac Nghia | 1970 | 35; 76 | 1 | 2006 |
| Ho Dang Phuc | 1955 | 60; 62 | 15 | 1980 |
| Ho Dinh Duan | 1947 | 16 | 2 | 1992 |
| Ho Duc Viet | 1947 | | 4 | 1975 |
| Ho Duy Binh | | | 1 | 2020 |
| Ho Huu Viet | 1952 | | 3 | 1981 |
| Ho Minh Toan | 1974 | 15; 46; 47 | 10 | 2009 |
| Ho Phi Tu | | 65; 90; 47 | 1 | 2020 |
| Ho Phu Quoc | | | 1 | 2014 |
| Ho Si Tung Lam | | 35 | 1 | 2011 |
| Ho Thuan | 1932 | 15; 65 | 24 | 1971 |
| Ho Thuc Quyen | | | 1 | 2011 |
| Ho Tu Bao | | 68; 03 | 7 | 2002 |
| Ho Van Hoa | 1943 | | 2 | 1978 |
| Ho Xuan Thang | 1973 | 13; 20; 18 | 1 | 2003 |
| Hoang Chi Thanh | 1952 | | 2 | 1991 |
| Hoang Dinh Dung | 1922 | 30; 76; 86 | 18 | 1967 |
| Hoang Dung | | | 1 | 1988 |
| Hoang Duong Tuan | 1964 | 90; 93; 49 | 12 | 1992 |
| Hoang Hai Hoc | | | 1 | 1979 |
| Hoang Hoa Trai | | 93; 55 | 2 | 2002 |
| Hoang Hong Son | | | 5 | 1982 |
| Hoang Huu Duong | 1936 | | 10 | 1964 |
| Hoang Huu Nhu | 1932 | | 1 | 1968 |
| Hoang Kiem | 1947 | | 4 | 1983 |
| Hoang Ky | 1934 | | 4 | 1969 |
| Hoang Le Minh | 1956 | 14; 81; 22 | 7 | 1980 |
| Hoang Le Truong | 1973 | 13 | 19 | 2008 |
| Hoang Luu Cam Vu | | 35; 47 | 1 | 2018 |
| Hoang Mai Le | 1968 | 26 | 6 | 1998 |
| Hoang Manh Tuan | | 65; 37 | 1 | 2020 |
| Hoang Manh Tuan | 1990 | 65; 37 | 2 | 2020 |
| Hoang Minh Chuong | | | 1 | 1985 |
| Hoang Minh Hai | | | 4 | 1985 |
| Hoang Nam | | 34; 47 | 5 | 2003 |
| Hoang Nam Dung | | 52; 51; 65 | 2 | 2015 |
| Hoang Ngoc Duong | | 65 | 1 | 2020 |
| Hoang Ngoc Long | 1952 | | 1 | 1992 |
| Hoang Ngoc Minh | | 11; 05; 37 | 13 | 2012 |
| Hoang Ngoc Tuan | | | 2 | 2013 |
| Hoang Ngoc Yen | 1993 | 13 | 1 | 2019 |
| Hoang Nhat Quy | 1979 | | 1 | 2013 |
| Hoang Quang Tuyen | | 90 | 4 | 2000 |
| Hoang Quoc Toan | 1945 | 35; 58 | 12 | 1979 |
| Hoang The Tuan | | 34; 26; 33 | 21 | 2012 |
| Hoang Thi Cam Thach | | 65; 90; 49 | 1 | 2020 |
| Hoang Thi Thao Phuong | 1987 | 65; 35; 74 | 9 | 2013 |
| Hoang Thi Vi | | | 1 | 1995 |
| Hoang Tien (Hoang, T. H) | | | 5 | 1984 |
| Hoang Tuy | 1927 | 90; 65; 78 | 140 | 1964 |
| Hoang Van Can | 1988 | 32 | 2 | 2018 |
| Hoang Van Lai | | | 9 | 1973 |
| Hoang Viet | | | 2 | 2016 |
| Hoang Viet | | | 1 | 2012 |
| Hoang Viet Ha | | 35 | 1 | 2020 |
| Hoang Viet Long | | 03; 46; 47 | 1 | 2017 |
| Hoang Xuan Huan | 1954 | 53 | 4 | 1991 |
| Hoang Xuan Phu | 1951 | 90; 52; 47 | 67 | 1984 |
| Hoang Xuan Sinh | 1933 | | 2 | 1978 |
| Hong Ngoc Binh | | 13 | 1 | 2015 |
| Huynh Ba Lan | 1956 | | 1 | 1993 |
| Huynh Minh Hien | | 34; 37; 93 | 1 | 2010 |
| Huynh Mong Giao | | 26 | 1 | 2002 |
| Huynh Mui | 1944 | | 7 | 1980 |
| Huynh The Phung | 1959 | 52; 46; 90 | 10 | 1989 |
| Huynh Thi Hoang Dung | | 45; 47; 65 | 1 | 2020 |
| Huynh Thi Hong Diem | | 90; 49; 91 | 5 | 2014 |
| Huynh Thi Thanh Binh | | | 3 | 2012 |
| Huynh Van Nam | 1978 | 03; 68; 62 | 12 | 1999 |
| Huynh Van Ngai | 1971 | 49; 90; 46 | 22 | 1999 |
| Huynh Viet Khanh | | 16 | 1 | 2020 |
| Hy Duc Manh | | 90; 65; 49 | 2 | 2020 |
| Khu Quoc Anh | | 32 | 1 | 2004 |
| Khuat Thi Binh | | 49; 47 | 1 | 2020 |
| Khuat Van Ninh | 1952 | 45; 47; 35 | 6 | 1990 |
| Khyng D. V | | | 1 | 1998 |
| Kieu Huu Dung | | 47; 42 | 1 | 2020 |
| Kieu Phuong Chi | | 46; 32; 30 | 5 | 2006 |
| Kieu The Duc | | | 1 | 1982 |
| Kieu Trung Thuy | | 65; 60 | 1 | 2020 |
| Kieu Van Hung | | 68; 94 | 9 | 2004 |
| L. V. Thuan | | | 1 | 2000 |
| La Huu Chuong | | | 1 | 2012 |
| La Thi Hong | | 47; 65; 91 | 1 | 2020 |
| Lam Hoang Nguyen | | | 1 | 2006 |
| Lam Quoc Anh | 1974 | 90; 49; 91 | 24 | 2004 |
| Lam Quoc Dung | | | 1 | 2011 |
| Le Anh Dung | | 47; 54; 53 | 5 | 2003 |
| Le Anh Minh | 1982 | 35 | 1 | 2020 |
| Le Anh Tuan | 1963 | 49; 90; 54 | 22 | 2003 |
| Le Anh Tuan | | 60; 39; 34 | 2 | 2018 |
| Le Anh Vinh | 1983 | 05; 11; 52 | 57 | 2007 |
| Le Anh Vu | 1957 | 22; 46; 20 | 7 | 1987 |
| Le Anh Vu | | 68; 05 | 1 | 2009 |
| Le Ba Long | | 81; 46 | 9 | 1990 |
| Le Bich Phuong | 1984 | | 1 | 2016 |
| Le Chi Ngoc | | | 1 | 2020 |
| Le Cong Loi | 1972 | 39; 34; 65 | 6 | 2001 |
| Le Cong Nhan | | 34; 35 | 4 | 2017 |
| Le Cong Son (Shon, Le Khong) | | | 1 | 2007 |
| Le Cong Thanh | 1951 | 68 | 7 | 1977 |
| Le Cong Trinh | 1980 | 14; 15; 13 | 11 | 2008 |
| Le Dao Hai An | | 34; 92 | 1 | 2017 |
| Le Dinh Long | | 35; 47; 26 | 18 | 2017 |
| Le Duc Thang | | 35; 47 | 4 | 2015 |
| Le Duc Thinh | 1979 | | 1 | 2007 |
| Le Duc Thoang | 1971 | 16 | 3 | 2002 |
| Le Dung | | | 1 | 1989 |
| Le Dung | | 35 | 6 | 2003 |
| Le Dung Muu | 1949 | 90; 65; 47 | 87 | 1978 |
| Le Dung Trang | 1947 | | 1 | 1984 |
| Le Gia Quoc Thong | | 47; 35 | 1 | 2017 |
| Le Giang | | 32; 35 | 1 | 2020 |
| Le H. An | 1978 | 49; 47 | 1 | 2004 |
| Le Hai An | 1978 | 35; 49; 58 | 1 | 2005 |
| Le Hai Khoi | 1960 | 47; 30; 32 | 39 | 1983 |
| Le Hai Yen | 1987 | 65; 90; 46 | 9 | 2012 |
| Le Hoan Hoa | 1947 | 45; 47; 34 | 6 | 1979 |
| Le Hoang Mai | | 18; 16 | 1 | 2017 |
| Le Hoang Tri | 1963 | 54; 57 | 5 | 1994 |
| Le Hong Lan | 1961 | | 3 | 1999 |
| Le Hong Son | 1979 | | 1 | 2006 |
| Le Hong Trang | 1983 | 52; 65; 68 | 6 | 2012 |
| Le Hong Van | 1962 | 51; 53 | 2 | 1993 |
| Le Hung Son | 1944 | 35; 30; 31 | 35 | 1977 |
| Le Hung Viet Bao | | | 1 | 2014 |
| Le Huu Dien | | | 1 | 1985 |
| Le Huy Chuan | 1978 | 45; 30 | 2 | 2003 |
| Le Huy Tien | 1977 | | 1 | 2007 |
| Le Huy Vu | 1984 | | 1 | 2015 |
| Le Huynh My Van | | 90 | 1 | 2020 |
| Le Khac Bao | | | 1 | 1965 |
| Le Khac Huynh | | 16 | 3 | 2004 |
| Le Khanh Chau | 1958 | 54; 74 | 15 | 1982 |
| Le Khanh Luan | | 35 | 1 | 2011 |
| Le Khoi Vy | 1959 | 35; 49; 74 | 18 | 1990 |
| Le Kim Luat | | | 1 | 1992 |
| Le Ky Vy | | | 3 | 1992 |
| Le Long Phi | | 35; 26; 46 | 1 | 2019 |
| Le Long Trieu | | 47; 46; 32 | 32 | 2007 |
| Le Manh Ha | 1979 | 68; 05; 91 | 5 | 2009 |
| Le Mau Hai | 1951 | 32; 46; 30 | 65 | 1986 |
| Le Minh Ha | 1972 | 55; 20; 18 | 8 | 2001 |
| Le Minh Hieu | 1986 | | 2 | 2020 |
| Le Minh Luu | | 49; 90; 47 | 10 | 2000 |
| Le Minh Tri | | | 1 | 2005 |
| Le Minh Triet | | 65; 35 | 1 | 2018 |
| Le Minh Tung | | 90; 26 | 2 | 2002 |
| Le Ngoc Lang | 1941 | | 7 | 1971 |
| Le Ngoc Quynh | | 32; 30 | 5 | 2012 |
| Le Ngoc Tuan (Le Ngok T'euen) | | | 4 | 1982 |
| Le Ngoc Xuan | 1977 | 65; 34 | 3 | 2003 |
| Le Nhat Huynh | | 35; 47; 65 | 9 | 2019 |
| Le Nhu Duong (Le N'y Dyong) | | | 2 | 1989 |
| Le Nhu Hung | | | 1 | 1994 |
| Le Phi Long | | 35; 49; 58 | 1 | 2008 |
| Le Phuoc Hai | | 90; 49 | 2 | 2019 |
| Le Phuong | | 35 | 2 | 2020 |
| Le Phuong Quynh | | 35 | 1 | 2019 |
| Le Quang Ham | 1982 | 11; 20 | 3 | 2017 |
| Le Quang Nam | 1980 | 74; 47; 65 | 4 | 2003 |
| Le Quang Ninh | | 32 | 1 | 2014 |
| Le Quang Thuan | | | 1 | 2007 |
| Le Quang Thuy | | 90; 49; 65 | 11 | 2011 |
| Le Quang Trung | | 47; 46; 58 | 9 | 1985 |
| Le Quoc Han | 1953 | | 2 | 1996 |
| Le Quy Thuong | 1982 | 14; 32; 03 | 3 | 2017 |
| Le Si Dong | | 34 | 3 | 2012 |
| Le Si Vinh | 1980 | | 1 | 2003 |
| Le Sy Dong | | | 2 | 1987 |
| Le Tai Thu | | 32 | 3 | 2002 |
| Le Thai Thanh | | | 1 | 1993 |
| Le Thanh | | | 1 | 1985 |
| Le Thanh Hai | | | 1 | 1987 |
| Le Thanh Hue | 1964 | 65; 90 | 2 | 2009 |
| Le Thanh Hung | | 41; 46 | 1 | 2017 |
| Le Thanh Manh | | | 1 | 1995 |
| Le Thanh Quang | | 34 | 3 | 2008 |
| Le Thanh Tung | | 90; 49; 46 | 10 | 2011 |
| Le Thi Diem Hang | | 35; 47 | 1 | 2020 |
| Le Thi Hoai An | 1958 | 90; 65; 49 | 17 | 1996 |
| Le Thi Hoai Thu | 1962 | 11 | 3 | 2003 |
| Le Thi Hong Thom | | 30; 47 | 1 | 2019 |
| Le Thi Minh Duc | | 26; 35 | 2 | 2020 |
| Le Thi Ngoc Quynh | | | 1 | 2021 |
| Le Thi Nhu Bich | | 42 | 1 | 2015 |
| Le Thi Phuong Ngoc | 1966 | 35; 37; 65 | 24 | 2006 |
| Le Thi Phuong Thuy (Le Thi Thuy) | | 35; 37; 45 | 9 | 2010 |
| Le Thi Thanh | | 68 | 1 | 1994 |
| Le Thi Thanh Nhan | 1970 | 13; 16 | 21 | 1999 |
| Le Thi Thu Giang | 1984 | 65; 35 | 1 | 2018 |
| Le Thi Thu Thuy | | 68; 62; 35 | 2 | 2010 |
| Le Thi Thu Thuy (UNCC) | | 35 | 1 | 2020 |
| Le Thi Tuyet | | 35 | 2 | 2013 |
| Le Thu Hoai | | 35 | 1 | 2006 |
| Le Tien Nam | 1990 | | 2 | 2015 |
| Le Tien Tam | | | 2 | 1984 |
| Le Tien Vuong | | | 4 | 1973 |
| Le Tran Tinh | | | 1 | 2018 |
| Le Trieu Phong | | | 1 | 2004 |
| Le Trong Lan | | 35; 20; 26 | 2 | 2015 |
| Le Trong Luc | 1952 | | 4 | 1991 |
| Le Trung Hieu | | 05; 68 | 1 | 2009 |
| Le Trung Kien | | 68; 05 | 1 | 2009 |
| Le Trung Nghia | | 46; 26; 35 | 1 | 2019 |
| Le Tu Luc | 1971 | 90; 00 | 4 | 1996 |
| Le Tuan Hoa | 1957 | 13; 14; 05 | 59 | 1986 |
| Le Tung | | 65 | 1 | 2020 |
| Le Tung | | 20 | 2 | 2013 |
| Le Tung Son | 1979 | | 1 | 2007 |
| Le Van Bao | | | 7 | 1984 |
| Le Van Chong | | | 1 | 1986 |
| Le Van Cuong | 1946 | | 2 | 2001 |
| Le Van Dien | | | 2 | 1987 |
| Le Van Dung | 1979 | 60 | 1 | 2010 |
| Le Van Hap | 1957 | 35; 26; 46 | 7 | 1988 |
| Le Van Hien | 1979 | 34; 93; 92 | 16 | 2008 |
| Le Van Hien | | 90; 49 | 3 | 2017 |
| Le Van Hieu | 1975 | 35; 37; 34 | 7 | 2010 |
| Le Van Hoi | | | 1 | 2009 |
| Le Van Hop | | | 2 | 1986 |
| Le Van Hot | 1951 | | 2 | 1985 |
| Le Van Thanh | 1943 | 32; 14 | 12 | 1982 |
| Le Van Thanh | 1979 | 60; 52 | 16 | 2005 |
| Le Van Thiem | 1918 | 30 | 13 | 1947 |
| Le Van Thuyet | 1970 | 16 | 18 | 1991 |
| Le Van Ut | | 35 | 1 | 2005 |
| Le Van Vy | | 65; 47; 91 | 2 | 2020 |
| Le Viet Cuong | | 34 | 2 | 2019 |
| Le Xuan Can (Suan Le Kan) | | | 2 | 1974 |
| Le Xuan Dung | | 13 | 5 | 2012 |
| Le Xuan Hung | | 05; 34; 60 | 4 | 2003 |
| Le Xuan Huy | | 33; 44; 45 | 1 | 2013 |
| Le Xuan Quang | | | 11 | 1971 |
| Le Xuan Son | 1972 | 30; 26 | 5 | 2003 |
| Le Xuan Thanh | 1985 | 90; 47 | 4 | 2016 |
| Le Xuan Truong | 1978 | 35; 34; 42 | 47 | 2007 |
| Le Xuan Viet | | 93; 90 | 1 | 2008 |
| Lu Hoang Chinh | 1985 | 32; 53; 35 | 20 | 2013 |
| Luong Dang Ky | | 42; 46; 47 | 37 | 2011 |
| Luong Duc Trong | | 60; 65 | 3 | 2017 |
| Luong Thai Hung | | 35 | 2 | 2018 |
| Luong The Dung | | | 1 | 2013 |
| Luong Thi Tuyet | | 15; 30; 35 | 2 | 2018 |
| Luong Viet Chuong | | 14; 32 | 1 | 2020 |
| Luu Hoang Duc | 1981 | 37; 34; 35 | 17 | 2003 |
| Luu Hong Phong | | 35; 65 | 1 | 2018 |
| Luu Phuong Thao | | 13 | 1 | 2019 |
| Luu Thi Hiep | | | 1 | 2019 |
| Luu Vu Cam Hoan | | 35; 47; 74 | 5 | 2017 |
| Luu Xuan Thang | | 45; 65; 35 | 1 | 2020 |
| Luu Xuan Truong | | | 1 | 2019 |
| Ly Kim Ha | | 32; 41; 42 | 18 | 2014 |
| Mach Nguyet Minh | | 35; 47; 42 | 5 | 2009 |
| Mai  Thi Thu | | 26; 46; 45 | 7 | 2000 |
| Mai Anh Duc | 1975 | 32 | 3 | 2013 |
| Mai Duc Thanh | 1973 | 35; 76; 65 | 22 | 1995 |
| Mai Hai An | 1980 | 41 | 2 | 2019 |
| Mai Hoang Bien | 1982 | 16; 20; 05 | 20 | 2012 |
| Mai Thanh Nhat Truong | | 65; 35; 47 | 2 | 2017 |
| Mai Thi Kim Dung | | 35 | 1 | 2020 |
| Mai Thi Ngoc Ha | | | 1 | 2020 |
| Mai Van Tu | | 32; 11 | 3 | 1995 |
| Mai Viet Thuan | 1985 | 34; 93; 37 | 12 | 2012 |
| Mai Xuan Thao | 1957 | 41; 35; 42 | 8 | 2002 |
| My Vinh Quang | 1961 | 11 | 5 | 1987 |
| N. B. Minh | | 49; 90 | 2 | 2008 |
| N. H. Loi | | 41; 40 | 7 | 1985 |
| N. M. Tuan | | | 3 | 1992 |
| N.V. Loi | | 17; 16 | 2 | 1985 |
| Nghiem Do Quyen | | | 1 | 2008 |
| Ngo Anh Tu | | | 4 | 1978 |
| Ngo Anh Tuan | | 55 | 2 | 2019 |
| Ngo Bao Chau | 1972 | 22; 14; 11 | 26 | 1997 |
| Ngo Dac Tan | 1952 | 05; 20 | 41 | 1976 |
| Ngo Dac Tuan | | 11 | 2 | 2008 |
| Ngo Hoang Long | 1981 | 60; 65; 62 | 6 | 2006 |
| Ngo Huy Can | 1941 | 76; 41; 80 | 19 | 1969 |
| Ngo Lam Xuan Chau | | 34 | 1 | 2015 |
| Ngo Manh Hung | | | 1 | 1991 |
| Ngo Manh Tuong | | 65 | 1 | 2020 |
| Ngo Minh Man | 1987 | 49; 91; 93 | 1 | 2016 |
| Ngo Quang Hung | | 05; 68; 94 | 1 | 2003 |
| Ngo Quoc Anh | 1978 | 35; 26; 46 | 15 | 2005 |
| Ngo Quoc Chung | | 34; 60 | 1 | 2007 |
| Ngo Quoc Hoan | | | 3 | 2015 |
| Ngo Quy Dang | | 35; 34 | 1 | 2018 |
| Ngo Sy Tung | 1958 | 16 | 5 | 1994 |
| Ngo Tan Phuc | 1985 | 18; 05; 16 | 2 | 2017 |
| Ngo Thi Hien | | 68; 94 | 1 | 2018 |
| Ngo Thi Ngoan | 1980 | 20; 11; 14 | 6 | 2014 |
| Ngo Thoi Nhan | 1986 | 65; 49; 39 | 1 | 2017 |
| Ngo Van Giang | | | 1 | 2019 |
| Ngo Van Hoa | | 34; 47; 26 | 17 | 2012 |
| Ngo Van Luoc | | 30 | 18 | 1974 |
| Ngo Viet Trung | 1953 | 13; 14; 05 | 101 | 1978 |
| Nguyen Ai Viet | 1952 | | 14 | 1983 |
| Nguyen An Khuong | | 34 | 1 | 2015 |
| Nguyen An Sum | | | 1 | 2004 |
| Nguyen Anh Minh | | | 2 | 1978 |
| Nguyen Anh Tam | | | 1 | 2010 |
| Nguyen Anh Triet | | 35; 47; 65 | 10 | 2016 |
| Nguyen Anh Tuan | 1962 | | 3 | 1996 |
| Nguyen Ba Minh | 1953 | 49; 90; 54 | 15 | 2000 |
| Nguyen Bac Van | | 62; 15 | 7 | 1988 |
| Nguyen Bao Tran | | 26; 46; 47 | 2 | 2020 |
| Nguyen Bich Huy | | 47; 35 | 1 | 2018 |
| Nguyen Bich Huy | 1956 | 47; 35; 34 | 18 | 1980 |
| Nguyen Bich Van | 1985 | 05; 15; 37 | 3 | 2013 |
| Nguyen Buong | 1949 | 47; 65; 49 | 44 | 1985 |
| Nguyen Cam | 1954 | 30; 35 | 3 | 1995 |
| Nguyen Cam | | | 1 | 1970 |
| Nguyen Cang | | | 11 | 1972 |
| Nguyen Canh Luong | 1961 | 35; 30 | 5 | 1997 |
| Nguyen Canh Nam | | 90; 65; 49 | 3 | 2010 |
| Nguyen Canh Toan | 1926 | | 7 | 1962 |
| Nguyen Cao Menh | | | 4 | 1985 |
| Nguyen Cao Tri | | 32; 34 | 1 | 2020 |
| Nguyen Cat Ho | 1941 | 68; 03; 90 | 14 | 1972 |
| Nguyen Chan | | 65; 34 | 12 | 1980 |
| Nguyen Chanh Tu | 1965 | 14; 32 | 5 | 2001 |
| Nguyen Chi Liem | 1971 | 39; 65; 34 | 1 | 2013 |
| Nguyen Chien | | 16 | 1 | 2003 |
| Nguyen Chu Gia Vuong | | 22; 11 | 2 | 2004 |
| Nguyen Chuong | | | 1 | 1984 |
| Nguyen Chuong Khue (Nguen Cyong Kue) | | | 2 | 1975 |
| Nguyen Cong Dieu | | | 2 | 1993 |
| Nguyen Cong Minh | 1980 | 13; 05 | 10 | 2005 |
| Nguyen Cong Phuc | | 35; 31; 42 | 5 | 2006 |
| Nguyen Cong Tam | | 35 | 3 | 1981 |
| Nguyen Cong Thuy (Nguen Kong Tui) | | | 3 | 1965 |
| Nguyen D Tinh | | 34; 35; 60 | 1 | 2001 |
| Nguyen Dac Liem | 1960 | 35 | 6 | 1994 |
| Nguyen Dai Duong | 1985 | 14; 16; 13 | 3 | 2017 |
| Nguyen Dang Hanh | | | 1 | 1987 |
| Nguyen Dang Ho Hai | 1981 | 20 | 2 | 2004 |
| Nguyen Dang Hop | 1982 | 13; 14; 18 | 12 | 2015 |
| Nguyen Dang Minh | | 47; 35; 26 | 9 | 2015 |
| Nguyen Dang Phat | | | 1 | 1988 |
| Nguyen Dang Tuan | 1958 | 34 | 4 | 1988 |
| Nguyen Dang Tuyen | | | 1 | 2018 |
| Nguyen Dinh | 1958 | 90; 49; 39 | 55 | 1992 |
| Nguyen Dinh Binh | | 35 | 9 | 2000 |
| Nguyen Dinh Cong | 1960 | 34; 37; 60 | 51 | 1984 |
| Nguyen Dinh Dan | 1944 | 49; 90 | 5 | 1979 |
| Nguyen Dinh Hoa | | | 6 | 1984 |
| Nguyen Dinh Hoang | | 90; 49 | 2 | 2006 |
| Nguyen Dinh Huy | 1956 | 39; 45; 93 | 7 | 1990 |
| Nguyen Dinh Lan | | 46; 32 | 6 | 1998 |
| Nguyen Dinh Liem | 1986 | 35; 78; 65 | 22 | 2012 |
| Nguyen Dinh Ngoc | 1932 | | 4 | 1960 |
| Nguyen Dinh Phu | 1954 | 34; 47; 93 | 31 | 2006 |
| Nguyen Dinh Sang | | | 6 | 1980 |
| Nguyen Dinh Thuc | | | 1 | 1997 |
| Nguyen Dinh Tri | 1915 | 35 | 13 | 1966 |
| Nguyen Dinh Tuan | | 49; 90; 47 | 13 | 2006 |
| Nguyen Doan Tien | | | 1 | 1982 |
| Nguyen Doan Tuan | | 30; 32; 14 | 10 | 1986 |
| Nguyen Dong Anh | 1954 | 70; 60; 34 | 44 | 1978 |
| Nguyen Dong Phuong | | | 1 | 2016 |
| Nguyen Dong Yen | 1959 | 49; 90; 26 | 111 | 1985 |
| Nguyen Du Vi Nhan | | 35; 26; 44 | 1 | 2008 |
| Nguyen Duc Dat | 1947 | 06 | 3 | 1996 |
| Nguyen Duc Duyet | | 42 | 1 | 2020 |
| Nguyen Duc Hau | | | 1 | 2008 |
| Nguyen Duc Hien | 1976 | 65; 90; 49 | 2 | 2014 |
| Nguyen Duc Hieu | | | 5 | 1985 |
| Nguyen Duc Hoang | | 13; 05; 14 | 4 | 2001 |
| Nguyen Duc Manh | 1982 | | 2 | 2018 |
| Nguyen Duc Minh | 1963 | 13; 14; 16 | 10 | 1992 |
| Nguyen Duc Nghia | 1951 | 90 | 9 | 1982 |
| Nguyen Duc Phuong | | 35; 26; 47 | 9 | 2019 |
| Nguyen Duc Tam | | 13 | 1 | 2010 |
| Nguyen Duc Tuan | | | 5 | 1988 |
| Nguyen Dung | | 65; 44; 35 | 3 | 2005 |
| Nguyen Duong Toan | | 35; 45; 76 | 11 | 2012 |
| Nguyen Duy Binh | | | 1 | 1997 |
| Nguyen Duy Cuong | | 49; 90 | 4 | 2020 |
| Nguyen Duy Phuong | 1978 | 52; 51; 05 | 3 | 2017 |
| Nguyen Duy Tan | 1981 | 12; 20; 11 | 20 | 2004 |
| Nguyen Duy Thai Son | 1963 | 35; 26; 58 | 17 | 1988 |
| Nguyen Duy Thanh | | 47; 49; 35 | 2 | 2002 |
| Nguyen Duy Thinh | 1995 | | 1 | 2020 |
| Nguyen Duy Thuan | | | 2 | 1980 |
| Nguyen Duy Tien | 1942 | 60; 28 | 36 | 1972 |
| Nguyen Duy Truong | | 65 | 1 | 2019 |
| Nguyen Gia Dinh | 1961 | | 2 | 1996 |
| Nguyen H. Lam (HCMUT) | | 35; 78; 65 | 3 | 2016 |
| Nguyen Ha (Nguen Kha) | | | 2 | 1973 |
| Nguyen Ha Thanh | 1957 | 32; 46 | 4 | 1999 |
| Nguyen Hac Hai | | | 4 | 1992 |
| Nguyen Hai Chau | 1978 | 91 | 1 | 1999 |
| Nguyen Hai Dang | 1986 | 92; 60; 34 | 3 | 2012 |
| Nguyen Hai Nam | | | 3 | 2019 |
| Nguyen Hai Son | | 49; 35; 65 | 3 | 2010 |
| Nguyen Hieu Thao | | 49; 47; 90 | 1 | 2015 |
| Nguyen Ho Minh Duy | | 65; 35 | 1 | 2017 |
| Nguyen Hoai Minh | | 35; 78; 46 | 62 | 2005 |
| Nguyen Hoang | 1956 | 35; 26; 49 | 11 | 1994 |
| Nguyen Hoang Anh | | | 1 | 2013 |
| Nguyen Hoang Giang | | | 1 | 2019 |
| Nguyen Hoang Loc | | 35; 65; 78 | 32 | 2004 |
| Nguyen Hoang Luc | | 35; 47; 26 | 18 | 2018 |
| Nguyen Hoang Son | 1973 | 68; 05 | 4 | 2004 |
| Nguyen Hoang Thach | 1984 | | 3 | 2010 |
| Nguyen Hoang Thanh | 1981 | 54; 57 | 2 | 2009 |
| Nguyen Hoang Tuan | | 62; 47; 35 | 2 | 2018 |
| Nguyen Hoi Nghia | 1958 | 34; 35; 46 | 5 | 1998 |
| Nguyen Hong Chuong | 1960 | | 5 | 1988 |
| Nguyen Hong Duc | 1970 | 14; 32; 34 | 6 | 2008 |
| Nguyen Hong Hai | | | 1 | 2005 |
| Nguyen Hong Quan | | 49; 54; 47 | 13 | 2009 |
| Nguyen Hong Thai | 1963 | | 1 | 1992 |
| Nguyen Hung Chinh | | | 1 | 2017 |
| Nguyen Hung Son | 1957 | 20; 16 | 4 | 1986 |
| Nguyen Huong Lam | | 68; 94; 20 | 19 | 1990 |
| Nguyen Huu Anh | 1930 | 22; 17; 20 | 12 | 1967 |
| Nguyen Huu Bao | 1950 | 60; 62 | 5 | 1995 |
| Nguyen Huu Bi | | | 1 | 2002 |
| Nguyen Huu Can | | 35; 47; 26 | 32 | 2018 |
| Nguyen Huu Cong | 1964 | | 1 | 2018 |
| Nguyen Huu Cong | 1949 | 65; 34 | 43 | 1991 |
| Nguyen Huu Danh | 1994 | 49; 90; 91 | 1 | 2020 |
| Nguyen Huu Dien | 1954 | 26 | 3 | 1986 |
| Nguyen Huu Du | 1957 | 34; 60; 93 | 42 | 1986 |
| Nguyen Huu Duc | | | 11 | 1977 |
| Nguyen Huu Hoi (Hoi H. Nguyen) | | 11; 60; 15 | 29 | 2008 |
| Nguyen Huu Khanh | 1965 | 47; 35 | 3 | 2000 |
| Nguyen Huu Ngu (Nguen Hyu Ngy) | | | 1 | 1966 |
| Nguyen Huu Quang | 1960 | 52; 49; 58 | 3 | 1994 |
| Nguyen Huu Sau | | 93; 34; 37 | 6 | 2014 |
| Nguyen Huu Tho | 1976 | | 1 | 2002 |
| Nguyen Huu Tho | 1967 | 35; 49 | 2 | 2003 |
| Nguyen Huu Tro | 1946 | | 2 | 1981 |
| Nguyen Huu Tron | | 49; 90; 58 | 3 | 2017 |
| Nguyen Huu Tuyen | 1966 | 32 | 1 | 2003 |
| Nguyen Huu Vien | | | 1 | 1989 |
| Nguyen Huu Viet Hung | | 55; 18; 16 | 36 | 1981 |
| Nguyen Huy Chieu | 1979 | 49; 90; 54 | 22 | 2008 |
| Nguyen Huy Duc | | 35; 37; 76 | 1 | 2020 |
| Nguyen Huy Hoang | 1975 | 37; 35 | 4 | 2000 |
| Nguyen Huy Hung | | 13; 16; 20 | 2 | 2011 |
| Nguyen Huy Tuan | 1983 | 35; 47; 65 | 61 | 2006 |
| Nguyen Huy Tuan | | 35; 47; 26 | 45 | 2014 |
| Nguyen Huy Viet | | 54; 47; 60 | 6 | 1982 |
| Nguyen Huyen Muoi | | 34; 93 | 4 | 2015 |
| Nguyen Huynh Phan | 1955 | 93; 58; 55 | 10 | 1987 |
| Nguyen Khac Tin | | | 1 | 2015 |
| Nguyen Khac Viet | | 14; 11 | 21 | 1987 |
| Nguyen Khoa Son | 1948 | 93; 34; 47 | 66 | 1978 |
| Nguyen Kieu Linh | | 65 | 1 | 2015 |
| Nguyen Kieu Linh | | 51; 52 | 1 | 2020 |
| Nguyen Kim Khoi | | 65 | 1 | 1998 |
| Nguyen Kim Ngoc | 1985 | 20 | 1 | 2016 |
| Nguyen Kim Tan | | | 3 | 1993 |
| Nguyen Kong Shi | | | 1 | 1982 |
| Nguyen Ky Nam | | 62 | 8 | 2015 |
| Nguyen Lam Tung | 1981 | 90 | 1 | 2018 |
| Nguyen Le Anh | 1955 | 55 | 7 | 1981 |
| Nguyen Le Hoang Anh | 1984 | 90; 49; 46 | 9 | 2011 |
| Nguyen Le Huong | | 32 | 3 | 1993 |
| Nguyen Le Luc | | 47; 74; 55 | 4 | 2003 |
| Nguyen Long | | | 1 | 2012 |
| Nguyen Lu Trong Khiem | | 35; 76 | 3 | 2013 |
| Nguyen Luong Thai Binh | | 17 | 1 | 2019 |
| Nguyen Luu Son | | 93; 60 | 1 | 2020 |
| Nguyen Manh Cuong | 1982 | 42; 41 | 2 | 2017 |
| Nguyen Manh Hung | | | 2 | 1990 |
| Nguyen Manh Hung | | 90; 46 | 3 | 2004 |
| Nguyen Manh Hung | 1956 | 35; 74; 52 | 26 | 1988 |
| Nguyen Manh Linh | | 34; 49; 93 | 4 | 2001 |
| Nguyen Mau Nam | 1976 | 49; 90; 58 | 47 | 2005 |
| Nguyen Minh Chuong | 1931 | 47; 35; 46 | 71 | 1969 |
| Nguyen Minh Cong | | 41; 46; 40 | 2 | 2005 |
| Nguyen Minh Dien | | 35; 65; 49 | 5 | 2011 |
| Nguyen Minh Duc | | 41; 54 | 9 | 1984 |
| Nguyen Minh Ha | 1955 | 51; 46; 22 | 8 | 1994 |
| Nguyen Minh Hai | | | 2 | 2013 |
| Nguyen Minh Hang | | | 2 | 1997 |
| Nguyen Minh Khoa | | 45; 42; 44 | 10 | 2004 |
| Nguyen Minh Man | 1953 | 34; 65; 47 | 7 | 2003 |
| Nguyen Minh Sang | 1984 | 05; 11 | 2 | 2016 |
| Nguyen Minh Ta | | | 1 | 1990 |
| Nguyen Minh Trang | | 47 | 1 | 2019 |
| Nguyen Minh Tri | | 13 | 2 | 2017 |
| Nguyen Minh Tri | 1964 | 35; 76; 47 | 50 | 1987 |
| Nguyen Minh Tuan | 1969 | 45; 47; 30 | 10 | 1992 |
| Nguyen Minh Tuan | | 90; 49 | 3 | 1993 |
| Nguyen Minh Tung | | 90; 49; 91 | 6 | 2015 |
| Nguyen Mong | | | 2 | 1979 |
| Nguyen Nam Hai | | | 1 | 1988 |
| Nguyen Nam Hong | | | 5 | 1982 |
| Nguyen Nang Tam | 1953 | 90; 49; 32 | 22 | 1999 |
| Nguyen Nang Thieu | | 49; 90 | 1 | 2016 |
| Nguyen Ngoc Chu | 1955 | | 5 | 1980 |
| Nguyen Ngoc Cuong | | | 2 | 1987 |
| Nguyen Ngoc Cuong | | 32 | 1 | 2020 |
| Nguyen Ngoc Dai | | | 1 | 2013 |
| Nguyen Ngoc Dat | | | 1 | 2012 |
| Nguyen Ngoc Diep | | | 1 | 1972 |
| Nguyen Ngoc Doanh | | 35; 78; 80 | 4 | 2010 |
| Nguyen Ngoc Hai | | 52; 90; 26 | 16 | 1996 |
| Nguyen Ngoc Hai | | 47; 49; 90 | 1 | 2020 |
| Nguyen Ngoc Hung | | 20 | 32 | 2008 |
| Nguyen Ngoc Huy | | 60 | 1 | 2008 |
| Nguyen Ngoc Khanh | | 58; 35; 32 | 2 | 2015 |
| Nguyen Ngoc Luan | 1984 | 90; 46; 49 | 3 | 2018 |
| Nguyen Ngoc Nhu | | 34; 60; 92 | 1 | 2019 |
| Nguyen Ngoc Phan | 1980 | | 1 | 2003 |
| Nguyen Ngoc Phung | | | 1 | 2012 |
| Nguyen Ngoc San | | | 1 | 1996 |
| Nguyen Ngoc Thuan (Tkhuan, N. N) | | | 1 | 1998 |
| Nguyen Ngoc Toi | | | 1 | 1983 |
| Nguyen Ngoc Trong | | 42; 47; 35 | 16 | 2015 |
| Nguyen Ngoc Tu | | 60; 52 | 1 | 2019 |
| Nguyen Ngoc Tuu (Nguen Ngok Tyu) | | 90; 65 | 2 | 1980 |
| Nguyen Nhu Lan | | 35; 46; 65 | 2 | 2016 |
| Nguyen Nhu Ngoc | | 35 | 1 | 2020 |
| Nguyen Nhu Thang | | 35 | 2 | 2008 |
| Nguyen Nhut Hung | | | 8 | 2018 |
| Nguyen Nhuy | 1950 | 54; 46; 47 | 16 | 1981 |
| Nguyen Nuu Du | | 49 | 1 | 2004 |
| Nguyen Phuoc Tai | | 76; 35 | 3 | 2019 |
| Nguyen Phuong Hoa | 1996 | | 1 | 2020 |
| Nguyen Phuong Mai | | 35 | 1 | 2020 |
| Nguyen Phuong Thao | | 35 | 1 | 2019 |
| Nguyen Quan Son | 1967 | | 1 | 2006 |
| Nguyen Quang A | | | 1 | 1986 |
| Nguyen Quang Dieu | 1973 | 32; 31; 46 | 32 | 1996 |
| Nguyen Quang Hoa | | 41 | 3 | 1991 |
| Nguyen Quang Huy | 1973 | 90; 49; 46 | 16 | 2002 |
| Nguyen Quang Khanh | 1979 | | 1 | 2012 |
| Nguyen Quang Minh | | | 1 | 1987 |
| Nguyen Quang Thai | | | 6 | 1967 |
| Nguyen Quang Thang | | 35 | 1 | 2011 |
| Nguyen Quoc Dan (Nguen Kuok Zan) | | | 2 | 1968 |
| Nguyen Quoc Hung | | 35 | 1 | 2004 |
| Nguyen Quoc Hung (Shanghai) | 1988 | 35; 31; 76 | 24 | 2014 |
| Nguyen Quoc Son | 1937 | | 1 | 2004 |
| Nguyen Quoc Thang | 1957 | 20; 11; 14 | 53 | 1989 |
| Nguyen Quoc Thi | | | 2 | 1971 |
| Nguyen Quoc Tho | 1973 | 19; 46; 55 | 2 | 1999 |
| Nguyen Quoc Toan | | | 2 | 1978 |
| Nguyen Quoc Tuan | | 34; 93 | 1 | 2009 |
| Nguyen Quoc Tuan | | 49; 90 | 1 | 2020 |
| Nguyen Quy | | | 1 | 1988 |
| Nguyen Quy Hy | 1939 | 60 | 10 | 1974 |
| Nguyen Quy Khang | | 68; 20; 94 | 4 | 1985 |
| Nguyen Quyen | | 76; 35; 82 | 1 | 2020 |
| Nguyen Quynh Hoa | 1989 | 91; 49; 90 | 3 | 2016 |
| Nguyen Quynh Lan | | | 2 | 1989 |
| Nguyen Quynh Nga | 1976 | 47; 54 | 5 | 2001 |
| Nguyen S. N | | | 1 | 1986 |
| Nguyen Si Hoang | 1978 | 65 | 2 | 2006 |
| Nguyen Si Hoang (Nguen Si Hong) | | | 1 | 1968 |
| Nguyen Si Minh | | | 4 | 1977 |
| Nguyen Sinh Bay | 1951 | 34; 15; 39 | 6 | 1999 |
| Nguyen Song Ha | | 47; 49 | 2 | 2016 |
| Nguyen Sum | 1961 | 55 | 14 | 1992 |
| Nguyen Sy Anh Tuan | 1957 | | 2 | 1995 |
| Nguyen Sy Thai Ha | | | 1 | 2012 |
| Nguyen Sy Tuan (Nguen Si Tuen) | | | 1 | 1971 |
| Nguyen T. T. | | | 3 | 1991 |
| Nguyen T. Trinh | | 35 | 3 | 2018 |
| Nguyen Tam | | | 1 | 1968 |
| Nguyen Tam Nhan (Nhan-Tam Nguyen) | | | 2 | 2014 |
| Nguyen Tan Hoa | 1968 | 45; 47 | 4 | 2000 |
| Nguyen Tat Dat | 1999 | | 1 | 2020 |
| Nguyen Tat Thang | 1982 | 14; 32; 57 | 9 | 2011 |
| Nguyen Thac Dung | 1980 | 32; 58; 53 | 5 | 2005 |
| Nguyen Thai An | 1984 | 49; 90; 58 | 11 | 2012 |
| Nguyen Thai Hoa | 1978 | 13; 16 | 4 | 1999 |
| Nguyen Thai Son | 1959 | 32; 58 | 4 | 1998 |
| Nguyen Thanh Anh | | 35 | 1 | 2016 |
| Nguyen Thanh Bang | | | 2 | 1974 |
| Nguyen Thanh Binh | | 60; 91 | 3 | 2001 |
| Nguyen Thanh Chung | | 35; 34; 47 | 3 | 2020 |
| Nguyen Thanh Dieu | | 60; 34; 92 | 12 | 2013 |
| Nguyen Thanh Hai | | 44; 33 | 3 | 1991 |
| Nguyen Thanh Hao | 1982 | 47; 49 | 1 | 2006 |
| Nguyen Thanh Hong | 1980 | 44; 33; 45 | 6 | 2007 |
| Nguyen Thanh Hung | | 90 | 3 | 2019 |
| Nguyen Thanh Huong | | | 1 | 1988 |
| Nguyen Thanh Long | | | 1 | 1993 |
| Nguyen Thanh Long | 1972 | 90; 93 | 1 | 2004 |
| Nguyen Thanh Long | 1956 | 35; 65; 45 | 79 | 1986 |
| Nguyen Thanh Nhan | | 35; 65; 76 | 16 | 2013 |
| Nguyen Thanh Quang | 1957 | 11; 32; 30 | 9 | 1998 |
| Nguyen Thanh Qui | | 49 | 5 | 2010 |
| Nguyen Thanh Son | 1976 | 11 | 1 | 2018 |
| Nguyen Thanh Thien | | 37 | 1 | 2017 |
| Nguyen Thanh Thuy | | | 2 | 1985 |
| Nguyen Thanh Tuan | | 14 | 1 | 2006 |
| Nguyen Thanh Tung | 1986 | 47; 42; 35 | 3 | 2016 |
| Nguyen Thanh Tung | | 47; 34; 35 | 1 | 2020 |
| Nguyen Thanh Tung (Nanyang) | 1993 | 68; 93 | 1 | 2015 |
| Nguyen Thanh Van | | 31; 35; 30 | 9 | 2000 |
| Nguyen Thanh Van | 1943 | 30; 31; 35 | 5 | 1997 |
| Nguyen Thanh Vu | 1960 | 58; 35 | 2 | 2005 |
| Nguyen The Cuong | | 18; 55 | 4 | 2015 |
| Nguyen The Hoan | 1941 | 34; 41; 65 | 11 | 1968 |
| Nguyen The Hung | | | 1 | 2015 |
| Nguyen The Long | | | 2 | 1975 |
| Nguyen The Vinh | 1980 | 47; 65; 68 | 10 | 2005 |
| Nguyen Thi Bach Kim | 1961 | 90 | 9 | 2000 |
| Nguyen Thi Dung | 1964 | 13 | 5 | 2004 |
| Nguyen Thi Hien | | 34; 93; 94 | 3 | 2013 |
| Nguyen Thi Hoa | | 60 | 1 | 1995 |
| Nguyen Thi Hoai Phuong | | 90; 78; 65 | 6 | 2002 |
| Nguyen Thi Hong | | | 3 | 1968 |
| Nguyen Thi Hong | | 42; 47 | 2 | 2016 |
| Nguyen Thi Hong | | 34; 47; 90 | 3 | 2015 |
| Nguyen Thi Hong Loan | | 13 | 5 | 1999 |
| Nguyen Thi Hong Minh | 1968 | 65; 34 | 3 | 2000 |
| Nguyen Thi Hong Nhung | | 35; 46 | 1 | 2020 |
| Nguyen Thi Kieu | 1986 | | 2 | 2017 |
| Nguyen Thi Kieu Nga | 1975 | 13 | 1 | 2010 |
| Nguyen Thi Kim Anh (N. K. Anh) | 1963 | | 1 | 1991 |
| Nguyen Thi Kim Chuc | | | 4 | 2009 |
| Nguyen Thi Kim Son | | 03; 35; 46 | 2 | 2008 |
| Nguyen Thi Lan Huong | 1979 | 34; 32 | 2 | 2019 |
| Nguyen Thi Le | | 54 | 1 | 2008 |
| Nguyen Thi Le Trang | | 90 | 1 | 2007 |
| Nguyen Thi Loan | | 35; 34 | 1 | 2020 |
| Nguyen Thi Minh | | 60 | 1 | 1992 |
| Nguyen Thi Minh Hue | | 90 | 1 | 2006 |
| Nguyen Thi Minh Khai | | 42 | 1 | 2015 |
| Nguyen Thi Minh Toai | | 35; 76 | 1 | 2020 |
| Nguyen Thi Nga | | 35; 53 | 3 | 2004 |
| Nguyen Thi Ngan | | 45; 46; 42 | 2 | 2011 |
| Nguyen Thi Ngan | | 76; 35; 37 | 2 | 2020 |
| Nguyen Thi Ngan | | 45; 47; 42 | 1 | 2010 |
| Nguyen Thi Ngoc Anh | | | 1 | 2010 |
| Nguyen Thi Ngoc Diep | 1984 | 14; 11; 12 | 2 | 2012 |
| Nguyen Thi Ngoc Oanh | 1985 | 65 | 4 | 2016 |
| Nguyen Thi Nhung | | 30; 32 | 3 | 2017 |
| Nguyen Thi Phong | | 35; 45; 78 | 1 | 2015 |
| Nguyen Thi Phuong Dung | 1972 | 17; 16; 20 | 4 | 2003 |
| Nguyen Thi Quynh Anh | 1981 | 91; 49; 90 | 2 | 2011 |
| Nguyen Thi Quynh Phuong | | | 1 | 2017 |
| Nguyen Thi Quynh Trang | | 49; 90 | 2 | 2012 |
| Nguyen Thi Tam Bach | | | 3 | 1969 |
| Nguyen Thi Thanh Ha | | 90; 65; 47 | 2 | 2020 |
| Nguyen Thi Thanh Ha | | 47 | 2 | 2008 |
| Nguyen Thi Thanh Hien | | 32; 30 | 1 | 2018 |
| Nguyen Thi Thanh Huyen | | 34; 93 | 3 | 2016 |
| Nguyen Thi Thanh Huyen | | 68 | 1 | 2006 |
| Nguyen Thi Thanh Tam | 1993 | 13 | 2 | 2019 |
| Nguyen Thi Thao | 1979 | 14; 11; 26 | 6 | 2009 |
| Nguyen Thi Thao Truc | | 35 | 2 | 2005 |
| Nguyen Thi The | | 34; 60; 93 | 3 | 2010 |
| Nguyen Thi Thu Hang | | 32; 30 | 1 | 2017 |
| Nguyen Thi Thu Hien | | 60 | 1 | 2020 |
| Nguyen Thi Thu Huong | 1975 | 90; 49 | 7 | 2011 |
| Nguyen Thi Thu Thuy | | 47; 41; 49 | 9 | 2008 |
| Nguyen Thi Thu Van | | 90; 26 | 2 | 2003 |
| Nguyen Thi Thuy | | 60 | 1 | 2014 |
| Nguyen Thi Thuy Loan | | | 1 | 2013 |
| Nguyen Thi Thuy Quynh | | 93; 60; 34 | 3 | 2010 |
| Nguyen Thi Tinh | | | 3 | 1995 |
| Nguyen Thi Toan | 1982 | 49; 90; 93 | 10 | 2010 |
| Nguyen Thi Tuyet Mai | | 32; 46 | 5 | 2000 |
| Nguyen Thi Van Anh | | 13 | 1 | 2018 |
| Nguyen Thi Van Hang | 1987 | 90; 49 | 4 | 2014 |
| Nguyen Thi Vinh | | | 1 | 2016 |
| Nguyen Thi Yen Ngoc | | 65; 35 | 1 | 2015 |
| Nguyen Thi Yen Nhi | | 49; 90 | 1 | 2014 |
| Nguyen Thien Luan | | | 2 | 1985 |
| Nguyen Thiep | | | 1 | 1979 |
| Nguyen Thieu Huy | 1975 | 34; 35; 47 | 21 | 2001 |
| Nguyen Thinh | 1980 | 47 | 4 | 2004 |
| Nguyen Thu Ha | 1983 | 34; 06; 65 | 1 | 2016 |
| Nguyen Thu Hang | 1982 | 13 | 3 | 2017 |
| Nguyen Thu Huong | 1979 | 90 | 1 | 2019 |
| Nguyen Thu Nga | | 46 | 5 | 1991 |
| Nguyen Thu Thuy | | 65; 60 | 1 | 2020 |
| Nguyen Thu Thuy | | 65 | 1 | 2012 |
| Nguyen Thua Hop | 1932 | 35 | 17 | 1964 |
| Nguyen Thuan Bang | | | 1 | 1993 |
| Nguyen Thuc Loan | | | 9 | 1970 |
| Nguyen Thuong Uan | | | 6 | 1975 |
| Nguyen Thuy Anh | 1974 | | 1 | 1996 |
| Nguyen Thuy Thanh | 1941 | 41; 30 | 2 | 1983 |
| Nguyen Tien Da | 1987 | | 2 | 2017 |
| Nguyen Tien Dai | 1952 | 35 | 5 | 1980 |
| Nguyen Tien Dung | | 37; 53; 58 | 15 | 1997 |
| Nguyen Tien Dung | | 60; 90; 93 | 2 | 2008 |
| Nguyen Tien Khai | 1984 | 49; 58; 35 | 29 | 2007 |
| Nguyen Tien Khiem | 1955 | | 11 | 1983 |
| Nguyen Tien Manh | 1964 | | 1 | 2006 |
| Nguyen Tien Quang | | 18 | 6 | 1987 |
| Nguyen Tien Tai | | 30 | 3 | 1981 |
| Nguyen Tien Trung | | 46; 32; 35 | 2 | 2004 |
| Nguyen Tien Yet | 1985 | 34; 37 | 2 | 2011 |
| Nguyen To Nhu | 1953 | 54; 46; 58 | 40 | 1979 |
| Nguyen Toan Thang | | | 1 | 1989 |
| Nguyen Tran Thuan | | 60; 65; 52 | 2 | 2014 |
| Nguyen Trong Hieu | 1982 | | 1 | 2006 |
| Nguyen Trong Hoa | | 32; 30 | 2 | 2006 |
| Nguyen Trong Toan | | | 2 | 1982 |
| Nguyen Trong Toan (Nguyen T. Toan) | 1979 | 35; 76; 82 | 43 | 2003 |
| Nguyen Trung Hieu | | 47 | 1 | 2020 |
| Nguyen Trung Hung | 1945 | 62 | 2 | 1998 |
| Nguyen Trung Kien | 1986 | 90 | 5 | 2019 |
| Nguyen Trung Thanh | 1980 | 35; 65; 80 | 11 | 2007 |
| Nguyen Trung Thanh | | 68 | 12 | 2012 |
| Nguyen Truong Thanh | 1980 | | 13 | 2005 |
| Nguyen Tu Cuong | 1951 | 13; 14; 16 | 58 | 1977 |
| Nguyen Tu Thanh | | | 4 | 1972 |
| Nguyen Tuan Duy | | 35; 26; 46 | 3 | 2013 |
| Nguyen Tuan Long | 1981 | 13 | 2 | 2015 |
| Nguyen Tuong | 1938 | 35; 65 | 5 | 1978 |
| Nguyen V Chan | | | 1 | 1991 |
| Nguyen V Kh | | | 2 | 1986 |
| Nguyen Van An | | 32; 30 | 2 | 2017 |
| Nguyen Van Bao | | | 1 | 1980 |
| Nguyen Van Chau | 1955 | 14; 32; 34 | 22 | 1985 |
| Nguyen Van Co | 1950 | 11; 47; 54 | 5 | 1999 |
| Nguyen Van Dac | | | 1 | 2017 |
| Nguyen Van Dao | 1937 | 34 | 28 | 1968 |
| Nguyen Van Dinh | | 34 | 4 | 2000 |
| Nguyen Van Dong | 1955 | 32 | 6 | 1986 |
| Nguyen Van Duc | 1981 | 35; 47; 65 | 11 | 2008 |
| Nguyen Van Duc | | | 1 | 2012 |
| Nguyen Van Dung | 1981 | 54; 47; 39 | 7 | 2006 |
| Nguyen Van Gia | 1930 | | 2 | 1976 |
| Nguyen Van Giang | | | 4 | 1988 |
| Nguyen Van Hanh | 1980 | 53 | 1 | 2002 |
| Nguyen Van Hao | 1970 | 46 | 4 | 1998 |
| Nguyen Van Hien | | 90; 49; 47 | 11 | 2004 |
| Nguyen Van Hieu | 1938 | 34; 81 | 33 | 1963 |
| Nguyen Van Hieu | | | 2 | 1994 |
| Nguyen Van Ho | 1947 | 60; 62 | 12 | 1976 |
| Nguyen Van Hoang | | | 2 | 1993 |
| Nguyen Van Hoang | 1976 | 13; 18 | 7 | 2005 |
| Nguyen Van Hoang | | 46; 26; 35 | 1 | 2020 |
| Nguyen Van Hoang (1985) | 1985 | 26; 46; 52 | 26 | 2011 |
| Nguyen Van Hong | | | 1 | 2020 |
| Nguyen Van Huan | | 60 | 2 | 2014 |
| Nguyen Van Hung | | | 8 | 1987 |
| Nguyen Van Hung | | 47; 49; 58 | 3 | 2020 |
| Nguyen Van Huu | 1941 | 62; 60 | 8 | 1972 |
| Nguyen Van Khai | | | 3 | 1992 |
| Nguyen Van Khiem | 1977 | 41 | 4 | 2005 |
| Nguyen Van Khue | 1944 | 32; 46; 30 | 58 | 1965 |
| Nguyen Van Kinh | | 47; 65 | 3 | 1989 |
| Nguyen Van Long | 1963 | 68 | 1 | 2007 |
| Nguyen Van Luong | | 55; 49; 54 | 2 | 2014 |
| Nguyen Van Man | | | 1 | 1991 |
| Nguyen Van Manh | 1954 | | 1 | 1990 |
| Nguyen Van Mau | 1951 | 45; 47; 34 | 39 | 1982 |
| Nguyen Van Minh | | 65; 34 | 2 | 2007 |
| Nguyen Van Minh | 1966 | 34; 47; 35 | 114 | 1987 |
| Nguyen Van Minh | | | 1 | 2017 |
| Nguyen Van Minh (FTU) | | 58; 37; 53 | 2 | 2013 |
| Nguyen Van Nghi | | | 1 | 2000 |
| Nguyen Van Ngoc | 1949 | 42; 46; 47 | 18 | 1979 |
| Nguyen Van Nhan | | 35; 30; 41 | 6 | 1998 |
| Nguyen Van Ninh | | 14; 52; 55 | 1 | 2017 |
| Nguyen Van Phu | | 32 | 4 | 2012 |
| Nguyen Van Quang | 1957 | 60; 46; 47 | 18 | 1992 |
| Nguyen Van Quang | | 35; 90 | 3 | 2010 |
| Nguyen Van Quy | | 90; 47; 65 | 9 | 2001 |
| Nguyen Van Sanh | | 16 | 16 | 1993 |
| Nguyen Van Son | 1977 | | 1 | 2003 |
| Nguyen Van Son | | | 1 | 2016 |
| Nguyen Van Su | | | 2 | 1984 |
| Nguyen Van Thang | 1987 | 35; 47; 65 | 4 | 2015 |
| Nguyen Van Thanh | | 35; 60; 37 | 5 | 2012 |
| Nguyen Van The | 1998 | | 1 | 2020 |
| Nguyen Van Thin | | 30; 35; 32 | 21 | 2014 |
| Nguyen Van Thinh | | 35; 47; 65 | 9 | 2013 |
| Nguyen Van Thoai | 1950 | 90; 65 | 29 | 1978 |
| Nguyen Van Thu | 1946 | 60; 19; 33 | 35 | 1976 |
| Nguyen Van Thuong | | | 3 | 1984 |
| Nguyen Van Tin | | | 1 | 1975 |
| Nguyen Van Toan | | 62; 60 | 7 | 1998 |
| Nguyen Van Trao | 1973 | 32 | 9 | 2000 |
| Nguyen Van Truyen | | 35; 22; 26 | 4 | 2005 |
| Nguyen Van Tuan | | | 5 | 1995 |
| Nguyen Van Tuyen | | 90; 49; 58 | 9 | 2012 |
| Nguyen Van Ty | | | 2 | 1977 |
| Nguyen Van Vu | | 49; 90 | 1 | 2020 |
| Nguyen Van Vy | 1944 | | 5 | 1977 |
| Nguyen Van Y | | 35; 74 | 2 | 2016 |
| Nguyen Viet Anh | 1974 | 32; 37; 60 | 9 | 2010 |
| Nguyen Viet Duc | 1959 | | 1 | 1994 |
| Nguyen Viet Dung | | 14; 16; 32 | 20 | 1983 |
| Nguyen Viet Dung | | 16; 18 | 23 | 1988 |
| Nguyen Viet Hai | | 57; 22 | 3 | 2001 |
| Nguyen Viet Hung | 1973 | 28 | 2 | 2006 |
| Nguyen Viet Linh | | 35; 53 | 18 | 2003 |
| Nguyen Viet Phu | | | 2 | 1974 |
| Nguyen Viet Phuong | | 30; 35 | 2 | 2017 |
| Nguyen Viet Trieu Tien | 1950 | | 1 | 1985 |
| Nguyen Viet Tuan | | 35 | 2 | 2017 |
| Nguyen Vu Duy Linh | | 26 | 2 | 2001 |
| Nguyen Vu Huy | | 44; 65; 47 | 3 | 2002 |
| Nguyen Vu Luong | 1954 | 45; 47; 39 | 2 | 1996 |
| Nguyen Vu Tien | 1951 | 49 | 2 | 1997 |
| Nguyen Xuan Ha | | 26; 90; 49 | 3 | 2000 |
| Nguyen Xuan Hai | | 49; 47; 91 | 7 | 2006 |
| Nguyen Xuan Hoai | | | 2 | 2017 |
| Nguyen Xuan Hong | 1983 | 32 | 16 | 2010 |
| Nguyen Xuan Hung | | 90 | 1 | 2001 |
| Nguyen Xuan Hung | | 65; 74 | 1 | 2015 |
| Nguyen Xuan Huy | 1957 | | 12 | 1985 |
| Nguyen Xuan Ky | | 41; 42 | 26 | 1972 |
| Nguyen Xuan Lai | | 11; 30 | 1 | 2012 |
| Nguyen Xuan Loc | 1935 | 60; 31 | 21 | 1967 |
| Nguyen Xuan My | 1940 | | 16 | 1980 |
| Nguyen Xuan Nguyen | | | 6 | 1968 |
| Nguyen Xuan Tan | 1950 | 49; 90; 47 | 63 | 1980 |
| Nguyen Xuan Thanh | 1971 | | 2 | 2009 |
| Nguyen Xuan Thanh | | 35; 76; 65 | 2 | 2020 |
| Nguyen Xuan Thao | 1957 | 44; 33; 45 | 28 | 1994 |
| Nguyen Xuan Thuan | | 60; 47 | 6 | 2001 |
| Nguyen Xuan Tu | | 35 | 1 | 2017 |
| Nguyen Xuan Tuyen | 1949 | 16; 18; 20 | 17 | 1976 |
| Ninh Quang Hai | 1952 | 70 | 2 | 1999 |
| Ninh Van Thu | 1980 | 32; 37; 51 | 11 | 2009 |
| Nong Quoc Chinh | | | 1 | 2005 |
| Oanh Nguyen | 1989 | 60; 05; 68 | 7 | 2016 |
| Ong Thanh Hai | | 65; 74; 35 | 3 | 2015 |
| P. H. Thoa | | | 1 | 1991 |
| P. T. Thuy | | 35 | 3 | 2009 |
| Pham An Vinh | 1983 | 13; 14 | 1 | 2016 |
| Pham Anh Minh | 1960 | 20; 55; 17 | 33 | 1982 |
| Pham Canh Duong | 1949 | 90; 65 | 7 | 1978 |
| Pham Chi Vinh | 1959 | | 2 | 2017 |
| Pham D. Dat | | 49; 90 | 2 | 2018 |
| Pham Dinh Huong | | 54; 32 | 1 | 2003 |
| Pham Dinh Tao | 1955 | 90; 65; 49 | 14 | 1995 |
| Pham Dinh Tung | | 62 | 6 | 2015 |
| Pham Du | 1976 | 65 | 1 | 2006 |
| Pham Duc Chinh | 1958 | 74 | 23 | 1991 |
| Pham Duc Hiep | 1984 | 35 | 2 | 2017 |
| Pham Duc Thoan | | 32; 14; 30 | 15 | 2011 |
| Pham Duy Dat | | | 1 | 2012 |
| Pham Duy Khanh | | 49; 47; 26 | 9 | 2013 |
| Pham Gia Hung | 1963 | 49; 90 | 3 | 2011 |
| Pham Gia Thu | 1944 | 62 | 3 | 2001 |
| Pham Hien Bang | 1955 | 46; 32; 31 | 6 | 1998 |
| Pham Hoang Giang | 1995 | | 2 | 2017 |
| Pham Hoang Ha | 1981 | 30; 32; 53 | 7 | 2010 |
| Pham Hoang Hiep | 1982 | 32; 31; 14 | 39 | 2005 |
| Pham Hoang Quan | 1966 | 35; 65; 44 | 13 | 2004 |
| Pham Hong Nam | | 13 | 2 | 2015 |
| Pham Hong Quang | 1966 | | 10 | 1982 |
| Pham Hung Quy | 1983 | 13; 14 | 26 | 2010 |
| Pham Huu Anh Ngoc | 1967 | 34; 93; 39 | 35 | 1998 |
| Pham Huu Khanh | 1975 | 13 | 2 | 2010 |
| Pham Huu Sach | 1941 | 49; 90; 26 | 80 | 1968 |
| Pham Huu Tiep | | 20; 11; 17 | 160 | 1986 |
| Pham Huu Tri | | 35 | 1 | 1995 |
| Pham Huy Dien | 1949 | 90; 49 | 18 | 1981 |
| Pham Huy Tung | 1977 | 26 | 1 | 2007 |
| Pham Khac Ban | | 58; 32 | 5 | 1991 |
| Pham Kim Quy | | 65; 47; 91 | 3 | 2020 |
| Pham Ky Anh | 1949 | 47; 65; 34 | 46 | 1974 |
| Pham Loi Vu | 1934 | 35; 37; 34 | 29 | 1971 |
| Pham Minh Hien | 1958 | 65; 35; 34 | 4 | 2002 |
| Pham Minh Thong | | 39; 34; 92 | 1 | 2011 |
| Pham Ngoc Anh | 1956 | 16; 13; 06 | 50 | 1978 |
| Pham Ngoc Anh | 1970 | 90; 65; 49 | 21 | 2004 |
| Pham Ngoc Anh Cuong | 1962 | 57; 55; 19 | 3 | 1985 |
| Pham Ngoc Boi | 1959 | 34; 41; 47 | 4 | 1999 |
| Pham Ngoc Can (Fam Ngok Can) | | | 1 | 1968 |
| Pham Ngoc Khoi | | | 2 | 1986 |
| Pham Ngoc Mai | | 32; 54 | 3 | 2003 |
| Pham Ngoc Thao | 1946 | | 2 | 1988 |
| Pham Nguyen Thu Trang | 1980 | 32; 54; 46 | 6 | 2003 |
| Pham Phu Phat | | 14; 32; 58 | 2 | 2018 |
| Pham Quang Dung | | | 3 | 2012 |
| Pham Quang Hung | 1965 | | 2 | 1995 |
| Pham Quang Trinh | | 46; 54; 18 | 4 | 1997 |
| Pham Quy Muoi | 1980 | 35; 65; 49 | 5 | 2013 |
| Pham Thanh Duoc | | 90; 49; 91 | 1 | 2020 |
| Pham Thanh Duong | | 65; 41; 35 | 3 | 2020 |
| Pham Thanh Hieu | | 47; 41 | 2 | 2016 |
| Pham Thanh Huyen | | | 1 | 2019 |
| Pham Thanh Son | | 35 | 1 | 2011 |
| Pham Thanh Tri | 1950 | | 1 | 1986 |
| Pham The Anh | 1985 | 37; 60; 47 | 6 | 2014 |
| Pham The Lai | | | 1 | 1978 |
| Pham The Long | 1954 | 90; 65 | 12 | 1981 |
| Pham The Que | | | 2 | 1987 |
| Pham Thi Bach Ngoc | 1974 | 47; 34; 15 | 4 | 2001 |
| Pham Thi Hoai | | 47; 49; 90 | 1 | 2020 |
| Pham Thi Lan | | | 2 | 2009 |
| Pham Thi Minh Tam | | 35; 47 | 2 | 2017 |
| Pham Thi Thu Hoai | | 47; 49 | 1 | 2020 |
| Pham Thi Thuy | | 35 | 1 | 2019 |
| Pham Thi Trang | 1987 | 37; 35 | 5 | 2013 |
| Pham Thi Vui | | 54; 49; 80 | 1 | 2020 |
| Pham Tien Son | 1964 | 32; 14; 90 | 47 | 1997 |
| Pham Tra An | 1944 | 68 | 9 | 1976 |
| Pham Tran Nhu | 1945 | | 19 | 1979 |
| Pham Tri Nguyen | 1980 | 60; 28 | 1 | 2015 |
| Pham Trieu Duong | 1971 | 35 | 3 | 2004 |
| Pham Trong Quat | 1950 | | 3 | 1984 |
| Pham Trong Tien | 1987 | 47; 32; 30 | 9 | 2017 |
| Pham Trung Kien | | 90 | 3 | 2000 |
| Pham Truong Hoang Nhan | 1996 | | 1 | 2020 |
| Pham Tuan A | | | 1 | 2019 |
| Pham Tuan Anh | | | 2 | 2017 |
| Pham Van Bang | | 34; 35 | 1 | 2017 |
| Pham Van Binh | | | 1 | 2012 |
| Pham Van Chien | | | 1 | 2019 |
| Pham Van Chung | | 11 | 5 | 1993 |
| Pham Van Chung | 1960 | 62; 60 | 1 | 2007 |
| Pham Van Hai | | | 2 | 2019 |
| Pham Van Hien | | 47; 34; 35 | 1 | 2020 |
| Pham Van Hoang | | 76; 42; 47 | 1 | 2016 |
| Pham Van Huy | | 90; 65; 49 | 3 | 2020 |
| Pham Van Loi | | 49; 58; 65 | 1 | 2004 |
| Pham Van Luong | 1988 | | 1 | 2012 |
| Pham Van Minh | | 16 | 1 | 2004 |
| Pham Van Quoc | | 49 | 1 | 2004 |
| Pham Van Son | | 90; 49; 47 | 2 | 2007 |
| Pham Van Thang | 1990 | 11; 05; 52 | 29 | 2012 |
| Pham Van Thao | | 68 | 2 | 1999 |
| Pham Van Trung | 1985 | 05; 68; 90 | 12 | 2011 |
| Pham Van Viet | | 35; 34 | 5 | 2004 |
| Pham Viet Duc | | 32; 30 | 9 | 1999 |
| Pham Viet Hai | 1986 | 47; 34; 30 | 19 | 2010 |
| Pham Viet Hung | 1986 | 60; 26; 30 | 9 | 2013 |
| Pham Viet Hung | 1955 | 16; 15; 20 | 10 | 1986 |
| Pham Vu Long | | | 1 | 2012 |
| Pham Xuan Binh | | 60; 62 | 3 | 1998 |
| Pham Xuan Du | 1976 | 65; 35; 34 | 4 | 2002 |
| Pham Xuan Huyen | 1980 | 60; 93; 91 | 77 | 1996 |
| Pham Xuan Trung | | 90 | 1 | 2006 |
| Phan Dan | | 16 | 9 | 1988 |
| Phan Dang Cau | | | 5 | 1980 |
| Phan Dinh Dieu | 1936 | | 28 | 1965 |
| Phan Dinh Phung | | 34; 26; 47 | 8 | 2011 |
| Phan Duc Chau | | | 3 | 1981 |
| Phan Duc Chinh | 1936 | | 2 | 1965 |
| Phan Duc Tinh | | | 2 | 1964 |
| Phan Duc Tuan | 1980 | 11; 30; 32 | 6 | 2004 |
| Phan H. Giang | 1963 | 68; 62; 91 | 3 | 1994 |
| Phan Hoang Chon | | 55; 57 | 2 | 2011 |
| Phan Hong Tin | 1981 | 16 | 2 | 2013 |
| Phan Huu San | 1950 | | 2 | 1992 |
| Phan Huy Khai | 1951 | | 20 | 1981 |
| Phan Huy Phu | 1951 | | 2 | 1986 |
| Phan Huy Thien | 1956 | 46 | 1 | 1994 |
| Phan Le Na | | 60; 34 | 3 | 1997 |
| Phan M Hung | | 49; 90; 65 | 23 | 2011 |
| Phan Minh Dung | 1954 | | 8 | 1986 |
| Phan Ngoc Vinh | | 76; 35 | 1 | 2002 |
| Phan Nhat Tinh | 1962 | 49; 90; 47 | 11 | 1998 |
| Phan Phien | | 53; 14 | 1 | 2014 |
| Phan Quoc Hung | 1984 | 35; 58; 53 | 9 | 2008 |
| Phan Quoc Khanh | 1946 | 90; 49; 47 | 127 | 1979 |
| Phan Tang Da | | | 3 | 1972 |
| Phan Thanh An | 1968 | 52; 26; 65 | 38 | 1996 |
| Phan Thanh Hong | | | 3 | 2018 |
| Phan Thanh Nam | 1974 | 34; 30; 37 | 15 | 2005 |
| Phan Thanh Tung | | 41 | 2 | 2019 |
| Phan Thanh Viet | | 46; 35 | 1 | 2020 |
| Phan Thi Ha Duong | 1973 | 05; 68; 06 | 32 | 2000 |
| Phan Thi Ha Trang | | | 1 | 2018 |
| Phan Thi Huong | | 60; 65; 34 | 4 | 2018 |
| Phan Thi Khanh Van | | 35 | 3 | 2019 |
| Phan Thi Thuy | | 05; 13 | 2 | 2019 |
| Phan Thien Danh | | | 3 | 1996 |
| Phan Thien Nhan | 1952 | | 2 | 1997 |
| Phan Thien Thach | | 90; 49; 55 | 40 | 1985 |
| Phan Thoai | | | 1 | 1986 |
| Phan Tri Kien | | 60 | 2 | 2020 |
| Phan Trong Tien | | 49; 35 | 1 | 2020 |
| Phan Trung Hieu | | 42; 35; 34 | 1 | 2011 |
| Phan Trung Huy | | 68; 20; 94 | 12 | 1992 |
| Phan Tu Vuong | 1983 | 47; 49; 65 | 6 | 2012 |
| Phan Van Chuong | 1934 | 45; 54; 47 | 26 | 1966 |
| Phan Van Hap | 1938 | 65; 46; 47 | 13 | 1965 |
| Phan Van Thien | 1964 | 13 | 3 | 1999 |
| Phan Van Tri | | | 4 | 2014 |
| Phan Van Tuoc | 1978 | 35; 47; 55 | 5 | 2004 |
| Phan Viet Thu | 1951 | 46; 81 | 4 | 1993 |
| Phan Xuan Thanh | 1981 | 65; 35 | 7 | 2012 |
| Phi Ha | | 34; 15; 39 | 2 | 2018 |
| Pho Duc Tai | 1979 | 14; 32 | 5 | 2001 |
| Pho Duc Tru | | | 1 | 1974 |
| Pho Huy Tai | 1972 | 14; 32 | 1 | 2000 |
| Phong Thi Thu Huyen | 1989 | 51; 65; 52 | 1 | 2019 |
| Phung Ho Hai | 1970 | 16; 14; 17 | 28 | 1999 |
| Phung Minh Duc | | 65; 90 | 2 | 2016 |
| Phung Van Manh | | 41; 32; 26 | 9 | 2006 |
| Pierre N.V. Tu | | | 1 | 1994 |
| Quach Van Chuong | | | 1 | 2020 |
| Quang Ha (Q.P. Ha or Q. Ha) | | | 3 | 2009 |
| Sa Thi Lan Anh | 1988 | | 1 | 2018 |
| Si Duc Quang | 1981 | 32; 30; 14 | 52 | 2004 |
| T. T. Khanh | | 35 | 1 | 2010 |
| Ta Duy Phuong | 1952 | 90; 49; 65 | 12 | 1985 |
| Ta Hong Quang | 1957 | | 2 | 1986 |
| Ta Khac Cu | | 54; 60; 58 | 2 | 1989 |
| Ta Le Loi | 1960 | 14; 32; 03 | 13 | 1994 |
| Ta Ngoc Anh | | 47; 60; 54 | 2 | 2010 |
| Ta Ngoc Cau | | | 1 | 1989 |
| Ta Ngoc Tri | | 42; 65; 44 | 3 | 2002 |
| Ta Quang Son | | 90; 65; 41 | 4 | 2007 |
| Ta Quoc Bao | | 60; 45 | 2 | 2003 |
| Ta Thi Hoai An | 1971 | 30; 11; 12 | 30 | 2001 |
| Ta Thi Hong Yen | | 35 | 1 | 2011 |
| Ta Thi Huyen Trang | | 34; 93; 49 | 2 | 2016 |
| Ta Thi Thanh Mai | | 41 | 1 | 2020 |
| Ta Van Dinh | 1949 | 65; 35 | 14 | 1964 |
| Ta Van Tu | | 90 | 4 | 2000 |
| Ta Viet Ton | | 47; 60; 92 | 1 | 2010 |
| Tang Quoc Bao | | 35; 37; 92 | 34 | 2010 |
| Tang Thi Ha Yen | | 90 | 1 | 2007 |
| Tang Van Long | 1976 | 32; 31; 46 | 7 | 2002 |
| Thai Doan Chuong | 1979 | 49; 90; 65 | 6 | 2010 |
| Thai Quynh Phong | 1953 | 90; 49 | 6 | 1990 |
| Thai Thi Kim Chung | 1983 | 35 | 2 | 2019 |
| Thai Thuan Quang | 1966 | 46; 32; 30 | 15 | 1995 |
| Thai Trung Hieu | | 20; 39 | 1 | 2011 |
| Than Quang Khoat | | 68 | 1 | 2011 |
| Than Van Dinh | | 49; 47 | 1 | 2020 |
| Thanh V. V | | | 3 | 1993 |
| Thieu Dinh Phong | 1980 | 13 | 1 | 2015 |
| Thieu Thi Kim Thoa (Thoa Thieu) | | 65; 35 | 1 | 2020 |
| Thuc Manh Le | | 37; 39; 92 | 2 | 2011 |
| Thuy Do | 1995 | | 2 | 2017 |
| Tit Bau | | 70 | 1 | 1997 |
| To Tat Dat | 1976 | 35; 32 | 2 | 2019 |
| Ton Quoc Binh | 1954 | | 1 | 1987 |
| Ton That Tri | 1956 | 20 | 8 | 1995 |
| Tong Thanh Trung | 1974 | 92; 93; 65 | 3 | 2007 |
| Tong Viet Phi Hung | 1978 | 20; 05 | 55 | 2003 |
| Tra Quoc Khanh | | 47; 65; 31 | 3 | 2020 |
| Tran An Hai | | 30; 32 | 4 | 2018 |
| Tran An Hai | | | 1 | 2008 |
| Tran Bao Anh | | 34; 47 | 1 | 2008 |
| Tran Bao Ngoc | | 35; 26; 47 | 16 | 2018 |
| Tran Cao Nguyen | | | 1 | 1986 |
| Tran Dan Thu | 1966 | 05 | 4 | 1993 |
| Tran Dang Hong | | | 1 | 1986 |
| Tran Dang Hung | | | 1 | 2013 |
| Tran Dang Phuc | 1997 | | 1 | 2016 |
| Tran Dao Dong | 1955 | 53; 22; 14 | 6 | 1989 |
| Tran Dinh Duc | | 11; 30 | 2 | 2008 |
| Tran Dinh Ke | 1972 | 35; 47; 34 | 14 | 2004 |
| Tran Dinh Khang | | 68; 91 | 2 | 1999 |
| Tran Dinh Long | 1963 | 47 | 1 | 2002 |
| Tran Dinh Luong | | 16; 55 | 1 | 2007 |
| Tran Dinh Phung | | 26 | 1 | 2017 |
| Tran Dinh Quoc | | 90; 65; 49 | 8 | 2004 |
| Tran Dinh Thanh | | 45; 49; 47 | 3 | 2002 |
| Tran Dinh Tuong | | 60; 37; 92 | 3 | 2017 |
| Tran Dong Xuan | | 35; 46; 65 | 2 | 2018 |
| Tran Duc Anh | 1987 | 32; 30 | 2 | 2016 |
| Tran Duc Dung | 1988 | 13 | 1 | 2019 |
| Tran Duc Khanh | | | 1 | 2013 |
| Tran Duc Van | 1951 | 35; 46; 34 | 60 | 1975 |
| Tran Gia Lich | 1946 | 35; 76 | 2 | 1992 |
| Tran Gia Loc | 1959 | 39; 33 | 1 | 2012 |
| Tran Giang Nam | 1982 | 16; 18; 06 | 19 | 2007 |
| Tran Hoang Yen | | 65 | 1 | 1990 |
| Tran Hong Mo | 1979 | 39; 46; 49 | 10 | 2012 |
| Tran Hong Thai | | 39 | 1 | 2014 |
| Tran Hue Minh | 1977 | 32 | 4 | 2004 |
| Tran Hue Nuong | 1959 | 49; 90 | 8 | 1974 |
| Tran Hung Cuong | | | 2 | 2019 |
| Tran Hung Thao | 1940 | 60; 93 | 15 | 1982 |
| Tran Huu Nam | | | 4 | 1996 |
| Tran Huy Ho | 1942 | 35 | 2 | 1985 |
| Tran Huy Hoang | | | 1 | 1978 |
| Tran Kim Chi | | | 4 | 1980 |
| Tran Kim Dung | | | 1 | 1995 |
| Tran Kim Thanh | 1973 | 60; 62 | 3 | 1996 |
| Tran Kim Tu (Chan Kim T'i) | | | 2 | 1987 |
| Tran Le Nam | 1983 | 53 | 1 | 2008 |
| Tran Loc Hung | 1954 | 60; 62; 68 | 9 | 2001 |
| Tran Loc Hung | | 60; 41 | 3 | 2019 |
| Tran Luu Cuong | | | 1 | 1993 |
| Tran Manh Cuong | | 60; 37 | 2 | 2009 |
| Tran Manh Tuan | | 13 | 1 | 2011 |
| Tran Minh Binh | | 35; 47; 82 | 5 | 2009 |
| Tran Minh Hoang | | 30; 35 | 1 | 2008 |
| Tran Minh Ngoc | | 60 | 3 | 2005 |
| Tran Minh Nguyet | | 49; 35; 76 | 4 | 2016 |
| Tran Minh Phuong | 1982 | 35; 65; 78 | 12 | 2017 |
| Tran Minh Thuyet | | 46; 35; 34 | 4 | 1999 |
| Tran Minh Tuoc | 1970 | 05 | 2 | 2003 |
| Tran Nam Trung | 1979 | 13; 05; 14 | 21 | 2008 |
| Tran Ngoc | | 47; 35 | 1 | 2020 |
| Tran Ngoc Danh | 1971 | 06; 05 | 4 | 1996 |
| Tran Ngoc Diem | | 35 | 5 | 1997 |
| Tran Ngoc Giao | 1949 | 32 | 5 | 1991 |
| Tran Ngoc Ha | | | 1 | 2014 |
| Tran Ngoc Hoi | | | 5 | 1987 |
| Tran Ngoc Lien | | 65; 30; 33 | 4 | 2003 |
| Tran Ngoc Minh | | | 1 | 1993 |
| Tran Ngoc Nam | 1970 | 55 | 4 | 2000 |
| Tran Ngoc Nhat Huyen | 1996 | | 1 | 2020 |
| Tran Ngoc Tam | | 91; 90; 49 | 6 | 2012 |
| Tran Ngoc Thach | | 35; 47; 62 | 14 | 2018 |
| Tran Ngoc Tuan | | 34; 93 | 1 | 2019 |
| Tran Nguyen An | | 13 | 1 | 2019 |
| Tran Nguyen Binh | 1984 | | 1 | 2020 |
| Tran Nhat Luan | | 31; 47; 65 | 3 | 2020 |
| Tran Ninh Hoa | | 90; 49 | 4 | 2005 |
| Tran Phuong Thuy Lam | | 45; 35 | 1 | 2020 |
| Tran Quan Ky | | 60; 34; 92 | 1 | 2020 |
| Tran Quang Hung | | 32 | 1 | 2019 |
| Tran Quang Vinh | 1973 | 60 | 3 | 2001 |
| Tran Quoc Binh | 1969 | 47; 65; 54 | 8 | 1996 |
| Tran Quoc Chien | 1950 | | 3 | 1989 |
| Tran Quoc Cong | | 52; 32 | 1 | 2014 |
| Tran Quoc Duy | | 49; 90 | 1 | 2016 |
| Tran Quoc Tuan | | 35; 34 | 1 | 2020 |
| Tran Quoc Viet | | | 1 | 1984 |
| Tran Quoc Viet (DTU) | | 35; 62; 65 | 10 | 2013 |
| Tran Quyet Thang | 1955 | | 5 | 1987 |
| Tran Tat Dat | 1982 | 34; 26 | 5 | 2005 |
| Tran Thai An Nghia | 1979 | 49; 90; 65 | 23 | 2008 |
| Tran Thai Bao | | 93 | 1 | 2007 |
| Tran Thai Duong | | 05 | 1 | 2015 |
| Tran Thai Son | | | 2 | 1986 |
| Tran Thanh | | | 1 | 1990 |
| Tran Thanh (UNSW) | | 35; 47; 65 | 4 | 2015 |
| Tran Thanh Binh | | 47; 35 | 2 | 2020 |
| Tran Thanh Binh (SGU) | | 35; 47; 92 | 12 | 2014 |
| Tran Thanh Tuan | 1980 | | 1 | 2017 |
| Tran Thanh Tung | | 34; 93 | 5 | 2006 |
| Tran The Anh | | 45; 34 | 1 | 2018 |
| Tran The Hung | | 35; 47; 42 | 4 | 2015 |
| Tran The Hung | | | 1 | 2015 |
| Tran Thi Hue | 1958 | 49; 90; 65 | 4 | 1993 |
| Tran Thi Huong | | 47; 49 | 1 | 2020 |
| Tran Thi Huong Anh | | 65; 90 | 1 | 2017 |
| Tran Thi Huyen Thanh | | 90; 65; 49 | 2 | 2020 |
| Tran Thi Khieu | | 31; 65; 47 | 2 | 2020 |
| Tran Thi Kim Chi | 1945 | 34 | 5 | 1991 |
| Tran Thi Lan Anh | | 47; 54 | 4 | 1999 |
| Tran Thi Le | 1949 | 35 | 8 | 1987 |
| Tran Thi Loan | | 34; 92; 35 | 9 | 1999 |
| Tran Thi Loan | | 35 | 1 | 2010 |
| Tran Thi Mai | | 49; 90; 91 | 3 | 2018 |
| Tran Thi Phuong | 1982 | 13 | 2 | 2004 |
| Tran Thi Thu Huong | 1984 | 91; 05; 68 | 4 | 2010 |
| Tran Thi Tu Trinh | | 49 | 1 | 2020 |
| Tran Thien Thanh | 1983 | 62; 60 | 1 | 2008 |
| Tran Thu Ha | | | 1 | 1992 |
| Tran Thu Le | | | 1 | 2020 |
| Tran Tin Kiet | 1950 | 93; 34 | 3 | 2000 |
| Tran Tri Dung | | 47; 42; 35 | 2 | 2020 |
| Tran Tri Kiet | 1965 | 46; 30 | 3 | 2004 |
| Tran Trinh Minh Son | | 90; 49; 91 | 2 | 2013 |
| Tran Trong Nguyen | 1972 | 60; 92 | 3 | 2001 |
| Tran Trung | | | 1 | 2020 |
| Tran Tuan Nam | 1965 | 13; 16 | 8 | 2000 |
| Tran Tuyen | | 49; 90 | 4 | 2017 |
| Tran Van An | 1965 | 54; 46 | 13 | 1993 |
| Tran Van Bang | 1974 | 35; 49 | 3 | 2006 |
| Tran Van Dung | 1955 | 20 | 4 | 1993 |
| Tran Van Hoai | 1972 | | 7 | 2006 |
| Tran Van Hung (Chan Van Khung) | | | 1 | 1994 |
| Tran Van Huyen (Tr\cfgrv an Huyen) | | | 1 | 1982 |
| Tran Van Lang | 1959 | | 1 | 1996 |
| Tran Van Nam | | 03; 15; 65 | 3 | 2019 |
| Tran Van Nghi | | 90 | 2 | 2017 |
| Tran Van Nhung | 1948 | 34; 60; 93 | 18 | 1979 |
| Tran Van Su | | 90; 49; 45 | 3 | 2020 |
| Tran Van Tan | 1976 | 30; 32; 11 | 14 | 2005 |
| Tran Van Thang | | 90; 49; 47 | 4 | 2011 |
| Tran Van Thuy | | 32 | 2 | 2018 |
| Tran Van Truong | | 47 | 1 | 2000 |
| Tran Viet Anh | | 90; 65; 49 | 6 | 2016 |
| Tran Viet Dung | 1961 | | 3 | 1990 |
| Tran Vinh Hien | | | 1 | 1974 |
| Tran Vinh Hung | 1984 | 58; 47; 35 | 3 | 2007 |
| Tran Vinh Linh | | | 3 | 2009 |
| Tran Vu Khanh | 1983 | 32; 47; 37 | 8 | 2007 |
| Tran Vu Thieu | 1941 | 90; 65; 49 | 17 | 1980 |
| Tran Vui | 1960 | 14; 22; 53 | 6 | 1989 |
| Tran Xuan Khoi | 1976 | | 1 | 2017 |
| Tran Xuan Loc | | 35; 78 | 2 | 2019 |
| Tran Xuan Sinh | 1945 | | 1 | 1997 |
| Tran Xuan Tiep | 1957 | | 1 | 1990 |
| Trieu Quynh Trang | 1984 | | 1 | 2018 |
| Trieu Van Dung | | 32 | 1 | 2020 |
| Trinh Anh Ngoc | 1963 | | 1 | 1995 |
| Trinh Cong Dieu | 1956 | 93; 47; 46 | 4 | 1989 |
| Trinh Dao Chien | 1962 | 30 | 1 | 2000 |
| Trinh Duc Tai | 1966 | 39; 58; 26 | 2 | 2012 |
| Trinh Khanh Duy | 1985 | 65; 37 | 1 | 2007 |
| Trinh Ngoc Hai | | 65; 90; 47 | 3 | 2017 |
| Trinh Ngoc Minh | | | 3 | 1985 |
| Trinh Thanh Deo | | 16 | 4 | 2012 |
| Trinh Thi Minh Hang | | 35 | 3 | 2009 |
| Trinh Tuan | | 45; 44; 33 | 5 | 2010 |
| Trinh Tuan | | 44; 33 | 3 | 2003 |
| Trinh Tuan Anh | 1971 | 92; 34; 39 | 14 | 1997 |
| Trinh Tuan Phong | 1986 | 35 | 2 | 2008 |
| Trinh Tung | | | 1 | 2015 |
| Trinh Viet Duoc | 1985 | 35; 34; 37 | 7 | 2014 |
| Trung Truong | 1989 | 35; 65; 78 | 1 | 2020 |
| Truong Chi Tin | | | 1 | 1989 |
| Truong Chi Trung | | 22; 20; 46 | 2 | 1997 |
| Truong Cong Quynh | 1981 | 16 | 6 | 2007 |
| Truong Dinh Tu | | 16 | 1 | 2020 |
| Truong Ha Hai | 1967 | 35; 65 | 1 | 2011 |
| Truong Hong Minh | | 37; 32; 53 | 1 | 2013 |
| Truong Minh Tuyen | | 47; 49; 90 | 4 | 2019 |
| Truong My Dung | | | 2 | 1987 |
| Truong Ngoc Son | | | 1 | 2002 |
| Truong Quang Bao | | 90; 49; 46 | 31 | 2003 |
| Truong Thi Hieu Hanh | | | 1 | 2020 |
| Truong Thi Hong Thanh | | 14; 13 | 1 | 2020 |
| Truong Thi Thanh Phuong | | 90 | 1 | 2020 |
| Truong Thi Thuy Duong | 1982 | 90; 49; 91 | 4 | 2010 |
| Truong Trong Nghia | | 47; 35 | 1 | 2015 |
| Truong Trung Tuyen | 1980 | 30; 74; 47 | 6 | 2003 |
| Truong Van Thuong | 1955 | 46 | 4 | 1999 |
| Truong Viet Dung | 1952 | | 1 | 2009 |
| Truong Vinh An | | 34; 47 | 4 | 2013 |
| Truong Xuan Duc Ha | 1957 | 49; 90; 34 | 38 | 1980 |
| Tu Nguyen | | | 1 | 2019 |
| Ung Ngoc Quang | | | 2 | 1994 |
| Van Duc Trung | | 13; 14 | 1 | 2019 |
| Van Duong Dinh | | 35 | 3 | 2020 |
| Vo Anh Khoa | | 35; 65; 47 | 27 | 2015 |
| Vo Bay | | | 1 | 2013 |
| Vo Dang Thao | 1954 | 30 | 8 | 1976 |
| Vo Duc Cam Hai | | 35; 65 | 1 | 2019 |
| Vo Giang Giai | 1972 | 35 | 4 | 2007 |
| Vo Hanh Nguyen | | 42 | 1 | 2020 |
| Vo Hoang Hung | | 35; 92 | 1 | 2020 |
| Vo Minh Pho | 1946 | 90; 52; 47 | 3 | 2010 |
| Vo Minh Tam | | 47; 49; 74 | 2 | 2020 |
| Vo Ngoc Thieu | | | 1 | 2020 |
| Vo Si Trong Long | 1981 | 47; 49; 90 | 6 | 2011 |
| Vo Thanh Tung | 1979 | | 2 | 2000 |
| Vo Thi Bich Khue | 1982 | 46; 47; 15 | 2 | 2018 |
| Vo Thi Le Hang | 1986 | 54; 47; 46 | 2 | 2015 |
| Vo Thi Ngoc Chau | | | 2 | 2016 |
| Vo Thi Nhat Minh | | 35; 93 | 1 | 2016 |
| Vo Thi Nhu Quynh | | 55 | 1 | 2007 |
| Vo Thi Thu Hien | 1977 | 35; 47 | 2 | 2008 |
| Vo Thi Truc Giang | | 62; 60 | 1 | 2020 |
| Vo Van Au | | 35; 47; 65 | 26 | 2017 |
| Vo Van Tai | | 62; 68 | 1 | 2020 |
| Vo Van Tan | | | 1 | 2002 |
| Vo Viet Can | | | 1 | 1980 |
| Vo Viet Tri | | 47; 35; 34 | 4 | 2018 |
| Vu A. Ha | | | 1 | 1998 |
| Vu Anh My | 1984 | 65; 60 | 2 | 2020 |
| Vu Chi Cuong | | | 1 | 2012 |
| Vu Cong Bang | 1983 | 90; 42; 47 | 2 | 2006 |
| Vu Dinh Hoa | 1956 | 05 | 11 | 1983 |
| Vu Duc Nghia | | 06 | 1 | 2004 |
| Vu Duc Thi | 1949 | 68; 05 | 32 | 1985 |
| Vu Duc Thinh | | 49; 90; 65 | 1 | 2020 |
| Vu Duc Viet | 1988 | | 2 | 2014 |
| Vu Duy Man | 1950 | | 1 | 1978 |
| Vu Duy Manh | | | 1 | 2019 |
| Vu Ha Van | 1970 | 05; 60; 11 | 144 | 1996 |
| Vu Hai Sam | 1979 | 60; 34; 92 | 2 | 2006 |
| Vu Ho | | 34 | 7 | 2012 |
| Vu Hoai An | 1960 | 30; 11; 32 | 12 | 2002 |
| Vu Hoang Linh | 1968 | 34; 93; 65 | 13 | 1996 |
| Vu Huu Nhu | 1983 | 49; 35; 65 | 5 | 2010 |
| Vu Huy Tinh | | | 1 | 1983 |
| Vu Kim Tuan | 1960 | 44; 65; 45 | 74 | 1981 |
| Vu Manh Toi | | 35; 93; 76 | 14 | 2011 |
| Vu Ngoan | | | 1 | 1965 |
| Vu Ngoc Phat | 1953 | 93; 34; 49 | 144 | 1980 |
| Vu Nguyen Son Tung | | | 1 | 2020 |
| Vu Nhat Huy | | 46; 26; 41 | 19 | 2009 |
| Vu Nhu Lan | 1951 | 90; 93 | 1 | 2008 |
| Vu Quang Bach | | 76; 35; 82 | 1 | 2020 |
| Vu Quang Hung | | 46; 47 | 1 | 2005 |
| Vu Quang Thanh (Thanh Vu) | 1986 | 13 | 6 | 2015 |
| Vu Quoc Phong | 1954 | 47; 34; 35 | 47 | 1977 |
| Vu Quoc Thanh | | 34; 41; 28 | 4 | 1990 |
| Vu The Khoi | 1972 | 57; 20; 33 | 14 | 1996 |
| Vu Thi Hien Anh | | 35 | 1 | 2020 |
| Vu Thi Hong Thanh | | 28; 26; 30 | 5 | 1999 |
| Vu Thi Huong | 1990 | 49; 46; 91 | 3 | 2017 |
| Vu Thi Mai | | 76; 35 | 2 | 2017 |
| Vu Thi Ngoc Anh | 1990 | | 1 | 2018 |
| Vu Thi Ngoc Anh | | 60 | 1 | 2020 |
| Vu Thi Ngoc Ha | | 35; 30; 34 | 7 | 2005 |
| Vu Thi Thu Ha | | 53; 57 | 1 | 2001 |
| Vu Thi Thu Huong | | | 1 | 2005 |
| Vu Thi Thuy Duong | | 35; 76 | 1 | 2020 |
| Vu Thien Ban | | | 3 | 1983 |
| Vu Tien Dung | 1981 | 47; 65; 68 | 4 | 2018 |
| Vu Tien Viet | 1955 | 37; 65 | 1 | 2007 |
| Vu Trong Luong | | 35; 34; 47 | 4 | 2016 |
| Vu Trong Tuan | 1935 | 58 | 2 | 1973 |
| Vu Tuan | 1935 | 34; 35 | 15 | 1963 |
| Vu Van Dat | 1950 | | 1 | 1982 |
| Vu Van Dong | | 90 | 2 | 2018 |
| Vu Van Khu | 1988 | 60; 92 | 1 | 2012 |
| Vu Van Khuong | | 39 | 1 | 2014 |
| Vu Van Truong | 1975 | | 1 | 2017 |
| Vu Viet Hung | 1983 | 32; 14 | 5 | 2012 |
| Vu Viet Yen | | 60; 54 | 7 | 1988 |
| Vu Vinh Quang | | 30; 35; 65 | 2 | 2004 |
| Vuong Mai Phuong | | | 3 | 2018 |
| Vuong Manh Son | | 17; 22 | 3 | 1979 |
| Vuong Nguyen Doan | | 65; 46; 47 | 2 | 2017 |
| Vuong Quan Hoang | 1972 | 60 | 3 | 2005 |
| Vuong Thi Thao Binh | 1974 | | 1 | 2009 |

**Table B.** Number of articles in the world's most prestigious journals by year

| Journal | Year | Number of articles | Author | Subject |
|---|---|---|---|---|
| *Annales scientifiques de l'École Normale Supérieure* | 1950 | 1 | Le Van Thiem | 30 |
| *Annals of Mathematics* | 1976 | 1 | Nguyen Huu Anh | |
| *Duke Mathematical Journal* | 1977 | 1 | Duong Hong Phong | 32,35,46 |
| *Duke Mathematical Journal* | 1983 | 1 | Ha Huy Khoai | |
| *Inventiones mathematicae* | 1986 | 1 | Duong Hong Phong | 42 |
| *Acta Mathematica* | 1986 | 1 | Duong Hong Phong | 42,44,47,58 |
| *Duke Mathematical Journal* | 1989 | 1 | Duong Hong Phong | 42,44 |
| *Advances in Mathematics* | 1992 | 1 | Ngo Viet Trung | |
| *Inventiones mathematicae* | 1992 | 1 | Duong Hong Phong | 35,42,47 |
| *Annals of Mathematics* | 1994 | 1 | Duong Hong Phong | 35,43 |
| *Journal für die reine und angewandte Mathematik* | 1997 | 1 | Ngo Viet Trung | |
| *Acta Mathematica* | 1997 | 1 | Duong Hong Phong | 35,47 |
| *Advances in Mathematics* | 1998 | 1 | Duong Hong Phong | 35,42,47 |
| *Annales scientifiques de l'École Normale Supérieure* | 1999 | 1 | Ngo Bao Chau | |
| *Duke Mathematical Journal* | 1999 | 1 | Ngo Bao Chau | |
| *Annals of Mathematics* | 2000 | 1 | Duong Hong Phong | 11,32 |
| *Duke Mathematical Journal* | 2000 | 1 | Vu Ha Van | 05,11 |
| *Advances in Mathematics* | 2004 | 1 | Vu Ha Van | 05,60,74 |
| *Annals of Mathematics* | 2005 | 1 | Dinh Tien Cuong | |
| *Duke Mathematical Journal* | 2005 | 1 | Vu Ha Van | 11 |
| *Annales scientifiques de l'École Normale Supérieure* | 2006 | 1 | Ngo Bao Chau | |
| *Advances in Mathematics* | 2006 | 1 | Vu Ha Van | |
| *Annals of Mathematics* | 2006 | 1 | Vu Ha Van | |
| *Duke Mathematical Journal* | 2006 | 1 | Vu Ha Van | 05 |
| *Inventiones mathematicae* | 2006 | 2 | Duong Hong Phong, Ngo Bao Chau | 32,53 |
| *Journal of the American Mathematical Society* | 2006 | 1 | Vu Ha Van | |
| *Advances in Mathematics* | 2007 | 1 | Ngo Viet Trung | 13 |
| *Journal of the American Mathematical Society* | 2007 | 1 | Vu Ha Van | |
| *Advances in Mathematics* | 2008 | 2 | Phung Ho Hai, Vu Ha Van | 11 |
| *Journal für die reine und angewandte Mathematik* | 2008 | 1 | Pham Ngoc Anh | 16,46 |
| *Annals of Mathematics* | 2008 | 1 | Ngo Bao Chau | 14,22 |
| *Inventiones mathematicae* | 2008 | 2 | Duong Hong Phong, Pham Huu Tiep | 20,32,53 |
| *Advances in Mathematics* | 2009 | 3 | Pham Hoang Hiep, Pham Huu Tiep, Vu Ha Van | 05,15,20 |
| *Annals of Mathematics* | 2009 | 1 | Vu Ha Van | 11 |
| *Duke Mathematical Journal* | 2009 | 1 | Nguyen Trong Toan (Nguyen T. Toan) | 35 |
| *Acta Mathematica* | 2009 | 1 | Dinh Tien Cuong | |
| *Advances in Mathematics* | 2010 | 3 | Dao Hai Long (Hailong Dao), Nguyen Sum, Pham Huu Tiep | 13,16,20,55 |
| *Advances in Mathematics* | 2011 | 3 | Ngo Viet Trung, Nguyen Cong Minh, Nguyen Huu Hoi (Hoi H. Nguyen), Vu Ha Van | 11 |
| *Annals of Mathematics* | 2011 | 3 | Nguyen Hoai Minh, Pham Huu Tiep | 20,46,58 |
| *Duke Mathematical Journal* | 2011 | 1 | Nguyen Hoai Minh | 26,28 |
| *Inventiones mathematicae* | 2011 | 2 | Nguyen Hoai Minh, Pham Huu Tiep | 26,46 |
| *Acta Mathematica* | 2011 | 1 | Vu Ha Van | |
| *Advances in Mathematics* | 2012 | 4 | Dinh Tien Cuong, Ngo Viet Trung, Nguyen Tien Dung, Vu Ha Van | 13,53 |
| *Journal für die reine und angewandte Mathematik* | 2012 | 1 | Pham Huu Tiep | |
| *Duke Mathematical Journal* | 2012 | 2 | Nguyen Huu Hoi (Hoi H. Nguyen), Pham Huu Tiep | 11,20 |
| *Annals of Mathematics* | 2013 | 2 | Ngo Bao Chau, Pham Huu Tiep | |
| *Advances in Mathematics* | 2014 | 1 | Pham Huu Tiep | 20 |
| *Acta Mathematica* | 2014 | 1 | Pham Hoang Hiep | 14,32 |
| *Advances in Mathematics* | 2015 | 4 | Dinh Tien Cuong, Nguyen Duy Tan, Nguyen Sum, Pham Huu Tiep | 12,20,55 |
| *Journal für die reine und angewandte Mathematik* | 2015 | 1 | Pham Huu Tiep | 20 |
| *Advances in Mathematics* | 2016 | 4 | Nguyen Trong Toan (Nguyen T. Toan), Nguyen Van Hoang (1985), Pham Huu Tiep, Vu Ha Van | 05,20,26,46,52,60 |
| *Duke Mathematical Journal* | 2016 | 1 | Nguyen Trong Toan (Nguyen T. Toan) | 35 |
| *Advances in Mathematics* | 2017 | 3 | Lu Hoang Chinh, Nguyen Duy Tan, Pham Hung Quy | 13,53 |
| *Journal für die reine und angewandte Mathematik* | 2017 | 1 | Lu Hoang Chinh | 32 |
| *Inventiones mathematicae* | 2017 | 1 | Duong Hong Phong | 35,53,58 |
| *Geometry and Topology* | 2017 | 1 | Lu Hoang Chinh | 32,53 |
| *Inventiones mathematicae* | 2018 | 2 | Dinh Tien Cuong, Pham Huu Tiep | 20 |
| *Acta Mathematica* | 2018 | 1 | Pham Huu Tiep | 20 |
| *Advances in Mathematics* | 2019 | 4 | Pham Hung Quy, Pham Huu Tiep, Vu Ha Van | 20 |
| *Journal für die reine und angewandte Mathematik* | 2019 | 1 | Duong Hong Phong | 35,58 |
| *Annals of Mathematics* | 2019 | 1 | Pham Huu Tiep | 11,20 |
| *Duke Mathematical Journal* | 2019 | 1 | Dinh Tien Cuong | |
| *Inventiones mathematicae* | 2019 | 1 | Ngo Viet Trung, Nguyen Dang Hop | |
| *Annales scientifiques de l'École Normale Supérieure* | 2020 | 1 | Lu Hoang Chinh | 32,58 |
| *Advances in Mathematics* | 2020 | 3 | Nguyen Hoai Minh, Pham Huu Tiep | 20,26,35,46,78,82 |
| *Journal für die reine und angewandte Mathematik* | 2020 | 1 | Lu Hoang Chinh | |

**Table C**. Mathematics authors: Articles in top mathematics journals

| **Author** | **Vietnamese name** | **Number of articles** | **Number of top journals** |
|---|---|---|---|
| Pham Huu Tiep | Phạm Hữu Tiệp | 21 | 6 |
| Vu Ha Van | Vũ Hà Văn | 17 | 5 |
| Duong Hong Phong | Dương Hồng Phong | 13 | 6 |
| Dinh Tien Cuong | Đinh Tiến Cường | 6 | 5 |
| Ngo Bao Chau | Ngô Bảo Châu | 6 | 4 |
| Ngo Viet Trung | Ngô Việt Chung | 6 | 3 |
| Lu Hoang Chinh | Lữ Hoàng Chinh | 5 | 4 |
| Nguyen Hoai Minh | Nguyễn Hoài Minh | 4 | 4 |
| Nguyen Trong Toan (Nguyen T. Toan) | Nguyễn Trọng Toán | 3 | 2 |
| Nguyen Huu Hoi (Hoi H. Nguyen) | Nguyễn Hữu Hội | 2 | 2 |
| Nguyen Sum | Nguyễn Sum | 2 | 1 |
| Nguyen Duy Tan | Nguyễn Duy Tân | 2 | 1 |
| Pham Hoang Hiep | Phạm Hoàng Hiệp | 2 | 2 |
| Pham Hung Quy | Phạm Hùng Quý | 2 | 1 |
| Pham Ngoc Anh | Phạm Ngọc Anh | 1 | 1 |
| Phung Ho Hai | Phùng Hồ Hải | 1 | 1 |
| Nguyen Huu Anh | Nguyễn Hữu Anh | 1 | 1 |
| Nguyen Tien Dung | Nguyễn Tiến Dũng | 1 | 1 |
| Nguyen Van Hoang (1985) | Nguyễn Văn Hoàng | 1 | 1 |
| Nguyen Cong Minh | Nguyễn Công Minh | 1 | 1 |
| Nguyen Dang Hop | Nguyễn Đăng Hợp | 1 | 1 |
| Dao Hai Long (Hailong Dao) | Đào Hải Long | 1 | 1 |
| Ha Huy Khoai | Hà Huy Khoái | 1 | 1 |
| Le Van Thiem | Lê Văn Thiêm | 1 | 1 |